\font\caps=cmcsc10 at 12pt
\font\eightrm=cmr8 at 8pt

\documentclass[12pt]{oxarticle}
\usepackage{amsmath, amssymb}

\newcommand{\ona}{{\ov \na}}

\newcommand{\dB}{\d_{\rm BRST}}

\newcommand{\GSB}{Gauge Symmetry Breaking from \sg \ to \bsg }

\newcommand{\sg}{$SU(3) \times SU(2) \times U(1)$}

\newcommand{\bsg}{$SU(3)  \times U(1)$}

\newcommand{\LT}{\LaTeX}

\newcommand{\bt}{\begin{tabular}{c}}
\newcommand{\et}{\end{tabular}}

\newcommand{\eb}{\ee\be } 
\newcommand{\ebp}{\rt.\ee\be\lt.} 
\newcommand{\bmat}{\lt ( \begin{array} }
\newcommand{\emat}{  \end{array} \rt )}

\newcommand{\oH}{{\ov H}}
\newcommand{\oP}{{\ov P}}
\newcommand{\op}{{\ov p}}
\newcommand{\oQ}{{\ov Q}}

\newcommand{\oR}{{\ov R}}
\newcommand{\oZ}{{\ov Z}}

\newcommand{\oB}{{\ov B}}

\newcommand{\ovD}{{\ov D}}

\newcommand{\oN}{{\ov N}}
\newcommand{\oJ}{{\ov J}}
\newcommand{\cd}{{\cdot}}

\newcommand{\ob}{{\ov b}}

\newcommand{\ovv}{{\ov v}}
\newcommand{\ot}{{\ov t}}

\newcommand{\oq}{{\ov \q}}
\newcommand{\oT}{{\ov T}}

\newcommand{\ovG}{{\ov G}}
\newcommand{\oK}{{\ov K}}

\newcommand{\ED}{\end{document}}
\newcommand{\of}{{\ov f}}

\newcommand{\oY}{{\ov Y}}
\newcommand{\og}{{\ov g}}
\newcommand{\oy}{{\ov \y}}

\newcommand{\oC}{{\ov C}}
\newcommand{\oL}{{\ov L}}

\newcommand{\oF}{{\ov F}}
\newcommand{\A}{{\ov A}}

\renewcommand{\a}{\alpha}	
\renewcommand{\b}{\beta}
\newcommand{\g}{\gamma}
\renewcommand{\d}{\delta}
\newcommand{\e}{\epsilon}
\newcommand{\ve}{\varepsilon}
\newcommand{\z}{\zeta}

\newcommand{\q}{\theta}

\newcommand{\m}{\mu}
	
\newcommand{\x}{\xi}

\newcommand{\s}{\sigma}

\newcommand{\y}{\psi}
\newcommand{\w}{\omega}
\newcommand{\G}{\Gamma}
\newcommand{\D}{\Delta}

\renewcommand{\P}{\Pi}	
\renewcommand{\S}{\Sigma}

\newcommand{\la}{\label}
\newcommand{\ci}{\cite}

\newcommand{\ds}{\documentstyle}	
\newcommand{\fr}{\frac}

\newcommand{\na}{\nabla}
\newcommand{\pa}{\partial}
\newcommand{\ov}{\overline}
\newcommand{\be}{\begin{equation}}
\newcommand{\ee}{\end{equation}}
\newcommand{\ba}{\begin{array}} 
\newcommand{\ea}{\end{array}}
\newcommand{\bea}{\begin{eqnarray}}
\newcommand{\eea}{\end{eqnarray}}
\newcommand{\ra}{\rightarrow}
\newcommand{\Ra}{\Rightarrow}
\newcommand{\lra}{\longrightarrow}

\newcommand{\lt}{\left}
\newcommand{\rt}{\right}

\newcommand{\ben}{\begin{enumerate}}
\newcommand{\een}{\end{enumerate}}
\newcommand{\bitem}{\begin{itemize}}
\newcommand{\eitem}{\end{itemize}}

\newcommand{\articlenumber}{\LT 3228SusyJumps}

\proofmodefalse
\begin{document}
\begin{center}

\vspace*{1in}
{\Huge  
SUSY Jumps Out of Superspace \\in 
 the \\
Supersymmetric Standard Model \\[0.5in]}

\renewcommand{\thefootnote}{\fnsymbol{footnote}}
\renewcommand{\thefootnote}{\arabic{footnote}}

{\caps John A. Dixon\footnote{cybersusy@gmail.com}\\Toronto, Canada}
\\[2in] 
{\bf Abstract}
\end{center}



The supersymmetric standard model (SSM) appears to be firmly grounded in superspace.   For example, it would be natural to assume that all the physically important composite operators can be made by combining superfields and superspace derivatives.  But even for the simplest possible, free, massless and unbroken SUSY theory in 3+1 dimensions, this is not true.

 This paper shows that there is a large set of physically important composite operators in the SSM that require explicit factors of the Grassmann odd `$\theta$' parameters of superspace.   These explicitly break superspace invariance.  These  composite operators  will be called `Outfields' here, because they are intrinsically `outside' of superspace. It is not possible to write the Outfields using only superfields and superspace derivatives.

These Outfields are present, and physically important, in all chiral SUSY theories in 3+1 dimensions.  However they are very well hidden.  They arise from a tricky mechanism involving the field equations.  The superspace violating part of the SUSY variation of an Outfield  is proportional to the field equations.  The field equations are `equivalent to zero', but they are not equal to zero, and that is why the Outfields have gone unnoticed for a long time.  

  This `field equation' property of the Outfields  means that the Outfields can be found by computing the local BRST cohomology of chiral SUSY in 3+1 dimensions.  An Outfield then (typically) consists of  the sum of two terms.  The first term is a field part which violates  the symmetry.  But the violation of the symmetry is very special: it is  proportional to the field equations.  The second term contains a Zinn source times a ghost.  The variation of the second term then cancels the variation of the first term, so that the combination is invariant under the BRST operator.  

 This explicit breaking of the initial symmetry, linked to a dependence on the Zinn sources through the field equations, is a feature that is quite rare  in the BRST cohomology of non-SUSY theories such as gauge theories and gravity.   However, for the rigid chiral SUSY theory in 3+1 dimensions, it is an essential ingredient of the cohomology. 

  In this paper, the masses are assumed to arise from the spontaneous breaking of internal symmetry with a Vacuum Expectation Value (VEV) for some scalar field.  In accord with the usual case for most chiral actions, it is assumed that  SUSY itself is not spontaneously broken by this VEV. So the present results can be utilized, for example, for the supersymmetric standard model (SSM),  where internal symmetry is spontaneously broken from $SU(2) \times U(1)$ down to  U(1), although SUSY itself is not spontaneously broken.   

 The calculation of the BRST cohomology space for these theories is performed in this paper using a spectral sequence analysis, starting with the free massless theory, and then adding interactions, and then masses.  There are nine nested differentials, with ten nested cohomology spaces.  The first three differentials relate to the free massless theory, and they establish the basic Outfields.  The next two differentials come from the coupling  terms. They give rise to many  constraint equations,  and the result is  that SUSY picks out various physical composite operators in remarkable ways that depend crucially on the details of the particle content and the couplings. The last four differentials come from the mass terms, and these impose further constraints, but they also give rise to new terms because the mass parameter is now present.

The constraints give rise to  a remarkable contrast between the cohomology of SUSY and the cohomology of gauge theories.  The cohomology of gauge theories and SUSY both start with the cohomology of the free massless theories.  Because the gauge theory cohomology does not involve the Zinn sources, the constraints that arise for the interacting or massive theories relate to the gauged Lie algebra that one starts with.  However, for the SUSY Outfields, the constraints  come from the  symmetries of the superpotential alone, not the whole action.  The most interesting solutions arise when the relevant symmetries do not extend to the whole action. These have nothing to do with the usual gauge-type symmetries of the action. These constraints  give rise to a new kind of mingling between the interactions and SUSY itself.  

 These general cohomology results are  illustrated with some  examples from a special version of the SSM, which we call the CSSM.  The CSSM requires right handed neutrinos and a Higgs singlet in addition to the usual SSM.  One can see that the CSSM  appears to have  a $ raison \; d'{\hat e}tre$ that is related to these SUSY constraints. For the Leptons, there is one  SU(2) doublet and two  SU(2) singlets.  For the Higgs, the field content is reversed.  There are two Higgs SU(2) doublets and one Higgs singlet.   This structure  gives rise to  simple  Outfields in the cohomology space for each of the Leptons.  The Quarks work the same way.

The symmetry that creates the Outfields is softly broken by the development of the VEV when spontaneous gauge symmetry breaking occurs.  This causes  the Quark and Lepton Outfields  to mix with the corresponding elementary Quark and Lepton superfields.  This means that these Outfields leave the cohomology space when the VEV turns on.


\normalsize

\section{Introduction: Composite Operators and BRST Cohomology}

\la{introchap}

The composite operators of a quantum field theory contain a great deal of information about the theory.  
A list of the physically important composite operators would be expected to satisfy the following  guidelines:
\ben
\item
Physically important composite operators should be invariant, or covariant, under the symmetries of the action, and
\item
Two physically important  composite operators that differ by the field equations should be equivalent, because the field equations should be equivalent to zero, in some sense. 
\een

At the start, it is not obvious how to put these two properties together in a sensible way. Fortunately, it has been discovered \ci{BRS,zinn,T} that the set of physically important composite operators can be naturally  organized by the construction of a nilpotent BRST  operator.  Then the BRST cohomology of the theory yields a complete and unique list of the local operators which incorporate the invariance, or the covariance, modulo the field equations. All quantum field theories have   invariances,  including invariance under the transformations of special relativity.  For any given choice of such invariances, one can construct the related nilpotent BRST operator. The BRST operator always has two parts, which match the guidelines above:\ben
\item
 A symmetry variation part,  and
\item
 A field equation part.  
\een

\subsection{ A Minimal BRST operator can point the way to New Directions in a Theory}

When a theory has a set of  symmetries, it is frequently  possible to write down a number of different BRST operators for it.  Some have more, and some fewer, of the  total set of symmetries. For example, one might decide not to include the Lorentz transformations in a given BRST operator, even though the theory has Lorentz invariance.  Such a   BRST operator for a theory would typically yield a cohomology space which contains   Lorentz covariant operators, whereas inclusion of Lorentz transformations in the BRST operator would be expected to restrict the cohomology space to operators which are Lorentz invariant. 

In this way, a minimal BRST operator which has a minimal subset of the invariances of the starting action, is likely to result in a cohomology space with operators which are covariant, rather than invariant, under the symmetries that are not included in the BRST operator. 
The resulting cohomology space is likely to be larger, and more interesting, than the more restricted cohomology space that would result from a BRST operator which incorporates all possible symmetries. 

For this reason, the BRST operator examined in the present paper is quite minimal.  Although  the symmetries of interest do include invariance under some gauge group, and under Lorentz transformations, we do not include the gauge group,  or the Lorentz transformations, in the BRST operator whose cohomology we will examine here.  The result is that we find a host of  `covariant' composite operators in the cohomology space.   If we had imposed invariance under Lorentz transformations, or an internal Lie algebra, these covariant  objects in the cohomology space would have been excluded.

So, in this paper, we work out the 
BRST cohomology for the simplest possible BRST operator for  chiral rigid SUSY  in 3+1 dimensions.  The  result is fairly complicated, and we find objects which are covariant under the Lorentz group (they have spinor indices) and under the internal Lie algebra (they have internal indices). 

The calculation of the cohomology  is accomplished using the mathematical machinery of spectral sequences.  This very detailed task constitutes the bulk of the technical part of this paper, and it is mostly contained in the Appendices.  The main body of the paper explains the results with as little technical detail as possible.  The results do suggest possible further developments, but in this paper, which is already very long,  we restrict the analysis to the cohomology, leaving interpretation and related issues for other papers.

\subsection{Plan of this Paper}

This is a  long paper, and it is hard to make it much shorter, because there are ten spaces in the spectral sequence, and it makes no sense to cut related matters  up into different papers.  Even at this length, the paper only touches on many subjects that require a fuller treatment.  The hope is that this paper introduces techniques of general applicability.
The paper is divided into eight  
  sections and nine Appendices.

Section
\ref{introchap} introduces the paper with remarks about the relation between composite operators and
BRST cohomology,  and then introduces the Sections and Appendices.

Section
\ref{brsintrochap}  introduces BRST cohomology in a simple general way, and explains how the field equations, through the Zinn sources, can lead to a violation of the initial symmetry  in certain cases. 

In Section
\ref{twoopssubsection}, we write down the BRST operator and action for the SUSY theory. There are two ways to do this (integrating the auxiliary or not), and we use them both as a check on each other.

In Section
\ref{cohomsum}, we write down the  Outfield  solutions for chiral SUSY, and compare them with results known from past work.  These Outfields  are defined by equation
 (\ref{canonicalformsupefield}), and they constitute one of the major results of this paper. To understand the composite Outfields, we need to use the fundamental Outfields 
in subsection \ref{gendiscusssect}.

  Section \ref{specsumchap} summarizes the result of the spectral sequence calculation in a summary way.  This leads to the Appendices which contain the details of the spectral sequence machinery.

Section
\ref{examplesofetoh}
 contains a discussion of the mapping from the spectral sequence to the Cohomology space, and also has a summary of the space $E_{\infty}$ and a 
detailed summary of  the spectral sequence space for the free massless case.

Section \ref{cssmexamples}
 contains some specific examples of solutions of the constraint equations for a specific version of the SSM, which we call the CSSM.

 Section  \ref{conclusionchap} is the conclusion. It comments on   the examples from the SSM which are worked out in sections
\ref{examplesofetoh} and
  \ref{cssmexamples}.  
The paper concludes with a discussion of   possible extensions of the results.

 Appendix 
\ref{derivofe3chap}
reviews some preliminary matters, including Counting operators.

 Appendix 
\ref{descripofe1subsection}
is a treatment of the   Differentials $d_0$  and  the Space   $E_{1}$  for the Massless Free  Chiral  SUSY Theory.

 Appendix 
 \ref{regchap}
is a treatment of the   Differentials $d_1$  and  the Space   $E_{2}$  for the Massless Free   Chiral  SUSY Theory.
 
 Appendix 
 \ref{d2chap}
is a treatment of the   Differentials $d_2$  and  the Space   $E_{3}$  for the Massless Free   Chiral  SUSY Theory.

Appendix 
\ref{simplifiedequas} discusses the generation and solution of the  separated   irregular equations  for   Chiral  SUSY Theory.

Appendix \ref{freemasslesscohom}
summarizes what is known and not known about the cohomology space of the free massless theory. 

 Appendix 
\ref{interactingspectralse}
is a treatment of 
the   Differentials $d_r, r=3,4$  and     Spaces   $E_{r}, r=4,5$   for the Interacting Massless Chiral SUSY Theory

 Appendix 
\ref{chapfore9}
is a treatment of the   Differentials $d_r, r=5,6,7,8$  and  Spaces   $E_{r}, r=6,7,8,9$     for the Massive Interacting  Chiral  SUSY Theory.

Appendix \ref{collapseappendix}
summarizes the situation   relating to the collapse of the three spectral sequences for the free massless, interacting and massive cases considered here.

Finally, there is a  {\bf Table of Contents}.

\section{A Simple Introduction to BRST Cohomology with Emphasis on  Three Kinds of Cohomology Terms}

\la{brsintrochap}

In this section we shall review the BRST formalism 
\ci{BRS, T} in a simple general way.  Our purpose here is to indicate how  the equations of motion fit into the BRST formalism through the Zinn sources, and how this can result in a violation of the original symmetry. 
This explanation is very important for the SUSY theory that we will look at next.

\subsection{A Simple Introduction to the BRST Operator}

\la{simpintro}

Any quantum field theory that possesses some kind of  invariance can be analyzed using BRST cohomology.
Here is how this works in the simplest case.  Suppose we have an action
depending on some bosonic fields $A^i$:
\be
{\cal A}_{\rm Invariant} = \int d^4 x \; {\cal L}_{\rm Invariant}  
\la{actionandlag}
\ee
where ${\cal L}_{\rm Invariant} = {\cal L}_{\rm Invariant} [A]$ is a local Lagrangian. Suppose that the action is invariant under some  transformation 
\be
\d_{\rm Field\; Variation} {\cal A}_{\rm Invariant} = \int d^4 x \; 
\lt \{
\d_{\rm Field\; Variation}{\cal L}_{\rm Invariant}  \rt \}
=0
\la{invovact}
\ee
where $\d_{\rm Field\; Variation}$ acts on the fields locally:
\be
A^i \ra A^i + \e \d_{\rm Field\; Variation} A^i  
\la{variationwithe}
\ee
We can always arrange for the parameter   $\e$ to be Grassmann odd, and for the  transformation $\d_{\rm Field\; Variation}$ to be Grassmann odd  and nilpotent\ci{BRS,T}:
\be
\d_{\rm Field\; Variation}^2 =0
\ee
The procedure now is to add the following new terms to the action:
\be
{\cal A}_{\rm Zinn} = \int d^4 x \; \lt \{
 \Lambda_i \d_{\rm Field\; Variation} A^i \rt \}
\la{actionandlagZinn} \ee
 Here  $\Lambda_i$ is a Grassmann odd `Zinn Justin source'   coupled \ci{zinn} to the variation 
in (\ref{variationwithe}). So now we have 
 a new action:
\be
{\cal A}_{\rm Total}
={\cal A}_{\rm Invariant}+
{\cal A}_{\rm Zinn}
\ee
 Then the following identity follows 
from (\ref{invovact}) for this new action:
\be
\int d^4 x \; 
\lt \{
\fr{\d {\cal A}_{\rm Total}}{\d A^i}  
 \fr{\d {\cal A}_{\rm Total}}{\d \Lambda_i}  
\rt \}= 0
\la{actionidentity}
\ee
It can be shown using the Feynman path integral formulation of the field theory that  (\ref{actionidentity}) is the lowest term  of the following identity:
\be
\int d^4 x \; 
\lt \{
\fr{\d  {\cal G} }{\d A^i}  
 \fr{\d  {\cal G} }{\d \Lambda_i}  
\rt \}= 0
\la{1PIidentity}
\ee
In the above, ${\cal G}[A, \Lambda]$ 
 is the one particle irreducible generating functional for the full quantum field theory.  It has a loop expansion in powers of $\hbar$: 
\be
{\cal G}[A, \Lambda]= {\cal A}_{\rm Total}
+ \hbar {\cal G}_1+ \hbar^2 
{\cal G}_2 \cdots \ee
 The one loop\footnote{Higher loops ${\cal G}_n$  are related to $\dB$ also through `canonical transformations' \ci{dixoncanontrans}.} functional ${\cal G}_1$ is governed by the cohomology\footnote{The cohomology of $\dB$ is defined below in equation 
(\ref{definofh})} of the 
BRST operator $\dB$, which  is defined by the `square root' of 
(\ref{actionidentity}):
\be
\dB = \int d^4 x \; 
\lt \{
\fr{\d {\cal A}_{\rm Total}}{\d A^i}  
 \fr{\d  }{\d \Lambda_i}  
+
\fr{\d {\cal A}_{\rm Total}}{\d \Lambda_i}  
 \fr{\d  }{\d A^i}  
\rt \}
\la{simpBRSop}
\ee
As a result of (\ref{actionidentity}), the  BRS operator in (\ref{simpBRSop}) is nilpotent
\be
\dB^2=0
\la{brsnilpot}
\ee
and this also carries through to more 
complicated\footnote{\la{noteaboutfermions} If there are Grassmann odd fermions $\y^i_{\a}$ as well as the bosons $A^i$, as happens in SUSY, the above carries through with appropriate changes. All the formulae get to be twice as big. Frequently one also needs some Zinn sources for variation of the ghosts to complete the nilpotence, which again increases the formulae in size. For   this introductory discussion  we shall imagine that we are dealing with the simplest case where the identity is simply (\ref{actionidentity}). It is easy, but cumbersome, to give this explanation for  the full SUSY theory below, but that interferes with the simplicity of the exposition of this part. } examples.

\subsection{Three Kinds of Terms in $\dB$ and the definition of the BRST cohomology space ${\cal H}$}

Given our   assumption that $\dB$ has the simple form (\ref{simpBRSop}), we  can write 
\be
\dB =  \d_{\rm Field\; Variation}
+\d_{\rm Zinn\; Variation}
+
\d_{\rm Field \; Equation} 
\la{decompofBRS}
\ee
where 
\be
\d_{\rm Field\; Variation}
 = \int d^4 x \; 
\fr{\d {\cal A}_{\rm Zinn}}{\d \Lambda_i}  
 \fr{\d  }{\d A^i}  
\la{defofBRSinv}
\ee
\be
 \d_{\rm Zinn\; Variation} 
 = \int d^4 x \; 
\fr{\d   {\cal A}_{\rm Zinn} }{\d A^i}  
 \fr{\d  }{\d \Lambda_i}  
\la{defofBRSzinn}
\ee
\be
\d_{\rm Field \; Equation}  
 = \int d^4 x \; 
\fr{\d {\cal A}_{\rm Invariant}}{\d A^i}  
 \fr{\d  }{\d \Lambda_i}  
\la{defofBRSfieldeq}
\ee
The two invariance transformations can be put together to define:
\be
\d_{\rm Total\; Variation}
=
\d_{\rm Field\; Variation}
+\d_{\rm Zinn\; Variation}
\la{defoftotalBRSinv}
\ee
and then we have
\be
\dB 
= \d_{\rm Total\; Variation}
+
\d_{\rm Field \; Equation} 
\la{decompofBRS2}
\ee

The above division of $\dB$ in (\ref{decompofBRS2}) 
 is quite general, and it applies to all kinds of $\dB$, including the one for SUSY\footnote{In general we might have the following instead of the form (\ref{decompofBRS}):
\be
\dB =  \d_{\rm Field\; Variation}
+\d_{\rm Zinn\; Variation}
+\d_{\rm Other\; Variations}
+
\d_{\rm Field \; Equation} 
\la{decompofBRSmoregeneral}
\ee
together with
\be
 \d_{\rm Total\; Variation}
=
\d_{\rm Field\; Variation}
+\d_{\rm Zinn\; Variation}
+
\d_{\rm Other\; Variations}
\ee
instead of (\ref{defoftotalBRSinv}).
For example, there might be ghosts which transform, and these could form part of $\d_{\rm Other\; Variations}
$. For our purposes here we do not need to use this more general form.}.

The local BRST cohomology space of  $ \dB$ is defined by
\be
{\cal H}=\fr{\ker \dB}{{\rm Im}\; \dB}
\la{definofh}
\ee
 in the space ${\cal P}$ of local integrated polynomials in the fields $A^i$ , sources $\Lambda_i$,  the `ghost'  parameters in 
$\d_{\rm Field\; Variation}$ and the derivative operator. 
In the above we define
\be
\ker \dB = \lt \{ {\cal P}\; {\rm such \; that}\; \dB {\cal P} =0 \rt \}
\ee
\be
{\rm Im}\; \dB = \lt \{ {\cal P}\; {\rm such \; that}\;  {\cal P} = \dB {\cal P}'  
 \; {\rm for \; some \; local } \; {\cal P}' \rt \}
\ee
  The space ${\cal P}$ is the space of  terms which are of the form
\be
{\cal P} = \int d^4 x \; {\cal P}_{\rm Local}
\la{localtointegral}
\ee
where ${\cal P}_{\rm Local}$ is a local polynomial in the fields, Zinns, ghosts and their derivatives.  So 
the Lagrangian in (\ref{actionandlag})
 is an example of ${\cal P}_{\rm Local}$ 
and the Actions in (\ref{actionandlag})  and 
(\ref{actionandlagZinn})  are examples of ${\cal P}$. 

The space $
{\cal H}$
 is a factor space, which means that for each class of elements $\lt \{ {\cal P}_i \rt \}$ which satisfy $\dB {\cal P}_i=0$,   we choose one representative ${\cal P}_1$  to be in $
{\cal H}$. Two elements ${\cal P}_1$ and ${\cal P}_2$ 
are defined to be in the same class if there exists a 
${\cal P}_3$ such that:
\be{\cal P}_1=   {\cal P}_2 +
\dB {\cal P}_3
\ee
\subsection{Three Kinds of Possible Terms in the Cohomology}
\la{threekinds}

There are three kinds of terms ${\cal P}$ that can  arise in the cohomology.  These three kinds of terms depend on  the division of the equations into the three parts in (\ref{decompofBRS}). This division is the same when one includes the complications of fermions properly. 
\ben
\item
{\bf Invariant Terms of Type ${\cal I}$ that do not use the Zinn sources:} The simplest situation occurs when we have invariants ${\cal I}$ which  satisfy:  
\be
\d_{\rm Field\; Variation} {\cal I}=0
\ee
Since ${\cal I}$ is assumed to be in the cohomology space, it also satisfies:
\be
 \d_{\rm BRST} {\cal I}
=0
\ee
Hence, because of (\ref{decompofBRS}), 
 it must also satisfy
 \be
\lt \{\d_{\rm Zinn\; Variation}
+ \d_{\rm Field \; Equation} \rt \}
 {\cal I}=0
\ee
This equation can be   satisfied in a trivial way by having ${\cal I}$  independent of $\Lambda_i$ and any other quantities than the fields, and then we have:
\be
\d_{\rm Zinn\; Variation}
 {\cal I}=0
\ee
and
\be
\d_{\rm Field \; Equation} 
 {\cal I}=0
\ee

\item
\la{noninvusesfield}

{\bf Non-invariant Terms of Type ${\cal N}$ that  use the field equations:}

Suppose that we have a term which is not invariant under the field variation operator: 
\be
\d_{\rm Field\; Variation} {\cal N}\neq0. 
\la{notinv}
\ee
Since ${\cal N}$ is assumed to be in the cohomology space, it satisfies:
\be
\d_{\rm BRST} {\cal N} =0,
\ee
This ${\cal N}$ 
cannot be  independent of $\Lambda_i$. Using  (\ref{decompofBRS}),  we see that it must also satisfy 
\be
\d_{\rm Field\; Variation} {\cal N}= -\lt (  \d_{\rm Zinn\; Variation} + \d_{\rm Field\; Equation} \rt )
{\cal N} 
\neq 0
\la{consofnoninv}
\ee
The most common way for this to happen is
\[
\d_{\rm Total\; Variation} {\cal N} 
= \lt ( 
\d_{\rm Field\; Variation}
+\d_{\rm Zinn\; Variation} \rt ) {\cal N} 
\]
\be
=
- \d_{\rm Field \; Equation} 
 {\cal N} 
\neq 0
\la{nontrivfieldeqpart}
\ee

\item
{\bf Non-invariant Terms of Type ${\cal N}$ that do not use the field equations:} 

 It is also possible for an non-invariant term in the cohomology space that satisfies (\ref{notinv}) 
and (\ref{consofnoninv}),
to satisfy:
\be
\d_{\rm Total\; Variation} {\cal N} = \d_{\rm Field \; Equation} 
 {\cal N} =0,
\la{trivfieldeqpart}
\ee 
instead of (\ref{nontrivfieldeqpart}). 
\een

It might appear that the above distinctions are not real, because the cohomology is defined only up to classes, as mentioned above.  
One can often add a boundary term 
$\dB {\cal B}$ to an invariant term ${\cal I}$ that makes it look like a non-invariant term ${\cal N}$.  However the distinction is real, because the inverse statement is not always true:

\ben\item
{\bf Invariant Terms of Type ${\cal I}$ that do not use the Zinn sources:}  If it is possible to find a boundary term $\dB{\cal B}$ that can be added to  a given non-invariant term ${\cal N}$ to make it into an invariant ${\cal I}$, then the relevant term in the cohomology  is a genuine invariant term ${\cal I}$.
This is the usual case for gauge theories or gravity where SUSY is not present.

\item
{\bf Non-invariant Terms of Type ${\cal N}$ that  use the field equations:} 
 If it is impossible to find any boundary $\dB{\cal B}$ that can be added to  a given non-invariant term ${\cal N}$ 
\ben
\item
 to make it into an invariant ${\cal I}$,
\item
 or a non-invariant term ${\cal N}$ satisfying equation 
(\ref{trivfieldeqpart}),
\een
 then the relevant term in the cohomology  is a genuine non-invariant term ${\cal N}$ of the type of the type that  uses the field equations.
These satisfy (\ref{nontrivfieldeqpart}). 
SUSY has many examples of this.

\item
{\bf Non-invariant Terms of Type ${\cal N}$ that do not use the field equations:} 
 If it is 
\ben
\item
impossible to find a boundary term $\dB{\cal B}$ that can be added to  a given non-invariant term ${\cal N}$ to make it into an invariant ${\cal I}$, but 
\item
possible to find a boundary term $\dB{\cal B}$ that can be added to  a given non-invariant term ${\cal N}$ 
to make it into a non-invariant term ${\cal N}$ satisfying the equation 
(\ref{trivfieldeqpart}), 
\een
then it is a genuine non-invariant term ${\cal N}$ of the type that does not use the field equations. This happens in several examples for SUSY, for operators that have   ghost charge ${\cal N}_{\rm Ghost} =-1$.
\een

We will see that SUSY has all three of these kinds of terms.  The second type of term, the {\bf Non-invariant Terms of Type ${\cal N}$ that use the field equations},  are the ones that generate the Outfields.  These non-invariant terms ${\cal N}$  have explicit factors of the parameters $\q,\oq$ of superspace in them.

\section{The Action and the $\dB$ operators for  Chiral SUSY with spontaneous breaking of internal symmetry}

\la{twoopssubsection}

There are two  possible $\dB$ operators for chiral SUSY in 3+1 dimensions, depending on whether one functionally integrates the auxiliary, or not.

 The formulation of $\dB$  where the {\bf auxiliary is not integrated} we will call $\d_{\rm Sup}$ (short for $\d_{\rm Superfield}$). This formulation where  the auxiliary is not integrated can be written in the usual superfield formulation, but we will need to write it out in components to solve the cohomology.  Then we can write the result again in terms of superfields, except that we will find there is some tricky business in doing that, so that manifest supersymmetry is not present for the cohomology. 

 The formulation of $\dB$  where the {\bf auxiliary is integrated} we will call $\d_{\rm Phys}$ (short for 
$\d_{\rm Physical}$). This formulation uses only the physical scalar and spinor particles. However, the superfields are lost at the beginning, because they require auxiliary fields.  We will see that  the cohomology of $\d_{\rm Phys}$ generates its own set of objects that behave much like superfields.  But again manifest supersymmetry is not present for the cohomology.

We will find that the two versions  
$ \d_{\rm Phys}$ and $ \d_{\rm Sup}$ have isomophic cohomology spaces:
\be
 {\cal H}_{\rm Sup}
\approx
{\cal H}_{\rm Phys}
\ee

This is easy to prove in fact, because they 
  yield exactly the same spectral sequence $E_{r}$, and so, as we shall explain later,    
\be
E_{\infty \;{\rm Sup}}= E_{\infty \;{\rm Phys}} \approx {\cal H}_{\rm Sup}
\approx
{\cal H}_{\rm Phys}
\ee
Now we shall set out the two operators and the corresponding actions.

\subsection{The Superfield Formulation}

\la{supformsec}

\subsubsection{Superfield Version of the Superfield Formulation}

We start with the superfield approach 
\ci{WB,superspace,west}, which has the following action:
\[
{\cal A}_{\rm Superspace}
= \int d^4 x \;d^4 \q \lt \{
 {\widehat A}^i
{\widehat \A}_i \rt \}
\]\[
+ \int d^4 x \;d^2 \q \lt \{  \fr{1}{3}g_{ijk} {\widehat A}^i{\widehat A}^j
{\widehat A}^k + m^2 g_k {\widehat A}^k
+ {\widehat \Lambda}_k \d_{\rm SS}  {\widehat A}^k
\rt \}
\]\be
+ \int d^4 x \;d^2 {\ov \q} \lt \{  
\fr{1}{3} {\ov g}^{ijk} {\widehat \A}_i{\widehat \A}_j
{\widehat \A}_k
+ m^2 {\ov g}^k {\widehat \A}_k
+ {\widehat {\ov \Lambda}}^k \d_{\rm SS}  {\widehat \A}_k
\rt \}
+ Z_{\a \dot \b} C^{\a} \oC^{\dot \b}
\la{supspaceaction}
\ee
The Matter superfields have the component forms:
\be
{\widehat  A}^{i}  = A^i  +
\q^{\a} 
\y^i_{\a }  
+ \fr{1}{2} \q^{\g} \q_{\g} 
F^i 
\la{startsup}
\ee
and the Zinn-Justin superfields have the component forms:\be
{\widehat  \Lambda}_{i}  = \Lambda_{i}   +
\q^{\a} 
Y_{i\a }  
+ \fr{1}{2} \q^{\g} \q_{\g} 
\G_i  
\la{startlamsup}
\ee

Here the supersymmety variation $\d_{\rm SS} {\widehat \A}_k=( C Q + \oC \oQ + \x \pa){\widehat \A}_k$.
Using standard methods\footnote{We   drop a term  $\lt \{ \int d^4 x  d^2 \q 
{\widehat \Lambda}_i \pa_{\a \dot \b} {\widehat A}^i \rt \}\fr{\pa}{\pa Z_{\a \dot \b} }+ *$ in this operator, and its analogues. This causes no problems.} we can derive the BRST operator in superspace form: 
\[
\d_{\rm Sup} =
\int d^4 x\;d^2 \q
\lt \{
( C Q + \oC \oQ + \x \pa) {\widehat \Lambda}_i +
\ovD^2 {\widehat \A}_i +   g_{ijk}  {\widehat A}^j
{\widehat A}^k  + m^2 g_k  \rt \}
\fr{\d  }{\d {\widehat \Lambda}_i} 
\]
\be
+\int d^4 x\;d^2 \q
( C Q + \oC \oQ + \x \pa)  {\widehat A}^i 
\fr{\d  }{\d {\widehat A}^i} 
+*
-  C^{\a} \oC^{\dot \b}
\x_{\a \dot \b}^{\dag} 
\la{transoflam}
\ee
The action and the BRST operator $\d_{\rm Sup}$ satisfy:
\be
 \d_{\rm Sup}^2
= \d_{\rm Sup} {\cal A}_{\rm Superspace}
=0
\ee
In this paper, it will be assumed \ci{Oraiff}
that there is a vacuum expectation value
\be
<A^i> = m v^i
\ee
which satisfies:
\be
 g_{ijk} v^j v^k + g_i =0
\ee
Then the shift:
\be
A^i \ra m v^i + A^i
\ee
serves to remove the $m^2 { g}_k {\widehat A}^k$ terms from the action and the operator $ \d_{\rm Sup}$.

\subsubsection{Component Version of the Superfield Formulation}

After this shift, the action  can be written in components as:
\[
{\cal A}_{\rm Superspace}
= \int d^4 x \;\lt \{
 F^i
\ov F_i  + g_{ijk} F^i \lt ( 2  m v^j A^k  + A^j  A^k\rt )
\rt.
\]
\[
\lt.
  +{\ov g}^{ijk} \ov F_i  
\lt ( 2 m \ov v_j \A_k+ \A_j \A_k\rt )
\rt.
\]
\[
  + g_{ijk}  \y^{i \a} \y^j_{\a} \lt ( m v^k +A^k \rt )   
  +
 {\ov g}^{ijk} {\ov \y}_{i }^{\dot \a} {\ov \y}_{j  \dot \a} 
\lt ( m \ov v_k + {\ov A}_k \rt )
\]
\[
-
\y^{i }_{\a  }     \pa^{\a \dot \b  }   
{\ov \y}_{i \dot \b}
+
\fr{1}{2} 
\pa_{ \a \dot \b  }   A^{i}    \pa^{\a \dot \b  } {\ov  A}_{k} 
+
\Lambda_i  \pa_{\a \dot \b}   \y^{i \a} {\oC}^{\dot   \b} 
+
{\ov \Lambda}^i  \pa_{\a \dot \b}    {\ov \y}_{i \dot  \b}  
{  C}^{\a} 
\]
\be
\lt.
+
\G_i    \y^{i}_{  \b} {  C}^{  \b} 
+
{\ov \G}^i  {\ov \y}_{i \dot  \b} {\ov C}^{\dot   \b}  
+
 Y_{i}^{ \a}  \lt (
 \pa_{ \a \dot \b }  A^{j} {\ov C}^{\dot \b}   
+ F^i C_{\a} 
\rt )
+
 \ov{Y}^{i \dot \b} 
\lt (
 \pa_{ \a \dot \b }  {\ov A}_{j} 
{  C}^{  \a}  
+ \ov F_i \oC_{\dot \b} 
\rt )
\rt \}
+ Z_{\a \dot \b} C^{\a} \oC^{\dot \b}
\la{Aphysical1}
\ee

The operator in components, using the language above in subsection (\ref{threekinds})  is:
\be
\d_{\rm Sup} =
\d_{\rm Field\; Variation} 
+
\d_{\rm Other\; Variation}
+
\d_{\rm Zinn\; Variation} 
+
\d_{\rm Field\; Equations} 
\la{supdivided}
\ee
where
\be
\d_{\rm Field\; Variation} =
\int d^4 x\;
  \y^{i}_{  \b} {C}^{  \b} 
\fr{\d  }{\d A^i} 
+
\int d^4 x\;
\lt \{
\pa_{ \a \dot \b }  A^{i} {\ov C}^{\dot \b}  
+
C_{\a}  F^i 
\rt \}
\fr{\d  }{\d \y_{\a}^i} 
+
\int d^4 x\;
\pa_{\a \dot \b} \y^{i \a} \oC^{\dot \b}
\fr{\d  }{\d F^i} 
\eb
+
\x_{\a \dot \b}  \pa^{\a \dot \b} 
\la{fieldvarsup}
\ee
and
\be
\d_{\rm Other\; Variation}
=
-  C^{\a} \oC^{\dot \b}\x_{\a \dot \b}^{\dag} 
\ee

\[
\d_{\rm Zinn\; Variation} 
+
\d_{\rm Field\; Equations} 
\]
\[=
\int d^4 x\;
\lt \{
 - \fr{1}{2} \pa_{ \a \dot \b  }       \pa^{ \a \dot \b  }        {\ov  A}_{i} 
-
\pa_{ \a \dot \b } Y_{i}^{ \a}    {\ov C}^{\dot \b}   
   + g_{ijk} \lt [
   2  A^{j}   
F^k - \y^{j \a} 
\y^{k}_{ \a}  
 \rt ]
 +
2 m g_{ijk} v^j  F^{k}     
\rt \}
 \fr{\d}{\d \G_i } 
\]
\[
+
\int d^4 x\;
\lt (
-
  \pa^{\a \dot \b  }   
{\ov \y}_{i   \dot \b}
+
\pa_{\a \dot \b} \Lambda_i      {\oC}^{\dot   \b} 
 +
2 g_{ijk}  \y^{j \a} A^k    
 +
2 m g_{ijk}  \y^{j \a} v^k    
-
\G_i  
 {C}^{  \a}
\rt )
 \fr{\d  }{\d Y_{i}^{ \a}} 
+ *
\]
\be
+
\int d^4 x\;
\lt (
\ov F_i  
-
\ov G_i
\rt )
 \fr{\d  }{\d \Lambda_{i}} 
+ *
\la{thebigoperatornotint}
\ee
where we use the abbreviation:
\be
\ov G_i =  - \lt (    
 { g}_{ijk} A^j A^k   +
2 m { g}_{ijk} v^j A^k    +
{ Y}_i^{  \b} { C}_{ \b } 
\rt ).
\la{termfromauxintfornotint}
\ee

It would be natural to add rigid internal symmetry transformations, or local internal symmetry transformations coupled to Yang-Mills supersymmetry, to the above.  Those additions   will not be considered in this paper because we already have enough complexity for the time being.  Also it should be noted that one can miss interesting developments if one is too restrictive in choosing the $\dB$ that one is looking at.

\subsection{The Physical Formulation}

\la{physformsec}

\subsubsection{Integration of the auxiliary fields}
The auxiliary fields in a supersymmetric theory are not physical. They appear in linear and quadratic terms in the action.  They do not propagate, because there are no derivatives in the quadratic terms $\int d^4 x F^i \oF_i$ containing them in the action. 
 As a result, it is possible to integrate them out of the theory in an exact non-perturbative way, by simply completing the square, performing a shift $F^i + G^i\Ra F^i$ in the  $F^i$ variable,  and then performing the Gaussian integration of $F^i$ in the Feynman path integral. The $F$ dependent terms disappear into a multiplicative constant and we are left with 
a new  quadratic term made of the  propagating fields (and sources). In the above case this new quadratic term is
\be
\int d^4 x 
\lt \{ \lt ( F^i + G^i\rt ) \lt ( \ov F_i + \ovG_i \rt )  
- 
\lt (   G^i\rt ) \lt (   \ovG_i \rt )  \rt \}
\Ra
- \int d^4 x 
G^i \ovG_i
\la{stufffromintf}
\ee
where $G^i$ is defined by (\ref{termfromauxint}). This new term  is the square of the term that multiplied the linear term in the auxiliary. 
 When this integration is done, supersymmetry ceases to be manifest.  For example, superfields  like
(\ref{startsup})
cease to be applicable, because they use the auxiliary field $F^i $.
The supersymmetry is still in the theory however, and one way to understand it  is by constructing the nilpotent  BRST operator 
$\d_{\rm Phys}$ below in (\ref{thebigoperator}),  and then solving for its local BRST cohomology.

Since for this case we do  not have the auxiliary fields in the generating functionals, we do not include the source $\Lambda_i$ for the variation  of the auxiliary either.  

\subsubsection{Component Version of the Physical Formulation}

This results in  the following action
\[
{\cal A}_{\rm Phys}
\]
\[
= \int d^4 x \;\lt \{
- G^i
\rt.
\ov G_i  -
\y^{i }_{\a  }     \pa^{\a \dot \b  }   
{\ov \y}_{i \dot \b}
  - g_{ijk}  \y^{i \a} \y^j_{\a} \lt ( m v^k +A^k \rt )   
  -
 {\ov g}^{ijk} {\ov \y}_{i }^{\dot \a} {\ov \y}_{j  \dot \a} 
\lt ( m \ov v_k + {\ov A}_k \rt )
\]
\be
+
\fr{1}{2} 
\pa_{ \a \dot \b  }   A^{i}    \pa^{\a \dot \b  } {\ov  A}_{k} 
+
\G_i    \y^{i}_{  \b} {  c}^{  \b} 
+
{\ov \G}^i  {\ov \y}_{i \dot  \b} {\ov C}^{\dot   \b}  
\lt.
+
 Y_{i}^{ \a}   \pa_{ \a \dot \b }  A^{j} {\ov C}^{\dot \b}   
+
 \ov{Y}^{i \dot \b}  \pa_{ \a \dot \b }  {\ov A}_{j} 
{  C}^{  \a}  
\rt \}
+  C^{\a} \oC^{\dot \b}
Z_{\a \dot \b}
\la{Aphysical2}
\ee
and the derived operator is
\[
\d_{\rm Phys} =
\int d^4 x\;
  \y^{i}_{  \b} {C}^{  \b} 
\fr{\d  }{\d A^i} 
+
\int d^4 x\;
\lt \{
\pa_{ \a \dot \b }  A^{i} {\ov C}^{\dot \b}  
+
C_{\a}  G^i 
\rt \}
\fr{\d  }{\d \y_{\a}^i} 
+
\int d^4 x\;
\]
\[
\lt \{
 - \fr{1}{2} \pa_{ \a \dot \b  }       \pa^{ \a \dot \b  }        {\ov  A}_{i} 
-
\pa_{ \a \dot \b } Y_{i}^{ \a}    {\ov C}^{\dot \b}   
   + g_{ijk} \lt [
   2  A^{j}   
G^k - \y^{j \a} 
\y^{k}_{ \a}  
 \rt ]
 +
2 m g_{ijk} v^j  G^{k}     
\rt \}
 \fr{\d}{\d \G_i } 
\]
\be
+
\int d^4 x\;
\lt (
-
  \pa^{\a \dot \b  }   
{\ov \y}_{i   \dot \b}
 +
2 g_{ijk}  \y^{j \a} A^k    
 +
2 m g_{ijk}  \y^{j \a} v^k    
-
\G_i  
 {C}^{  \a}
\rt )
 \fr{\d  }{\d Y_{i}^{ \a}} 
+ *
-  C^{\a} \oC^{\dot \b}
\x_{\a \dot \b}^{\dag} 
+
\x^{\a \dot \b}  \pa_{\a \dot \b} 
\la{thebigoperator}
\ee
The composite field $G^i$ is the same as the value of the equation of motion above in 
(\ref{termfromauxintfornotint}), with complex conjugate:
\be
G^i =  - \lt (    
 {\ov g}^{ijk} {\ov A}_j  {\ov A}_k  +
2
m {\ov g}^{ijk} {\ov A}_j  {\ov v}_k  +
{\ov Y}^{i \dot \b} {\ov C}_{\dot \b } 
\rt ).
\la{termfromauxint}
\ee
and we have
\be
 \d_{\rm Phys}^2 
= \d_{\rm Phys} 
 {\cal A}_{\rm Phys}
=0
\ee

\section{Quick Summary of the Old and New Results for the Cohomology Space
${\cal H}$ for Chiral SUSY in 3+1 Dimensions}

\la{cohomsum}

\subsection{Some Old Results and Some New Results}
\la{oldandnewsubsection}

In other papers \ci{holes,dixmin,dixminram}, it was shown that the BRS cohomology of the chiral superfield, without the Zinn terms, contained  terms\footnote{There are also terms with derivatives described in those papers, and the spin must be maximized, as shown in those papers.  In addition there are non-chiral terms that we will discuss briefly later.}  that look like

\be
\int d^4 x d^2 {\ov \q} \; {\widehat R}_{(\a_1 \cdots \a_{p})}
= \int d^4 x \; d^2 \oq T^{(j_1\cdots j_m)}\;
{\widehat \A}_{j_1}\cdots {\widehat \A}_{j_m} C_{\a_1}\cdots C_{\a_p} \in {\cal H} 
\la{canonicalfornozinn}
\ee
The antichiral superspace integral in (\ref{canonicalfornozinn}) just picks out supersymmetric invariant $\ov F$ type terms.  In this paper, we find that the generalization\footnote{Again there  are also terms with derivatives, and the spin must be maximized, as proved below.}  of the result (\ref{canonicalfornozinn}) to the case where the Zinn sources are included takes the very similar form:
\be
\int d^4 x d^2 {\ov \q} \; {\widehat R}_{(\a_1 \cdots \a_{n+p})}
= \int d^4 x d^2 {\ov \q} \;T_{[i_1\cdots i_n]}^{(j_1\cdots j_m)}{\widehat \A} _{j_1}\cdots {\widehat \A}_{j_m}
{\widehat \y}^{i_1}_{(\a_1} \cdots 
{\widehat \y}^{i_n}_{\a_n}
C_{\a_{n+1}}\cdots C_{\a_{n+p})} 
\la{canonicalformsupefield}
\ee
The only difference from (\ref{canonicalfornozinn}) is the addition of the terms 
${\widehat \y}^{i_1}_{(\a_1} \cdots 
{\widehat \y}^{i_n}_{\a_n}$ in the middle of
(\ref{canonicalformsupefield}). 
These are the terms which make (\ref{canonicalformsupefield}) into an Outfield, because they contain explicit factors of $\q_{\a}$. 
 These are examples of the 
{\bf Non-invariant Terms of Type ${\cal N}$ that  use the field equations} 
referred to in section (\ref{threekinds}).
  This is one of the fundamental results of the present paper. It is established by using spectral sequences, as will be explained below.

 Our first task will be to explain what is meant by the   ${\widehat \y}^{i}_{\a}$
in (\ref{canonicalformsupefield}).
These are   antichiral pseudosuperfields, and sometimes we call them dotspinors or fundamental Outfields too.
They will be explained  in subsection \ref{gendiscusssect}.
  The expression inside the integral in 
(\ref{canonicalformsupefield}) transforms under the appropriate $\dB$ like an antichiral superfield:
\be
\d_{\rm Phys}  {\widehat R}_{(\a_1 \cdots \a_{n+p})}=(C Q + \oC \oQ + \x \pa) 
{\widehat R}_{(\a_1 \cdots \a_{n+p})}
\ee
\be
D_{\a} {\widehat R}_{(\a_1 \cdots \a_{n+p})}=0
\ee

The superspace integral in (\ref{canonicalformsupefield}) just picks out pseudosupersymmetric invariant $\ov F$ type terms.  
The highest component transforms as a total derivative, and the integral of that highest component generates a class  of   the integrated cohomology space ${\cal H}$.
\be
 \int d^4 x \; d^2  \oq   \; {\widehat R}_{(\a_1 \cdots \a_{n+p})} \equiv 
\int d^4 x \; \lt (  \ovD^2 {\widehat R}_{(\a_1 \cdots \a_{n+p})} \rt )_| \in {\cal H}
\ee
The ghost number of these terms in the action ranges from zero to positive infinity.  The complex conjugate terms are also in ${\cal H} $ for both
(\ref{canonicalfornozinn}) and (\ref{canonicalformsupefield}). 
  
The general form (\ref{canonicalformsupefield}) is only  true for the free massless theory.  For the case where there are interactions, there are constraints which remove some of the  terms in 
(\ref{canonicalformsupefield}), or else combine them together. 
When masses are added in addition, there are new terms here proportional to masses, and there are new constraints too.  The details of the constraints will be derived using the spectral sequence.

\subsection{The Fundamental Pseudosuperfields}

\la{gendiscusssect}

In the next subsubsections we explain what is meant by these expressions when used in (\ref{canonicalformsupefield}). 
It should be mentioned that this is one of the tricky steps when using the spectral sequence.  The forms given in the next subsections are not easy to find, and the best way to find them seems to be to simply guess what they have to be, given the information that one has about them from the spectral sequence, which will be explained below.

\subsubsection{ Fundamental Expressions in the superfield approach, with  $\d_{\rm Sup}$}

\la{supsubsect}

In the superfield approach, the fundamental expressions to be used in 
(\ref{canonicalformsupefield}), 
 for construction of ${\cal H}$  are made from 
the superfields that we start with: 
 \be
{\widehat  A}^i \lra  {\widehat  A}^i_{\rm Sup} = {\widehat  A}^i
\la{supbosonnew}
\ee
\be
  {\widehat \oy}_{\; i\dot \a} \lra 
 {\widehat \oy}_{{\rm Sup}\; i\dot \a} = {\widehat \Lambda}_i \oC_{\dot \a} + \ovD^2 \lt ( 
{\widehat \A}_i \oq_{\dot \a} \rt )
\equiv 
{\widehat \Lambda}_i \oC_{\dot \a} + 
\lt ( \ovD^2  
{\widehat \A}_i  \rt ) \oq_{\dot \a}+ 2 \ovD_{\dot \a}  
{\widehat \A}_i  
\la{supfermion}
\ee

Both of these are chiral:
 \be
{\ov D}_{\dot \b} {\widehat  A}^i  =0
\ee
\be
{\ov D}_{\dot \b} {\widehat \oy}_{{\rm Sup}\; i\dot \a} = 0
\ee

  The first expression ${\widehat  A}^i$ is the usual superfield. 
 The second pseudosuperfield (\ref{supfermion}) is new.  It is constructed out of the  Zinn field and a rather strange combination of the chiral projector on ${\widehat \A}_i$ with an explicit factor of the superspace coordinate $\oq_{\dot \a}$. It is natural to wonder why an explicit factor of $\oq_{\dot \a}$ arises here.  This happens because the combination of the variation of ${\widehat \Lambda}_i$, which brings in the equation of motion term $\ovD^2  
{\widehat \A}_i$ is accompanied by an explicit factor of the ghost $\oC_{\dot \a}$, and this gets compensated by the factor of $\oq_{\dot \a}$, as we shall see in detail below.

It is remarkable however that the cohomology is giving rise to these explicit factors of $\oq_{\dot \a}$, because they do  remove the manifest superspace invariance of the theory, even in this approach that strives to keep the superspace invariance by keeping the superfields, and not integrating the auxiliary field.
Of course, this explicit superspace breaking does not affect the action that we started with, because these objects ${\widehat \oy}_{{\rm Sup}\; i\dot \a} $ do not occur in the action. That raises a question of course--can we put them into a new action of some kind?  That question will be dealt with in another paper.

The transformations induced by $\d_{\rm Sup} $ are summarized\footnote{The equation 
(\ref{freedotspieqsup}) is valid only for the free massless theory. The more general case is a little more complicated.  See equation (\ref{bigtransfordotspinsup}).}   by the following equations:
\be
\d_{\rm Sup}   {\widehat  A}_{\rm Sup}^{i}(x)=   \d_{\rm SS} {\widehat  A}_{\rm Sup}^{i}(x) 
\la{notmuchere}
\ee
\be
\d_{\rm Sup}  {\widehat  {\oy}}_{{\rm Sup}\;i \dot \a}(x)=   \d_{\rm SS} {\widehat  \oy}_{{\rm Sup}\;i \dot \a }(x)  
\la{freedotspieqsup}
\ee 
The latter equation is easy to verify explicitly, and this is important.  One gets
\[
\d_{\rm Sup}  {\widehat  {\oy}}_{{\rm Sup}\;i \dot \a}(x)=    (C Q + \oC \oQ + \x \pa) {\widehat \Lambda}_i \oC_{\dot \a}  +
 \ovD^2  {\widehat \A}_i \oC_{\dot \a} + \ovD^2 \lt \{ 
\lt [ (C Q + \oC \oQ + \x \pa)  {\widehat \A}_i \rt ] \oq_{\dot \a} \rt \}
\]\[
=    (C Q + \oC \oQ + \x \pa) {\widehat \Lambda}_i \oC_{\dot \a} 
 +
 \ovD^2  {\widehat \A}_i \oC_{\dot \a} +
\ovD^2 
 \lt \{
(C Q + \oC \oQ + \x \pa)
 \lt ( 
 {\widehat \A}_i \oq_{\dot \a} \rt )  
-  {\widehat \A}_i \oC_{\dot \a}  
\rt \}
\]\be
=    (C Q + \oC \oQ + \x \pa) 
\lt \{
{\widehat \Lambda}_i \oC_{\dot \a} 
 +
\ovD^2
 \lt ( 
 {\widehat \A}_i \oq_{\dot \a} \rt )  
 \rt \}  
=    (C Q + \oC \oQ + \x \pa) 
{\widehat  {\oy}}_{{\rm Sup}\;i \dot \a}(x)
\la{fundderviation}
\ee

The chiral derivatives $ D_{\a} $ and $ \ovD_{\dot \a} $ both anticommute with both the $Q$ and $\oQ$ superspace translations.  This property is used in going from $  \ovD^2  
  (C Q + \oC \oQ + \x \pa) $ to $ (C Q + \oC \oQ + \x \pa) \ovD^2  
   $ in the above derivation  
(\ref{fundderviation}).  So this form is a simple example of a 
{\bf Non-invariant Term of Type ${\cal N}$ that  uses the field equations} 
referred to in section (\ref{threekinds}).  It generates the others by multiplication. 

\subsubsection{Superspace translation operator}
\la{suopstranssubsec}

We use the superspace translation operator in the previous section.  It is  
\be
\d_{\rm SS} =
C^{\a} Q_{\a} 
+ \ov C^{\dot \a} \ov Q_{\dot \a} 
+ \x^{\a \dot \a} \pa_{\a \dot \a} 
\la{defofsuptrans}
\ee
 The supertranslations are: 
\be
 Q_{\a} = \fr{\pa}{\pa \q^{\a}} -  \fr{1}{2} \pa_{\a \dot \b} \oq^{\dot \b} ,  \oQ_{\dot \a} = \fr{\pa}{\pa \oq^{\dot \a}} -  \fr{1}{2} \pa_{\b \dot \a} \q^{\b}
\ee
The chiral derivatives in this notation are: 
\be
 D_{\a} = \fr{\pa}{\pa \q^{\a}} +  \fr{1}{2} \pa_{\a \dot \b} \oq^{\dot \b} ,  \ovD_{\dot \a} = \fr{\pa}{\pa \oq^{\dot \a}} +  \fr{1}{2} \pa_{\b \dot \a} \q^{\b}
\la{chiralderivdef}
\ee

 The  relations 
(\ref{notmuchere})
and (\ref{freedotspieqsup})  mean that the effect of  $\d_{\rm Sup}$ on either of these particular combinations is the same as the effect of the superspace operator $\d_{\rm SS}$.   Of course ${\widehat  A}_{\rm Sup}^{i}(x)$ here is simply a superfield. But $ {\widehat  {\oy}}_{{\rm Sup}\;i \dot \a}(x)$ is not a superfield in the conventional sense, though it acts like one for some purposes.

\subsubsection{Fundamental Pseudosuperfields in the  physical  approach, with $\d_{\rm Phys}$}

In the physical approach, the fundamental pseudosuperfields for construction of ${\cal H}$ are made from: 
\be
{\widehat  A}^{i}(x)\lra   {\widehat  A}_{\rm Phys}^{i}(x) 
\la{supbosonphys}
\ee
\be
{\widehat  {\ov \y}}_{ i \dot \a}(x)
\lra{\widehat  {\ov \y}}_{{\rm Phys}\; i \dot \a}(x)
\la{supfermionphys}
 \ee
 
As in the superfield approach, both of these are chiral:
 \be
{\ov D}_{\dot \b} {\widehat  A}_{\rm Phys}^{i}(x)  =0
\ee
\be
{\ov D}_{\dot \b} 
{\widehat  {\ov \y}}_{{\rm Phys}\; i \dot \a}(x) = 0
\ee

Recall that in the physical approach, both the  auxiliary $F^i$ and the Zinn source for its variation $\Lambda_i$ are gone. However the theory constructs new superfields out of the remaining fields and Zinns as follows:

The first is the  {\bf Physical chiral scalar pseudosuperfield}.    It has the form:
\be
{\widehat  A}_{\rm Phys}^{i}(x) = A^i(y) +
\q^{\a} 
\y^i_{\a } (y) 
+ \fr{1}{2} \q^{\g} \q_{\g} 
G^i(x)
\la{firstcompsupsup}
\ee
 where the translated spacetime variable is $y_{\a \dot \b}  = x_{\a \dot \b} + \fr{1}{2} \q_{\a } \ov \q_{\dot\b} 
$. 
The only difference between this and the  superfield
that appears in the superfield approach is the replacement of $G^i \ra F^i$:
\be
{\widehat  A}^{i}(x) = A^i(y) +
\q^{\a} 
\y^i_{\a } (y) 
+ \fr{1}{2} \q^{\g} \q_{\g} 
F^i(x)
\ee
The second is the  {\bf Physical chiral dotted spinor pseudosuperfield}: 
\be
 {\widehat  {\ov \y}}_{{\rm Phys}\; i \dot \a}(x)
 = \ov \y_{i \dot \a }(y) 
+
\q^{\b} 
\pa_{\b \dot \a} \A_i(y) 
+
\q^{\b} 
  Y_{i  \b}(y) \ov C_{\dot \a} 
- \fr{1}{2} \q^{\g} \q_{\g} 
\G_{i}(x)  
\ov C_{\dot \a} 
\la{secondcompsup}\ee

This should be compared to  the related but rather different   superfield that appears in the superfield approach in equation 
(\ref{supfermion}):
\be
{\widehat \oy}_{{\rm Sup}\; i\dot \a} = {\widehat \Lambda}_i \oC_{\dot \a} + \ovD^2 \lt ( 
{\widehat \A}_i \oq_{\dot \a} \rt )
\eb
 = \lt \{
\Lambda_{i}(y)   
+
\q^{\b} 
  Y_{i  \b}(y)  
- \fr{1}{2} \q^{\g} \q_{\g} 
\G_{i}(x)  \rt \}
\ov C_{\dot \a} 
+ \ovD^2 \lt ( 
{\widehat \A}_i \oq_{\dot \a} \rt )
\la{secondcompsup2}
\ee
The expressions (\ref{secondcompsup}) and 
(\ref{secondcompsup2}) perform the same role for the two different formulations with $\d_{\rm Phys}$ and $\d_{\rm Sup}$, but they are quite different in appearance because the two formulations are related in a complicated way.

The transformations induced by $\d_{\rm Phys} $ are summarized\footnote{The equation 
(\ref{freedotspieq}) is valid only for the free massless theory. The more general case is a little more complicated.  See equation (\ref{bigtransfordotspin}).}
 by the following equations:
\be
\d_{\rm Phys}   {\widehat  A}_{\rm Phys}^{i}(x)=   \d_{\rm SS} {\widehat  A}_{\rm Phys}^{i}(x) 
\ee
\be
\d_{\rm Phys}  {\widehat  {\oy}}_{{\rm Phys}\;i \dot \a}(x)=   \d_{\rm SS} {\widehat  \oy}_{{\rm Phys }\;i \dot \a }(x)  
\la{freedotspieq}
\ee 
 where the superspace operator is $\d_{\rm SS} =
C^{\a} Q_{\a} 
+ \ov C^{\dot \a} \ov Q_{\dot \a} + \x^{\a \dot \a} \pa_{\a \dot \a} 
$ with the usual definitions above in subsubsection
 \ref{supsubsect}.  These relations mean that the effect of  $\d_{\rm Phys}$ on either of these particular combinations is the same as the effect of the superspace operator $\d_{\rm SS}$.  So they act like superfields for some purposes.

It is not so obvious in this notation that the superspace symmetry is being violated, since we have already lost superspace by integrating the auxiliary field.  This formulation does not seem so useful as the superspace formulation, but it does help to confirm the results.

\subsection{Interactions and Masses in the 
  Chiral SUSY Theory}

\la{sumofthreespaces}

In this subsection we want to briefly discuss what happens when we go beyond the free massless case to the case where the theory is interacting and massive.  

The following formulae can be derived from the above forms in subsection \ref{gendiscusssect}, and they are   useful for writing down the transformation of the full form of the Outfields when there are masses and interactions in the theory:
\ben
\item
 In general, the transformation of ${\widehat  {\oy}}_{{\rm Sup }\;i \dot \a}(x)$ under the action of  
$ \d_{\rm Phys} $ is: 
\be
\d_{\rm Sup }   {\widehat  {\oy}}_{{\rm Sup }\;i \dot \a}(x)=   \d_{\rm SS} {\widehat  \oy}_{{\rm Sup  }\;i \dot \a }(x)  -
 g_{ijk} 
  {\widehat  A}_{\rm Sup }^j {\widehat  A}_{\rm Sup }^k 
\ov C_{\dot \a} 
-
2m g_{ijk} 
  v^j {\widehat  A}_{\rm Sup }^k 
\ov C_{\dot \a} 
\la{bigtransfordotspinsup}
\ee 
This is quite easy to see from the transformation of ${\widehat  \Lambda}_i$ and the form of ${\widehat  {\oy}}_{{\rm Sup }\;i \dot \a}(x)$. 

\item

 In general, the transformation of ${\widehat  {\oy}}_{{\rm Phys}\;i \dot \a}(x)$ under the action of  
$ \d_{\rm Phys} $ is: 
\be
\d_{\rm Phys}   {\widehat  {\oy}}_{{\rm Phys}\;i \dot \a}(x)=   \d_{\rm SS} {\widehat  \oy}_{{\rm Phys }\;i \dot \a }(x)  -
 g_{ijk} 
  {\widehat  A}_{\rm Phys}^j {\widehat  A}_{\rm Phys}^k 
\ov C_{\dot \a} 
-
2m g_{ijk} 
  v^j {\widehat  A}_{\rm Phys}^k 
\ov C_{\dot \a} 
\la{bigtransfordotspin}
\ee 
Deriving this is not so simple, and requires some work.  
\item
These transformations make it clear that the above expression 
(\ref{canonicalformsupefield}),
 for construction of ${\cal H}$  
will not yield objects that transform simply as pseudosuperfields for the interacting or massive cases. The extra terms in 
(\ref{bigtransfordotspinsup})
and (\ref{bigtransfordotspin}) are the origin of constraints that we will analyze later.

\item
The necessary restrictions to accomplish that are the subject of the next sections. These restrictions will intertwine the internal symmetry with the supersymmetry to yield physically interesting restrictions, which show up in the SSM or the CSSM for example.

\een

\section{Introduction to the Spectral Sequence for Chiral SUSY in 3+1 Dimensions}

\la{specsumchap}

\subsection{Why do we use the Spectral Sequence Method?}

In section \ref{cohomsum}, we recorded some results for the old theory without Zinns and for the new theory with Zinns. We  explained how to construct a certain  sector of the cohomology space using the fundamental pseudosuperfields, but we have not explained where these come from, and we have not proved that they are really in the cohomology space.   Now, we need to explain how the fundamental pseudosuperfields arise, by introducing the spectral sequence methods needed to solve the problem in a general way.

Spectral sequences allow us to reduce a complicated cohomology problem to a set of simpler ones.  It is evident for example that the theory without mass or interactions is a simpler problem than the one with mass or interactions.  It is also evident that even for the massless free theory, it is possible to divide up $\dB$ into sets of coupled operators some of which are nilpotent by themselves.  The method of spectral sequences makes this into a rigorous mathematical technique.   Without the spectral sequences, the problem of finding the cohomology of $\dB$ is terrifically hard.    It is not easy even using spectral sequences.

\subsection{Quick Review of the Spectral Sequence Method}
The spectral sequence method \ci{McCleary}, as used here \ci{general},  requires us to choose a counting operator  
$N_{\rm Grading}$ that splits up any operator $\d$ (such as $\dB$) into some finite set of sub-operators that satisfy the  equations 
(\ref{sumofdrfromgrading}) and (\ref{rdrfromgrading}) below.

Then the machinery of the spectral sequence generates a series of nested subspaces $E_r$, each with a differential $d_r$, and an adjoint differential $d_r^{\dag}$, satisfying
\be
d_r^2=d_r^{\dag 2}= 0
; \;  d_r E_r \subset E_r
; \;  d_r^{\dag} E_r \subset E_r
\ee
Then we must solve  the cohomology $E_{r+1} =   \fr{\ker d_r}{{\rm Im}\; d_r}\approx E_{r} \cap \ker d_r \cap \ker d_r^{\dag} $  for each of the operators $d_r$ in turn for $r=0,1,2\cdots $.  Each new cohomology problem for $d_{r+1}$ is evaluated in the space $E_{r+1}$ defined by the cohomology of the previous operator.  For our method \ci{general}, which combines the spectral sequence with a Fock space, the orthogonal projection operators $\P_{r+1}$ at each stage project onto the space $E_{r+1} = \ker d_r \cap \ker d_r^{\dag} \cap  E_r$. If there are no more $d_r$ for $r \geq r_{\rm final}$ in some particular spectral sequence, then the spectral sequence is said to collapse  at $r = r_{\rm final}$.  In that case the space $E_{r_{\rm final}}  = E_{\infty} \approx {\cal H}$ is isomorphic to  the cohomology space ${\cal H}$  of the original $\d$ that we are interested in.

\subsection{Grading for the spectral sequence}

The spectral sequence is entirely determined by the Grading chosen. 
 The following Grading\footnote{These are counting operators.  More detail and other useful formulae can be found in subsection \ref{countingopssubsection}.  } is used for this paper: 
\be
{N}_{\rm  Grading} 
= 
{N}_C
+
{ {N}}_{\ov C}
+
2  N_{\x}
+
2  N_{\rm Fields}
+
2  N_{\rm Zinn}
+
4  N_{\rm m}
\la{NewGrading}
\ee

How does one choose a grading?  The desirable qualities are:
\ben
\item
The grading should ensure that the equations
\be
\d= \sum_{r=0}^{n} \d_r
\la{sumofdrfromgrading}
\ee
and
\be
\lt [N_{\rm Grading}, \d_r\rt ]
= r  \d_r.  
\la{rdrfromgrading}
\ee
are satisfied. In particular, there must not be any negative $r$ in the sum.
\item
A good grading should give rise to  $d_r$ for low values of r, for which the $E_{r}$ are  fully and easily solved. 
\item
A good grading should generate easily solved equations, or alternately a simple Elizabethan Drama, for the higher $E_r$.
\een

A search for a good grading is a form of art, as far as I know.  It depends on the details of the $\dB$ that one starts with. Trial and error seems to be the only way to proceed. Different gradings for the same problem make different parts of the problem easier or harder, or downright impossible.  For example, if one uses a grading that simply commutes with $\dB$, then the spectral sequence reduces to using $\dB=d_0$, with no other $d_r$ at all.  In other words $d_0$ is  the whole problem, and that looks impossible to do with one step. 

 The grading we use here is pretty good.  However it might be possible to find a grading that makes the unsolved problems in this paper easier, while making the solved ones harder.  The unsolved problems in this paper involve negative and zero ghost number, but they do not affect the results we quote. If there is a grading that helps to solve these, then the combination of the two gradings would solve the problem entirely.  

It is not obvious what the differentials $d_r$ are, and this question becomes more difficult as r increases.  This is a question which requires one to look at various possibilities and test the results, all as set out in \ci{general}.  If the results are looking wrong, then it is likely that one has missed a differential. So the process can take a long time.  For example, finding $d_7$ in the present case took a long time.  It is quite possible that there are other, so far unknown, differentials that affect the unsolved problems here for negative and zero ghost number. Errors have a way of showing up as one tries to map the space $E_{\infty}\lra{\cal H}$.

\subsection{Results for this Grading}

It turns out that the same grading is useful for both the superfield  operator $\d_{\rm Sup}$  and the physical operator $\d_{\rm Phys}$, and that the spectral sequence generated is the same for these two cases, although it arises a bit differently for the two cases.  So what we are going to find is that
\be
E_{\infty \;{\rm Phys}} =E_{\infty \;{\rm Sup}}
\approx {\cal H}_{\rm Phys}
\approx {\cal H}_{\rm Sup}
\ee

The above grading (\ref{NewGrading}) generates  nine differentials $d_r, r=0,\cdots 8$ and 
  ten cohomology spaces $E_r, r=0,1,\cdots 9$ for the chiral supersymmetry theory in 3 +1 dimensions.  The final space  $E_9= E_{\infty}$  is isomorphic to the cohomology space\footnote{As will be seen later, there are some unsolved issues here, but they do not affect the main results of this paper.} that we are looking for.

\subsection{Three Different Theories: Free Massless, Interacting Massless,  and Interacting Massive}

We will analyze the cohomology here in  three stages, which are the {\bf free and massless theory}, the {\bf interacting  massless theory} 
and the 
{\bf interacting massive  theory}.  The grading (\ref{NewGrading}) 
  ensures that the massive stage comes after the  interacting stage, and that they  both  come after the free massless theory.  It also means that the `structure constant operator' and the `kinetic terms' appear in $\d_0$ and that the `exterior derivative operator' appears later in $\d_2$.  These are all important features that allow progress at the various stages of $E_r$.

 The following table summarizes   the form of the $d_r$ as recited above.  These operators will all be discussed and used in the following Sections and Appendices.

\be
\begin{tabular}{|c|c|c|}
\hline
\multicolumn{3}
{|c|}{Table \ref{summaryofdiffs}: Summary of Differentials $d_r$ from $\d_{\rm Phys}$ or  $\d_{\rm Sup}$}
\\
\hline
$d_r$ &  Form of Operator
&theory 
\\
\hline
$d_0$ & $ C \oC \x^{\dag} + \d_{\rm Kinetic} $
&\\
$d_1$ &$\P_1 \lt \{
C^{\a} \na_{\a} + {\ov C}^{\dot \b}  {\ov \na}_{\dot \b} 
\rt \}
\P_1$
&
\begin{tabular}{c}
 free\\
massless
\end{tabular} 
 \\
$d_2 $ & $\P_2 \lt \{ \x \pa -
\lt (C \na + {\ov C} {\ov \na} \rt )
\fr{\d_{\rm Str}^{\dag}}{\D_{\rm Str}} 
\lt (C \na + {\ov C} {\ov \na} \rt )
 \rt \} \P_2$
&
\\
\hline
$d_3$ & $\P_3 \lt \{
 C_{\a}
\lt (  {\ov g}^{ijk} \A_j \A_k 
\rt )
\y_{\a}^{i \dag}   \rt \} \P_3 + *$
&\begin{tabular}{c}
interacting
\end{tabular} 
\\
$d_4$ &
$\P_4 \lt \{
( C \x {\ov C}^{\dag} ) 
  {\ov g}^{ijk} {\ov A}_j {\ov A}_k
 A^{i \dag} 
\rt \}
\P_4 + *
$
&
\begin{tabular}{c}
massless
\end{tabular} 
\\
\hline
$d_5$ & $\P_5 \lt \{
 C_{\a}
\lt (  {\ov g}^{ijk} m v_j \A_k 
\rt )
\y_{\a}^{i \dag}   \rt \} \P_5 + *$
&\\
$d_6$ & $\P_6 ( C \x {\ov C}^{\dag} ) 
\lt \{
  {\ov g}^{ijk} m {\ov v}_j {\ov A}_k
\rt \}
 A^{i \dag} \P_6 + *$
&\begin{tabular}{c}
interacting
\end{tabular} 
\\
$d_7$ & $\P_7 \lt \{
 C_{\a}
\lt (  e_s^{ijkl} m^2 \ovv_i \ovv_j \A_k 
\rt )
\A_s^{\dag}  \y_{\a}^{l \dag}   \rt \} \P_7 + *$
&\begin{tabular}{c}
massive
\end{tabular} 
\\
$d_8$ & $\P_8 \lt \{
 ( C \x {\ov C}^{\dag} ) 
\lt (  e_s^{ijkl} m^2 \ovv_i \ovv_j \A_k 
\rt )
\A_s^{\dag}  A^{l \dag}   \rt \} \P_8 + *$
&
\\
\hline
\end{tabular}
\la{summaryofdiffs}
\ee

\subsection{Elizabethan Drama}

\la{elizadrama}

Essentially, the higher  differentials\footnote{The tensor $ e_s^{ijkl}$ in $d_7$ and $d_8$ is made from the other tensors $g_{ijk},g_i,\og^{ijk},\og^i$ in the theory, and it will be discussed in subsection \ref{dsevensubsection} and other subsections referred to there. These operators $d_r, r=3\cdots 8$, as written here, are really only the lowest terms of expressions that contain arbitrary numbers of derivatives.}  $d_r, r = 3,4,5,6,7,8 $ perform an Elizabethan Drama,   using the results for the free massless case.  Here is an amusing and insightful quote from  one of the pioneers of the spectral sequence (F. J. Adams ), as recited in one of the classic texts on the spectral sequence\ci{Mclearyp180}:

{\em ...the behavior of this spectral sequence... is a bit like an Elizabethan drama, full of action, in which the business of each character is to kill at least one other character, so that at the end of the play one has the stage strewn with corpses and only one actor left alive (namely the one who has to speak the last few lines)... }

The remaining actors form the cohomology space.  For the present formulation of the spectral sequence, whenever one has the  equations:
\be
d_r U = V
\ee
and 
\be
d_r^{\dag} V =U
\ee
the actors $U$ and $V$  kill each other, and neither of them  survives to live in the space $E_{r+1}$.   In the present case there are lots of actors left alive, but there are plenty of corpses too.

\subsection{Nice Clean Research}

  Here is another relevant quote from Adams\ci{McCleary}: 

 {\em Whenever a chance has arisen to show that a differential $d_r$ is non-zero, the experts have fallen on it with shouts of joy--`Here is an interesting phenomenon! Here is a chance to do some nice clean research!--and they have solved the problem in short order.}

The present problem is a very complicated one, because there are many differentials doing different things, and the cohomology space is large and has a lot of structure, which depends on the details of the interaction and mass terms.  There are still lots of differentials to examine in SUSY theories\footnote{\la{joyfootnote} Experts (and Novices too) in the area of spectral sequences are invited to do some nice clean research here!  The happy remarks of Adams have not  yet been put into effect for this case. The present paper is still somewhat incomplete, as is explained below. Supersymmetric gauge theories have a similar structure to the present chiral theories, but there are plenty of differences too, and that paper is in very rough draft form with lots of unsolved issues.  Then there is the whole panoply of other supersymmetric theories, both rigid and local, in various dimensions.  Exercises for the student:  Are the 10 dimensional supersymmetric gauge and gravitational theories actually consistent? Or are there subtle hidden supersymmetry anomalies in them? What about the superstring?}.

However the above quote is still on point, because the higher differentials $d_r$ really do simplify the problem immensely.  A glance at  Table (\ref{summaryofdiffs}) shows that all of the $d_r$ for $r=3\cdots 8$ depend on the interaction coefficients $g_{ijk}$.  These result in a host of quite simple, and similar, equations by way of the Elizabethan drama.  

The equations from the differentials are so simple in fact that we can solve them for the SSM quite easily, and this is done in section \ref{cssmexamples}.  In that context it is even obvious that the SSM could use some improvement, resulting in the CSSM.  Part of that improvement is to put in right handed neutrinos.  Fortunately these are  also experimentally viable and useful \ci{neutrinos}.

\section{Some Examples of the Mappings $E_{\infty} \ra {\cal H}$}

\la{examplesofetoh}

\subsection{An introduction to the practical use of the Spectral Sequence}

In section \ref{specsumchap}, we introduced the spectral sequence  in  a general way.  In the next section, we will apply these results to the specific case of the SSM.
Here we will discuss some general notions, and summarize the normal space $E_3=E_{\infty}$ for  the free massless theory, and relate it back to the cohomology described earlier.

  The following quantum numbers are preserved by the grading ${N}_{\rm  Grading} $ and by the operators $\d_{\rm Phys}$ and 
$\d_{\rm Sup}$, and therefore also  by the isomorphism Map that takes $e \in E_{\infty}$ to its image $ h\in {\cal H}_{\rm Phys}$ or $ h\in {\cal H}_{\rm Sup}$.  We shall write ${\cal H} \equiv {\cal H}_{\rm Integrated}$ to indicate either ${\cal H}_{\rm Phys}$ or $  {\cal H}_{\rm Sup}$ when we do not want to specify which one we mean (the two are isomorphic but not identical):
\be
e \in E_{\infty} \stackrel{\rm Map}{\lra} h\in  
{\cal H} 
\ee
\ben
\item
The mass dimension of the  expressions $e$ and $h$ must be the same.  
\item
The spin types $J(n,m)$ of the  expressions $e$ and $h$ must be the same.  If we write the indices explicitly, this means that:
\be
e_{(\a_1 \cdots \a_n), (\dot \b_1 \cdots \dot \b_m)}  \in E_{\infty} \ra h_{(\a_1 \cdots \a_n), (\dot \b_1 \cdots \dot \b_m)}\in {\cal H}
\ee
\item
Any other conserved quantum number, such as Lepton number or Baryon number or Charge or Ghost number of the  expressions $e$ and $h$ must be the same.
\item For the free massless theory, the number of fields plus the number of Zinns 
${\cal N}_{\rm Field}+{\cal N}_{\rm Zinn}$   is conserved too, and so, for the free massless theory, this is also preserved by the mapping.
 Moreover, since the interacting and massive theories are built from the free massless theory, this conservation is useful for those Mappings also. 
\een

 Now we will turn to the solutions for $E_{\infty}$ and the mapping $E_{\infty}\ra {\cal H}$ for the three stages.  The form of $E_{\infty}$ for the free massless theory will be  summarized below in 
subsection \ref{breifsummsec}.

\subsection{Brief Summary of the Results from the Spectral Sequence: The Normal Solutions}

\la{breifsummsec}

The spectral sequence calculations are done in detail in the Appendices. The plan of this paper is to try to make the main body of the new results accessible without referring very much to the Appendices.  So most of the application to actual superspace is in the body of the paper, and the spectral sequence spaces are mostly discussed in the Appendices.

 Here is a very brief summary of the results from the spectral sequence.  For the free massless theory we will find that an important part of the result of the spectral sequence has the following form.  We call these the Normal Solutions:

\be
E_{3 \;{\rm Normal}
} = 
  ( C \x  {\ov C} )
S_{0\;{\rm Normal}}
\eb
\oplus
\sum_{n=0}^{\infty} 
\lt \{
(C \x^2 C) {R}_{n\;{\rm Normal}}
+
(\oC \x^2 \oC) {\oR}_{n\;{\rm Normal}}
\rt \}
\la{normalfor3n}
\ee
where 
\be
{R}_{p\;{\rm Normal}}
=
{ R}_{(\a_1 \cdots \a_{n+p})}
=
 T_{[i_1\cdots i_n]}^{(j_1\cdots j_m)}\A_{j_1}\cdots \A_{j_m}
\y^{i_1}_{(\a_1} \cdots 
\y^{i_n}_{\a_n}  C_{\a_{n+1}}\cdots C_{\a_{n+p})} 
\la{canonicalform3n}
\ee
The complex conjugate of (\ref{canonicalform3n}) is:
\be
 {\oR}_{p\;{\rm Normal}} =  \oR_{(\dot \a_1 \cdots \dot \a_{n+p})} =   \oT^{[i_1\cdots i_n]}_{(j_1\cdots j_m)} A^{j_1}\cdots A^{j_m}
\oy_{i_1(\dot \a_1} \cdots 
\oy_{i_n \dot \a_n}  \oC_{\dot \a_{n+1}}\cdots \oC_{\dot \a_{n+p})} 
\la{canonicalformccn}
\ee
and
\be
S_{0 \;{\rm Normal}}
= 
\lt \{
f_i A^i
+
\of^i \A_i 
+
f^i_j A^j \A_i 
+ f^i_{2,j}\lt (
A^i  \A_{j\g \dot \d}-   A^i_{\g \dot \d} \A_j
-
\y^i_{\g}
\oy_{j \dot \d}
\rt )+ \cdots
\rt \}
\la{defiverern}
\ee
There are also terms with more derivatives, and there is also another sector (the exceptional sector)  but the above gives a good idea of an important part of what is present, including the large `Outfield' sector.  Note the close resemblance between (\ref{canonicalform3n}) and (\ref{canonicalformccn}) and the form (\ref{canonicalformsupefield}) and its complex conjugate.  The form (\ref{canonicalformsupefield}) is in fact deduced using the spectral sequence result above.

 This result is taken from 
Appendix \ref{e3stuffassem} below.  The development of the spectral sequence takes place in Appendices 
\ref{derivofe3chap} to \ref{interactingspectralse},  with further material in all the Appendices up to 
\ref{collapseappendix}, with commentary found there.

For the interacting and massive theories, there are constraints on the above.  These are explained in 
Appendices \ref{interactingspectralse}
 and \ref{chapfore9}.  Now  we shall apply these results to  a specific variant of the SSM in section \ref{cssmexamples}.

\section{Some Examples of $E_{\infty}$ and ${\cal H}$ from the CSSM}

\la{cssmexamples}

Sometimes in this section we will refer to the various $d_r$.  Their forms can be found in
 Table (\ref{summaryofdiffs}) and in the various Appendices, where they are discussed at length.

\subsection{Trilinear Symmetric Interaction Terms}

The superpotential $P$ for a chiral theory has the general form (for a renormalizable theory):
\be
P = \int d^4 x d^2 \q 
\lt \{
m^2 g_{i} {\widehat  A}^i + m g_{ij} {\widehat  A}^i {\widehat  A}^j 
+  g_{ijk} {\widehat  A}^i {\widehat  A}^j {\widehat  A}^k  \rt \}
\la{superstuff}
\ee
This appears to have symmetric mass and interaction terms:
\be
g_{ij}=g_{(ij)};\;\; g_{ijk}=g_{(ijk)}
\la{symnform}
\ee

In the foregoing analysis, we have assumed that the superpotential $P$ has a symmetric interaction term $g_{ijk}=g_{(ijk)}$:
\be
P = \int d^4 x d^2 \q 
\lt \{
m^2 g_{i} {\hat A}^i 
+  g_{ijk} {\hat A}^i {\hat A}^j {\hat A}^k  \rt \}
\la{superstuff2}
\ee

In the present paper we started with $g_{ij}=0$, but the mass term $g_{ij}\neq 0$ arises after symmetry breaking, and at that time we have assumed that we can and do choose a VEV such that $m^2 g_{i} {\widehat  A}^i$ disappears. 

These symmetric forms (\ref{symnform})
are generally not   a good way
to try to solve the equations that arise from the spectral sequence operators $d_r, r = 3 \cdots 8$.  It is   better  to use several different representations ${\widehat  A}_{p}^{i}$, and then the matrices and tensors  $g^{pq}_{ij},g^{pqr}_{ijk} $ have more complicated symmetries. In the SSM there are, of course, several indices (colour, flavour, weak isospin, hypercharge, Lepton number, Baryon number), not just two.
  The natural notation uses irreducible representations of the symmetries, which depends on the quantum numbers, and it contains no artificial symmetrization, which arises if we try to write all the representations in one reducible way like the above using just one superfield index ${\widehat  A}^i$.  
What this means in practice is that one should regard the operators  $d_r, r = 3 \cdots 8$, as given above,  as a   shorthand version, but we need  to write down their non-symmetric versions for any given theory.

These symmetrization issues are more subtle than they appear to be. There is a folkloric tendency to believe that the unified theory of everything must be based on an irreducible model with one huge representation of some huge group.  This results from an aesthetic notion that irreducibility is equivalent to simplicity.  However, it turns out  that   certain combinations of irreducible representations can mingle with each other in a remarkable way through the constraint equations for SUSY.  We shall see how this works below. One could get the impression that SUSY and internal symmetries are linked in some non-group-like way here that gets around various no-go theorems, and a careful study on that would be worth trying.
The next section summarizes this situation for the simplest interesting example.

\subsection{An Important Simple Example}

At this point we must look at the example examined in 
section \ref{importexamp}.

Let us first consider the  constraint equation which looks like this
\be
{\cal L}_{f}  P_3 
=0
\la{constraint}
\ee
where ${\cal L}_f$ is a Lie algebra generator, made of scalars, of the form
\be
{\cal L}_{f}  = f_{i}^{j} A^i \fr{\pa}{\pa A^j}  
\la{eqpotent3}
\ee
and $P_3$ is the unintegrated trilinear scalar part of the superpotential, extracted from (\ref{superstuff}):
\be
P_3 = g_{pqr} A^p A^q A^r
\la{potent3}
\ee
These equations are written in the symmetric form, and they typically involve redundant and artificial symmetrization, as discussed above. So at this point, we look at a non-symmetric and interesting example--the supersymmetric standard model (SSM).  

\subsection{Introduction to the CSSM and the Quark and Lepton Outfields}

\la{funnyinvariancesincssm}

Some  curious solutions of this equation arise when one writes these equations in the non-symmetric form, using the non-symmetric form of the superpotential for the SSM, as follows. We will augment this to the CSSM which has right neutrinos and a singlet Higgs in addition to the usual minimal  SSM. The scalar version of the trilinear term of the superpotential for  the CSSM  has the following non-symmetric form (here we mean it does not have the symmetry $g_{(ijk)}$  discussed above, because it is expressed in terms of irreducible multiplets):

\[
P_{{\rm CSSM} }
=
g \ve_{ij} H^i K^j J
+
p_{p q} \ve_{ij} L^{p i} H^j P^{ q} 
+
r_{p q} \ve_{ij} L^{p i} K^j R^{ q}
\]
\be
+
t_{p  q} \ve_{ij} Q^{c p i} K^j T_c^{ q}
+
b_{p  q} \ve_{ij} Q^{c p i} H^j B_c^{ q}
\la{PSSM}
 \ee

Here the fields $J,H^i,K^i$ are Higgs/Goldstone  scalar fields from the respective supermultiplets, with hypercharge $Y=0,-1,+1$ respectively.  $ Q^{c p i}$ is the scalar from the Left Quark supermultiplet with hypercharge $Y=\fr{1}{3}$. $T_c^{ q}$ and $B_c^{ q}$
are the scalars from the right handed up and down antiQuark supermultiplets with hypercharge 
$Y=-\fr{4}{3}, \fr{2}{3}$ respectively.  
$ L^{ p i}$ is the scalar from the Left Lepton supermultiplet with hypercharge $Y=-1$.  $P^{ q} $ and   $R^{ q} $ are the scalars from the right handed antipositron and antineutrino  supermultiplets, with hypercharge $Y=2,0$ respectively. The indices $i,j=1,2$ are weak SU(2) indices.

 CSSM is an acronym for `Cybersusy Supersymmetric Standard Model'. The non-minimal terms that we add are the    right handed neutrinos $R^p$, and the  singlet Higgs field J.  The latter is designed to spontaneously break $SU(2) \times U(1) $ down to $U(1)$, when one includes a term $+m^2 J$ in the potential.  Note that this $J$ singlet also plays an important role in the Lie algebra generators for the Outfields below.  The reason for the choice of the CSSM is that the $R^p$ makes the Leptons behave similarly to the Quarks, and the $J$ allows construction of the Quark and Lepton Outfields, as shown below.

One could examine a great number of objects here, but we will concentrate on objects with non-zero quantum numbers for Baryon and Lepton number.  There are two reasons for this.  The first is that these sectors do not mix with the gauge sector, because the gauge theory does not have Baryon or Lepton number. So we can validly examine these sectors without worrying about the gauge theory. The second reason is that there are interesting things happening in these sectors.

The  following physically interesting Lie algebra operators exist
for the Leptons (provided that the term $- g_J m^2 J$ is absent):
\be
{\cal L}^{pi}_{L}
=
g^{-1} L^{p i} \fr{\pa}{\pa J}
+
(p^{-1})^{qp} K^{ i} \fr{\pa}{\pa P^q}
-
(r^{-1})^{qp} H^{ i} \fr{\pa}{\pa R^q}
\la{thetrickyoneforLeptons}
\ee
\be
{\cal L}^p_{P}
=
g^{-1} P^{p } \fr{\pa}{\pa J}
+
(p^{-1})^{pq} K^{ i} \fr{\pa}{\pa L^{iq}}
\la{rightelect}
\ee
\be
{\cal L}^p_{R}
=
g^{-1} R^{p } \fr{\pa}{\pa J}
-
(r^{-1})^{pq} H^{ i} \fr{\pa}{\pa L^{iq}}
\ee
where  the inverse matrices are defined in the following way:
\be
p_{s q}   (p^{-1})^{qp} 
=(p^{-1})^{pq}  
p_{q s}=
\d^{p}_{s};
\ee
\be
r_{s q}   (r^{-1})^{qp} 
=
   (r^{-1})^{pq} r_{q s}
=
\d^{p}_{s}.
\ee
Similarly,  the following  curious Lie algebra operators exist
for the Quarks:
\be
{\cal L}^{cpi}_{Q}
=
g^{-1} Q^{c p i} \fr{\pa}{\pa J}
-
(t^{-1})^{qp} H^{ i} \fr{\pa}{\pa T^q_c}
+
(b^{-1})^{qp} K^{ i} \fr{\pa}{\pa B^q_c}
\la{thetrickyoneforQuarks}
\ee
\be
{\cal L}^p_{T c}
=
g^{-1} T^p_c \fr{\pa}{\pa J}
-
(t^{-1})^{pq} H^{ i} \fr{\pa}{\pa Q^{icq}}
\ee
\be
{\cal L}^p_{Bc}
=
g^{-1} B^p_c \fr{\pa}{\pa J}
+
(b^{-1})^{pq} K^{ i} \fr{\pa}{\pa Q^{icq}}
\la{rdQuark}
\ee
where  the inverse matrices are defined in the following way:
\be
 (t^{-1})^{pq} t_{qs}
=
 t_{sq}(t^{-1})^{qp} 
=
\d^{p}_{s}
\ee
\be
(b^{-1})^{pq} b_{qs}
=
 b_{sq}(b^{-1})^{qp}
=
\d^{p}_{s}
\ee
Using the above forms  
(\ref{PSSM}) and 
(\ref{thetrickyoneforLeptons}), for example,  
it is easy to verify that:
\be
{\cal L}^{pi}_{L}
P_{{\rm CSSM} }
=0
\ee
as follows:
\[
{\cal L}^{pi}_{L}
P_{{\rm CSSM} }
=
\lt \{
g^{-1} L^{p i}g \e_{jk} H^j K^k  
+
(p^{-1})^{qp} K^{ i} p_{s q} \e_{jk} L^{s j} H^k
-
(r^{-1})^{qp} H^{ i} r_{s q} \e_{jk} L^{s j} K^k
\rt \}
\]\be
=
\lt \{
  L^{p i}  \e_{jk} H^j K^k  
+
     \e_{jk} L^{p j}
\lt ( K^{ i} H^k
-
 H^{ i}   K^k
\rt )
\rt \}
\la{firstveriflep}
\ee

Now use
\be
 K^{ i}     H^k
- H^{ i}     K^k
= \ve^{ik} \lt ( \ve_{lm}K^{l} H^m
\rt )
\ee
and we get 

\be
{\cal L}^{pi}_{L}
P_{{\rm CSSM} }
=
\lt \{
  L^{p i}  \e_{jk} H^j K^k  
+
     \e_{jk} L^{p j}
\ve^{ik} \lt ( \ve_{lm}K^{l} H^m
\rt )
\rt \}
=0
\ee
We also have, using (\ref{PSSM}) and (\ref{rightelect}), 

\be
{\cal L}^p_{P}P_{{\rm CSSM} }
=
\eb
g^{-1} P^{p } g \e_{ij} H^i K^j
+
(p^{-1})^{pq} K^{ i} \lt ( 
p_{q s} \e_{ij}   H^j P^{ s} 
+
r_{q s} \e_{ij}   K^j R^{ s}
\rt )
\ee
Now observe that
\be
K^{ i} 
 \e_{ij}   K^j 
=0
\ee
So we get
\be
{\cal L}^p_{P}P_{{\rm CSSM} }
=
\eb
 P^{p }  \e_{ij} H^i K^j
+
 K^{ i}  
  \e_{ij}   H^j P^{ p} 
=0
\ee

The other four Lie algebra operators work in a similar way. 
Observe the intertwining of the left doublets and right singlets here, and the crucial role of the singlet and the two SU(2) Higgs doublets. This all seems quite specific to the CSSM.  In the CSSM, 
each of these six Lie algebra invariance generators has a Lepton (Quark) scalar multiplied by $\fr{\pa}{\pa J}$, added to terms made from the Higgs $H^i,K^i$ multiplied by the derivative of an  AntiLepton (AntiQuark) scalar  (or vice versa).  Each of them is in a representation  of the gauge groups $U(1)$, $SU(2)$ and $SU(3)$.  Each of them is in an eigenstate of Quark and Lepton number.  These six operators form an invariant Abelian subalgebra of the invariances of the term (\ref{PSSM}). This invariance algebra includes the generators of $SU(3) \times SU(2) \times U(1)$ as well as Baryon and Lepton number.

These invariances would not exist if the standard model did not have its peculiar left-right asymmetry, which is also carried through to the Higgs sector in this supersymmetric version of the SM.  Also note that this invariance is far from obvious if one writes the superpotential in an artificially symmetrized and reducible way.

In accord with the analysis above, and the discussion in  
section \ref{importexamp}, we can write the following solutions for the $d_3$ constraints:

{\bf Lepton Outfields}: {\em Chiral Dotted Spinor  Pseudosuperfields with  
Quantum Numbers of the Leptons:}

\be
{\widehat \w}^{pi}_{L \dot \a}
=
g^{-1} {\widehat  L}^{p i} {\widehat  \oy}_{J\dot \a}
+
(p^{-1})^{qp} {\widehat K}^{ i}  {\widehat \oy}_{P q \dot \a}  
-
(r^{-1})^{qp} {\widehat H}^{ i} {\widehat \oy}_{R q \dot \a}   
\la{thetrickyoneforLeptonsdotspinor}
\ee
\be
{\widehat \w}^{p}_{P \dot \a}
=
g^{-1} {\widehat P}^{p } {\widehat  \oy}_{J\dot \a}
+
(p^{-1})^{pq} {\widehat K}^{ i} 
{\widehat \oy}_{L i q \dot \a}  
\la{rightelectdotspinor}
\ee
\be
{\widehat \w}^{p}_{R \dot \a}
=
g^{-1} {\widehat R}^{p }  {\widehat  \oy}_{J\dot \a}
-
(r^{-1})^{pq} {\widehat H}^{ i}  {\widehat \oy}_{L i q \dot \a}  
\la{leftelectdotspinor}
\ee

{\bf Quark  Outfields}: {\em Chiral Dotted Spinor
 Pseudosuperfields with  
Quantum Numbers of the Quarks:} 

\be
{\widehat \w}^{cpi}_{Q \dot \a}
=
g^{-1} {\widehat  Q}^{c p i} {\widehat  \oy}_{J\dot \a}
-
(t^{-1})^{qp} {\widehat H}^{ i} 
{\widehat  \oy}^c_{T q\dot \a} 
+
(b^{-1})^{qp} {\widehat K}^{ i} 
{\widehat  \oy}^c_{B q \dot \a}
 \la{thetrickyoneforQuarksdots}
\ee
\be
{\widehat \w}^{p}_{T c\dot \a}
=
g^{-1} {\widehat  T}^{ p }_c {\widehat  \oy}_{J\dot \a}
-
(t^{-1})^{pq} 
{\widehat  H}^{ i }  {\widehat  \oy}_{Q icq\dot \a}
\la{rdQuarksupTdots}
\ee
\be
{\widehat \w}^{p}_{B c\dot \a}
=
g^{-1} {\widehat  B}^{ p }_c {\widehat  \oy}_{J\dot \a}
+
(b^{-1})^{pq} 
{\widehat  K}^{ i }  {\widehat  \oy}_{Q icq\dot \a}
\la{rdQuarksup}
\ee

\subsection{The Constraints from $d_3$ in the CSSM}

We need the form of $d_3$ in the spectral sequence here:
\[
\d_3 = \oC_{\dot \a} \fr{\pa P_{\rm CSSM}}{\pa A^i}
\oy_{i \dot \a}^{\dag} + *
\]
\be=
g \ve_{ij} H^i K^j \oC_{\dot \a}
\oy_{J\dot \a}^{\dag}
+
\lt ( 
p_{p q} \ve_{ij} L^{p i} H^j  
  \rt )
\oC_{\dot \a} \oy_{Pq\dot \a}^{\dag}
+
\lt ( 
 r_{p q} \ve_{ij} L^{p i} K^j  
 \rt )
\oC_{\dot \a} \oy_{Rq\dot \a}^{\dag}
\eb
+
\lt ( 
t_{p  q} \ve_{ij} Q^{c p i} K^j 
 \rt )
\oC_{\dot \a} \oy_{T q\dot \a}^{c \dag}
+
\lt ( 
b_{p  q} \ve_{ij} Q^{c p i} H^j  
 \rt )
\oC_{\dot \a} \oy_{B q \dot \a}^{c \dag}
\eb
+
\lt ( t_{p  q} \ve_{ij}  K^j T_c^{ q}
+
b_{p  q} \ve_{ij}   H^j B_c^{ q}
 \rt )
\oC_{\dot \a} \oy_{Qcpi\dot \a}^{\dag}
\eb+
\lt ( 
p_{p q} \ve_{ij}   H^j P^{ q} 
+
r_{p q} \ve_{ij}   K^j R^{ q}
 \rt )
\oC_{\dot \a} \oy_{Lpi\dot \a}^{\dag}
\eb+
\lt ( 
g \ve_{ij}   K^j J
+
p_{p q} \ve_{ji} L^{p j}  P^{ q} 
+
b_{p  q} \ve_{ji} Q^{c p j}  B_c^{ q}
 \rt )
\oC_{\dot \a} \oy_{Hi\dot \a}^{\dag}
\eb+
\lt ( 
g \ve_{ji} H^j  J
+
r_{p q} \ve_{ji} L^{p j} R^{ q}
+
t_{p  q} \ve_{ji} Q^{c p j}   T_c^{ q}
 \rt )
\oC_{\dot \a} \oy_{Ki\dot \a}^{\dag}
+*
\ee
The constraint equation  for  (\ref{thetrickyoneforLeptonsdotspinor}), which is:
\be
d_3 
{\widehat \w}^{pi}_{L \dot \a}
=0
\ee
is a  direct consequence of the equations  
(\ref{firstveriflep}) above. The others work in exactly the same way. 

\subsection{The Constraints from $d_4$  in the CSSM}

In the context of the regular and unseparated irregular equations, $d_4$ takes a form that is closely related to the above $d_3$:
\[
d_4 = (\oC \x C^{\dag}) \fr{\pa P_{\rm CSSM}}{\pa A^i}
\A_{i }^{\dag} + *
\]
\[=
(\oC \x C^{\dag}) \lt \{
g \ve_{ij} H^i K^j  
\oJ^{\dag}
+
\lt ( 
p_{p q} \ve_{ij} L^{p i} H^j  
  \rt )
 \oP_{q}^{\dag}
+
\lt ( 
 r_{p q} \ve_{ij} L^{p i} K^j  
 \rt )
  \oR_{q}^{\dag}
\rt.\]\[
+
\lt ( 
t_{p  q} \ve_{ij} Q^{c p i} K^j 
 \rt )
  \oT_{ q}^{c \dag}
+
\lt ( 
b_{p  q} \ve_{ij} Q^{c p i} H^j  
 \rt )
  \oB_{ q }^{c \dag}
\]\[
+
\lt ( t_{p  q} \ve_{ij}  K^j T_c^{ q}
+
b_{p  q} \ve_{ij}   H^j B_c^{ q}
 \rt )
 \oQ_{cpi}^{\dag}
\]\[
+
\lt ( 
p_{p q} \ve_{ij}   H^j P^{ q} 
+
r_{p q} \ve_{ij}   K^j R^{ q}
 \rt )
 \oL_{pi}^{\dag}
\]\[
+
\lt ( 
g \ve_{ij}   K^j J
+
p_{p q} \ve_{ji} L^{p j}  P^{ q} 
+
b_{p  q} \ve_{ji} Q^{c p j}  B_c^{ q}
 \rt )
 \oH_{i}^{\dag}
\]\be \lt.
+
\lt ( 
g \ve_{ji} H^j  J
+
r_{p q} \ve_{ji} L^{p j} R^{ q}
+
t_{p  q} \ve_{ji} Q^{c p j}   T_c^{ q}
 \rt )
 \oK_{i}^{\dag}
\rt \}
+*
\la{d4ssm}
\ee
The adjoint of this is
\[
d_4^{\dag}
= (C \x^{\dag} \oC^{\dag}) \lt \{
\og \oJ \ve^{ij} H^{i\dag} K^{j\dag}  
+
\op^{p q}  \oP_{q}
 \ve^{ij} L^{p i\dag} H^{j \dag} 
+
 \ov r^{p q} \oR_{q} \ve^{ij} L^{p i\dag} K^{j\dag}  
\rt.
\]
\[
+
\ot^{p  q}   \oT_{ q}^{c }
\ve^{ij} Q^{c p i \dag} K^{j \dag} 
+
\ob^{p  q}   \oB_{ q }^{c }
\ve^{ij} Q^{c p i \dag} H^{j\dag}  
\]
\[
+
\oQ_{cpi}
\lt (
\ot^{p  q} \ve^{ij}  K^{j\dag} T_c^{ q\dag}
+
\ob^{p  q} \ve^{ij}   H^{j\dag} B_c^{ q\dag}
 \rt )
\]
\[
+
 \oL_{pi} \lt ( 
p_{p q} \ve_{ij}   H^j P^{ q} 
+
r_{p q} \ve_{ij}   K^j R^{ q}
 \rt )
^{\dag}
\]
\[
+
\oH_{i}\lt ( 
g \ve_{ij}   K^j J
+
p_{p q} \ve_{ji} L^{p j}  P^{ q} 
+
b_{p  q} \ve_{ji} Q^{c p j}  B_c^{ q}
 \rt )^{\dag}
\]
\be
\lt.+
 \oK_{i}\lt ( 
g \ve_{ji} H^j  J
+
r_{p q} \ve_{ji} L^{p j} R^{ q}
+
t_{p  q} \ve_{ji} Q^{c p j}   T_c^{ q}
 \rt )
^{\dag}
\rt \}
+*
\la{d4fromssdag}
\ee
This plays a role for the ghost charge zero and minus one sectors, as do $d_6$ and $d_8$.  Application of these differentials  will not be discussed here.

\subsection{Differentials $d_5$ and $d_6$ involving the mass in the SSM}

\la{symbkincssm}

To get \GSB \ we need to add a term to the superpotential of the form $
m^2 g_J J
$, and then shift the fields by
\be
H^i\lra( m h^i +H^i)
\ee
and
\be
K^i \lra( m k^i +K^i)
\ee
where
\be
g \ve_{ij}  h^i   k^j +
m^2 g_J =0
\ee
The new differentials $d_5$ and $d_6$  arise from the term in the following that is linear in $m$:
\[
P_{{\rm CSSM\;m} }
=
g \ve_{ij} ( m h^i +H^i) ( m k^j + K^j) J
+
m^2 g_J J
\]
\[
+
p_{p q} \ve_{ij} L^{p i} ( m h^j + H^j) P^{ q} 
+
r_{p q} \ve_{ij} L^{p i} ( m k^j + K^j) R^{ q}
\]
\be
+
t_{p  q} \ve_{ij} Q^{c p i} ( m k^j + K^j) T_c^{ q}
+
b_{p  q} \ve_{ij} Q^{c p i}( m h^j + H^j) B_c^{ q}
\la{PSSM2}
 \ee

\subsubsection{The operator $d_5$ for the CSSM} 

\la{dfiveforssmsubsection}

So we get, using $\d_3$ as a template:
\[
\d_5 =  
m  
g \ve_{ij} (  h^i  K^j + H^i k^j) 
 \oC_{\dot \a}
\oy_{J\dot \a}^{\dag}
\]\[
+
m p_{p q} \ve_{ij} L^{p i} h^j  
\oC_{\dot \a} \oy_{Pq\dot \a}^{\dag}
+
\lt ( 
 r_{p q} \ve_{ij} L^{p i} k^j  
 \rt )
\oC_{\dot \a} \oy_{Rq\dot \a}^{\dag}
\]
\be
+
m t_{p  q} \ve_{ij} Q^{c p i} k^j 
\oC_{\dot \a} \oy_{T q\dot \a}^{c \dag}
+
m b_{p  q} \ve_{ij} Q^{c p i} h^j  
\oC_{\dot \a} \oy_{B q \dot \a}^{c \dag}
\eb
+
m
\lt (
 t_{p  q} \ve_{ij}  k^j T_c^{ q}
+
b_{p  q} \ve_{ij}   h^j B_c^{ q}
 \rt )
\oC_{\dot \a} \oy_{Qcpi\dot \a}^{\dag}
\eb+
m \lt ( 
p_{p q} \ve_{ij}   h^j P^{ q} 
+
r_{p q} \ve_{ij}   k^j R^{ q}
 \rt )
\oC_{\dot \a} \oy_{Lpi\dot \a}^{\dag}
\eb+
m \lt ( 
g \ve_{ij}   k^j J
 \rt )
\oC_{\dot \a} \oy_{Hi\dot \a}^{\dag}
\eb+
m
\lt ( 
g \ve_{ji} h^j  J
 \rt )
\oC_{\dot \a} \oy_{Ki\dot \a}^{\dag}
+*
\ee

Now let us use this on one of our solutions at level $E_4$. Take for example (\ref{rightelectdotspinor}).
 First we   revert to the form in $E_4$ without the superfields:
\be
    {\w}^{p}_{  \; P \dot \a}
=
(\oC \x^2 \oC)
\lt \{
g^{-1} {P}^{p } {\oy}_{J\dot \a}
+
(p^{-1})^{pq}   {K}^{ i}  
{\oy}_{L i q \dot \a}  
\rt \}
\la{omegaP}
\ee
We can ignore the factor $(\oC \x^2 \oC)
$ here. Now we have
\be
d_5  = \P_5 \d_5 \P_5
\ee
and so
\be
d_5     {\w}^{p}_{ P \dot \a}
=
\P_5 \lt \{g^{-1} {P}^{p } g \ve_{ij} m (  h^i  K^j + H^i k^j) 
 \oC_{\dot \a}
\ebp
+
(p^{-1})^{pq} {K}^{ i}  
m \lt ( 
p_{p q} \ve_{ij}   h^j P^{ q} 
+
r_{p q} \ve_{ij}   k^j R^{ q}
 \rt )
\oC_{\dot \a}
\rt \}
\eb
=
\P_5 \lt \{   m    H^i \ve_{ij}   k^j 
{P}^{p }  \oC_{\dot \a}
+
m  
  {K}^{ i} \ve_{ij}  k^j 
(p^{-1})^{pq} r_{qr} R^{ r} 
\oC_{\dot \a}
\rt \}
\ee

Now consider

\be
d_3  m   {k}^{ i}  
(p^{-1})^{pq}{\oy}_{L i q \dot \a}  
=
(p^{-1})^{pq} m   {k}^{ i}  \lt ( 
p_{qr} \ve_{ij}   H^j P^{ r} 
+
r_{q	r} \ve_{ij}   K^j R^{ r}
 \rt )
\oC_{\dot \a}
\eb
= \lt \{
m
 {k}^{ i}  \ve_{ij}   H^j P^{ p} 
\oC_{\dot \a}
+m
{k}^{ i} \ve_{ij}   K^j (p^{-1})^{pq} r_{q	r}  R^{ r}
\oC_{\dot \a}
\rt \}
\ee

So we see that the image of $ {\w}^{p}_{  \; P \dot \a}$ under the action of  $d_5$ is the same as the image of 
$ m   {k}^{ i}  (p^{-1})^{pq}{\oy}_{L i q \dot \a} $
under the action of  $d_3$. 
Now anything which is a boundary of an object in $E_3$ using the differential $d_3$ does not survive to $E_4$.  So 
\be
d_3     m   {k}^{ i}  (p^{-1})^{pq}{\oy}_{L i q \dot \a} 
\cap E_4 =0
\ee
so such a term certainly does not survive to $E_5\subset E_4$.  So it follows that
\be
\P_{5} 
d_5     {\w}^{p}_{ P \dot \a}
\cap E_5 =0
\ee
which implies that
\be
\P_5 
d_5     {\w}^{p}_{ P \dot \a}
=0
\ee
So this imposes no new condition and we have
\be
    {\w}^{p}_{ P \dot \a}
\in E_6
\ee

\subsection{Differential   $d_6$ involving the mass in the CSSM}

\la{d6ssmsubsection}

The differential $d_6$ affects the ghost charge minus one sector and the chiral ghost charge zero solutions without spinor indices, and we shall not analyze it here.
This is better considered when one also has the gauge theory present.

\subsection{The operator $d_7$ for the CSSM} 

\la{d7ssmsubsection}

The reader should refer to Appendix \ref{dsevensubsection} in connection with this section.

We will see that $d_7$ kills  the chiral dotted spinor superfield Lepton and Quark Outfields set out above. 
Here is the detailed form of the adjoint of $\d_3$ in the physical formulation for the CSSM:
\be
\d_3^{\dag}=
\oy_{J\dot \a} \lt ( g \ve_{ij} H^i K^j \oC_{\dot \a}
\rt )^{\dag}
\eb
+
\oy_{Pq\dot \a}
\lt ( 
p_{p q} \ve_{ij} L^{p i} H^j  
\oC_{\dot \a} 
  \rt )^{\dag}
+
\oy_{Rq\dot \a}\lt ( 
 r_{p q} \ve_{ij} L^{p i} K^j  
\oC_{\dot \a}
 \rt )^{\dag}
\eb
+
 \oy_{T q\dot \a}^{c} \lt ( 
t_{p  q} \ve_{ij} Q^{c p i} K^j 
\oC_{\dot \a} \rt )^{ \dag}
+
\oy_{B q \dot \a}^{c}
\lt ( 
b_{p  q} \ve_{ij} Q^{c p i} H^j  
\oC_{\dot \a} 
 \rt )^{\dag}
\eb
+
\oy_{Qcpi\dot \a}
\lt ( t_{p  q} \ve_{ij}  K^j T_c^{ q}
\oC_{\dot \a} 
+
b_{p  q} \ve_{ij}   H^j B_c^{ q}
\oC_{\dot \a} 
 \rt )^{\dag}
\eb+
\oy_{Lpi\dot \a}\lt ( 
p_{p q} \ve_{ij}   H^j P^{ q} 
\oC_{\dot \a} 
+
r_{p q} \ve_{ij}   K^j R^{ q}
\oC_{\dot \a} 
 \rt )^{\dag}
\eb+
\oy_{Hi\dot \a}
\lt ( 
g \ve_{ij}   K^j J
\oC_{\dot \a} +
p_{p q} \ve_{ji} L^{p j}  P^{ q} 
\oC_{\dot \a} +
b_{p  q} \ve_{ji} Q^{c p j}  B_c^{ q}
\oC_{\dot \a} 
 \rt )^{\dag}
\eb+
\oy_{Ki\dot \a}\lt ( 
g \ve_{ji} H^j  J
\oC_{\dot \a} 
+
r_{p q} \ve_{ji} L^{p j} R^{ q}
\oC_{\dot \a} 
+
t_{p  q} \ve_{ji} Q^{c p j}   T_c^{ q}
\oC_{\dot \a} 
 \rt )^{\dag}
+*
\ee

The differential $d_7$ in the CSSM has a lot of terms.  We will pick out the terms that affect one particular example here, so that we can see how this works.  Let us continue with the example of (\ref{omegaP})
\be
    {\w}^{p}_{  \; P \dot \a}
=
(\oC \x^2 \oC)\lt \{
g^{-1} {P}^{p } {\oy}_{J\dot \a}
+
(p^{-1})^{pq}   {K}^{ i}  
{\oy}_{L i q \dot \a}  
\rt \}
\la{theexamplepjkl}
\ee
Again, we can ignore the factor $(\oC \x^2 \oC)
$ here.   We need to collect the relevant terms in 
\be
\d_5 \d_3^{\dag}\d_5 
    {\w}^{p}_{  \; P \dot \a}
\ee 
This yields  (using identities like $k^i k_i=0$) 
\be
\d_5 \d_3^{\dag}\d_5 
    {\w}^{s}_{  \; P \dot \a}
=
m \lt ( 
p_{p q} \ve_{ij}   h^j P^{ q} 
 \rt )
\oC_{\dot \d} \oy_{Lpi\dot \d}^{\dag}
\oy_{Lrk\dot \g}
\lt ( 
p_{ru} \ve_{kl}   H^l P^{u} 
\oC_{\dot \g} 
 \rt )^{\dag}
\eb
m  
g \ve_{mn} (
 H^m k^n) 
 \oC_{\dot \b}
\oy_{J\dot \b}^{\dag}
g^{-1} {P}^{s } {\oy}_{J\dot \a}
\eb
+
m \lt ( 
p_{p q} \ve_{ij}   h^j P^{ q} 
 \rt )
\oC_{\dot \d} \oy_{Lpi\dot \d}^{\dag}
\oy_{Lrk\dot \g}
\lt ( 
r_{ru} \ve_{kl}   K^l R^{u}
\oC_{\dot \g} 
 \rt )^{\dag}
\eb
m \lt ( 
r_{vw} \ve_{i_1i_2}   k^{i_2} R^{ w}
 \rt )
\oC_{\dot \e} \oy_{Lvi_1\dot \e}^{\dag}
(p^{-1})^{st}   {K}^{ i_3}  
{\oy}_{L  t i_3\dot \a}  
\ee
which reduces to
\[
\d_5 \d_3^{\dag}\d_5 
    {\w}^{s}_{  \; P \dot \a}
=
m^2   \oC_{\dot \a} 
\op^{rs}  p_{r q}   
   P^{ q}  
(
h^j    
 \ve_{jn}   k^n 
)
\]\[
-
m^2 
\oC_{\dot \a} 
(p^{-1})^{st}   
r_{tu} 
{\ov r}^{ru}    
p_{r q} 
 P^{ q}     k^{i}  
\ve_{ij}   h^j 
\]\be
=
m^2   \oC_{\dot \a} 
\lt \{
\op^{rs}  
+
(p^{-1})^{st}   
r_{tu} 
{\ov r}^{ru}    
\rt \}
p_{r q} 
 P^{ q}  
(
h^j    
 \ve_{jn}   k^n 
)
\la{resultofd7}
 \ee
Consider the matrix $\lt \{
\op^{rs}  
+
(p^{-1})^{st}   
r_{tu} 
{\ov r}^{ru}    
\rt \}
$ in (\ref{resultofd7}) and multiply it by $p_{zs}$.  We get:
\be
p_{zs}
\lt \{
\op^{rs}  
+
(p^{-1})^{st}   
r_{tu} 
{\ov r}^{ru}    
\rt \}
=
\lt \{
p_{zs}\op^{rs}  
+
r_{zs} 
{\ov r}^{rs}    
\rt \}
\ee
This is a sum of two positive definite matrices, assuming that the matrices $r$ and $p$ are non-singular, so it is also positive definite, 
and so this implies that 
\be
d_7    {\w}^{r}_{  \; P \dot \a}
\neq 0
\ee
Since this is not zero, we get 
\be
  {\w}^{r}_{  \; P \dot \a}
\not \in E_8
\ee
and this operator is removed from $E_{\infty}$ and the corresponding superfield integral expression is removed from ${\cal H}$.  

We have assumed here that  the target in 
(\ref{resultofd7}) is actually in the space $E_7$.  In the next subsubsection we show that this is so.

\subsubsection{An Important Example with ghost charge ${\cal N}_{\rm Ghost} =1$}

Here we will verify that the target in 
(\ref{resultofd7}) is actually in the space $E_7$.

This target has the form
\be
(\oC \x^2 \oC) m^2 q_k A^k 
\oC_{\dot \a} \ee

How can we see that 

\be
(\oC \x^2 \oC) m^2 q_k A^k 
\oC_{\dot \a} \in E_7 ?
\ee

Note that if   $f^i  g_{ijk}\neq 0$, then  
$(\oC \x^2 \oC) m f^i \oy_{i \dot \a} $ would  be missing from $E_4$, because:
\be
(\oC \x^2 \oC) m f^i \oy_{i \dot \a} 
\stackrel{d_3}{\ra}
(\oC \x^2 \oC)f^i  g_{ijk} m A^j A^k 
\oC_{\dot \a} \neq 0
\ee
In other words, then
\be
(\oC \x^2 \oC) m f^i \oy_{i \dot \a} \not \in E_4
\ee

If this were not true, then one could have had 
\be
(\oC \x^2 \oC) m f^i \oy_{i \dot \a} 
\stackrel{d_5}{\ra}
(\oC \x^2 \oC) g_{ijk} f^i m^2 v^j A^k 
\oC_{\dot \a} 
\la{latternever}
\ee
and one would need to worry whether
\be
q_k = g_{ijk} f^i v^j
\ee
but since   $f^i  g_{ijk}\neq 0$, this latter equation (\ref{latternever})
can never arise.  It follows that
\be
(\oC \x^2 \oC) m^2 q_k A^k 
\oC_{\dot \a} \in E_7
\ee
for any vector $q_k$. Of course the consequence of equation 
(\ref{resultofd7}) is that there are some vectors  $q_k$ for which this does not survive to $E_8$ or $E_9= E_{\infty}$.  We have 
\be
(\oC \x^2 \oC) m^2 q_k A^k 
\oC_{\dot \a} \not \in  E_{\infty}
\ee
for some vectors $q_k$, namely those which correspond to the Quarks and Leptons.

\subsection{Superfield Version}

 The Quark and Lepton Outfields are quite viable before the VEV appears, and then they mix with the elementary Quarks and Leptons when the VEV appears. 
The reason why this takes place is easy to see in superfield language.  If we define a slightly improved\footnote{This redefinition avoids some complications that are reflected in the discussion of  $d_5$ above in subsubsection \ref{dfiveforssmsubsection}} version as follows:

\be
{\widehat \w}^{p}_{P \dot \a}
=
g^{-1} {\widehat P}^{p } {\widehat  \oy}_{J\dot \a}
+
(p^{-1})^{pq}  \lt ( m k^i+  {\widehat K}^{ i} \rt )
{\widehat \oy}_{L i q \dot \a}  
\la{theexamplepjklnew}
\ee
then  the resulting  superfield  satisfies the simple  relation:
\be
\dB   {\widehat \w}^{p}_{  \; P \dot \a}
=(C Q+ \oC \oQ)  {\widehat \w}^{p}_{  \; P \dot \a}
+ m^2  h \cd k \oC_{\dot \a} 
  {\widehat P}^{p} 
\la{ssmmixingaftergaugebreaking}
\ee
The last term ensures that this is not a superfield, and so 
\be
   \int d^4 x \;d^2 \q \; {\w}^{p}_{  \; P \dot \a}
\not \in {\cal H}
\la{jumpsoutofcohom}
\ee
for the massive interacting CSSM, although this is in ${\cal H}$ for the massless interacting CSSM. In other words, this is removed from the cohomology space by the spontaneous breaking of the internal symmetry, which is what generates the mass. The other Quark and Lepton composite chiral dotted spinor superfields mentioned above work the same way.

\subsection{The operator $d_8$ for the CSSM} 

\la{d8ssmsubsection}

We will not attempt to find examples for $d_8$ in this paper.  In particular, this calls for a treatment of the gauge theory.

\section{Conclusion}

\la{conclusionchap}

We summarized our   general result for the cohomology space ${\cal H}$ for the free massless theory in 
section \ref{cohomsum}. One needs to substitute into that formula the expressions in section \ref{gendiscusssect} for the Outfields and the scalar Superfields. Since the superfield formalism and the physical formalism have isomorphic cohomology spaces, one can do this with either set of expressions in section \ref{gendiscusssect}.

We noted that this result was derived and proved in the Appendices, which lead up to the  $E_{\infty}$ space for the free massless theory in section \ref{breifsummsec}. 

When interactions or masses are present, one needs to subject the $E_{\infty}$ space in section \ref{breifsummsec}   to the Elizabethan Drama, which kills various expressions according to the $d_r$ summarized in 
Table \ref{summaryofdiffs}. The details of how these work, and how they are derived, are in the Appendices.

Our most interesting results so far are in section
\ref{cssmexamples}. In that section we wrote down the Outfields that correspond to Leptons and Quarks in the CSSM, and showed that they were in the Cohomology space for the interacting massless theory, by satisfying the $d_3$ constraint, and how they were removed from that space by the $d_7$ constraint when the VEV arises. 

These methods can be used to do much more cohomology work for all sorts of examples, both in the CSSM and in other theories.  It is not at all simple to summarize the results for the interacting or massive theories, since they are very tied to the details of the superpotential, through the constraints in Table 
\ref{summaryofdiffs}.  We have not attempted any sort of summary, but for any given case one can apply the constraints by applying the differentials and working out the solutions, as we have done for the examples in section
\ref{cssmexamples}.

We have not attempted to solve the unseparated irregular parts of the cohomology, even for the free massless theory, in this paper.  That is not an easy task.  It may well have interesting implications. 

We have learned a few things of a general nature:
\ben
\item
The transition from $E_{\infty}$ to ${\cal H}$ can be quite strange.  One is given the dimension and spin and other quantum numbers, and then there is a leap involved to deduce a form for ${\cal H}$ , given the results for 
$E_{\infty}$.  In the present case that leap results in a jump out of superspace.  
\item
The derivation of the $d_r$, given the form of the grading, is another task that requires some serious labour.  Even when one has found the $d_r$, and they become more hard to find as r increases, one has the puzzle of what they mean.  This search can be helped by finding the meaning of the expressions back in the starting space ${\cal P}$ to some extent.
\item
In the present case, the $d_r, r=3,4...8$ involve the coefficients in the superpotential.  Solving the resulting constraints has brought us into contact with Outfields that look like the Quarks and Leptons, except that they are composite chiral dotted spinor pseudosuperfields.  Again, the meaning, or usefulness, of these expressions is not obvious.  We will not try to discuss those possibilities here.
\item
These Quark and Lepton Outfields point to a number of remarkable things that are happening:
\ben
\item
The constraint equations from $d_3$ allow one to contruct Outfields from any invariance of the superpotential.  For the case of the Quark and Lepton Outfields, this invariance does not survive the introduction of the term 
$m^2 J$ into the superpotential. Indeed it is not even a symmetry of the rest of the action, even before the introduction of $m^2 J$.
\item
This means that the Quark and Lepton Outfields do not survive when the VEV develops, so there is some sort of a transition that takes place at that time. 
\item
For that matter there is the question of what the meaning of the peculiar invariances are that allow the Quark and Lepton Outfields to exist for the free interacting theory.  These have nothing to do with gauge invariance, but a great deal to do with the structure of the CSSM.
\een
\een

\subsection{Envoi}

The methods used here can be used with advantage for other problems, such as the supersymmetric gauge theories.  Preliminary  results there are rather similar to those here, and it is clear that the Higgs sector mixes with the gauge sector through various constraint equations.

Supergravity for 3+1 dimensions would be interesting to analyze in this way.  There are  a number of papers that deal with chiral supersymmetry and the gauge theory and supergravity \ci{Brandt}.  Those papers do not use the present methods, and they do not appear to  include the parts of the cohomology space that has Lorentz or internal indices, so none of the present results about Outfields appear in those papers. It is not a simple task to compare the results of those papers and the present paper, and I will not attempt to do so. 

Higher dimension supersymmetric theories, and  supersymmetric theories with $N\geq 2$ provide an interesting arena also. 
It is clear that the results in this paper depend crucially on the Weyl spinor formulation of this $3+1$ dimensional supersymmetric theory. It is not at all obvious what happens in other dimensions, or for supergravity.  Clearly one could try to analyze D=10 Yang-Mills theory or supergravity.  Absence of auxiliary fields is no problem, because the analog of the operator (\ref{thebigoperator}) still exists for those theories.

The interpretation of the results in this paper is another difficult issue that requires attention.  For example, it is an interesting and peculiar fact that chiral dotted spinor superfield cannot be coupled to supergravity \ci{westpage138}, and yet, as we have seen, the BRST cohomology of the rigid chiral theory contains  a complicated set of these.

The lesson to be learned from the spectral sequence here is that it is very easy to make errors and leave things out, but gradually the problem solves itself, if one looks at the machinery long enough. Even the mathematicians find the spectral sequence obscure and tricky to work with.  So it is a gradual process to get the entire result correct, especially for something as complicated as the present problem.  Spectral sequences have a lot of power, but they are  not at all obvious or easy to see through.  Consider the example of $d_7$ in  equation \ref{whatadiff}. There are a great many possible ways to construct such operators, most of which yield nothing.  As Samuel Johnson might have said, if he had the opportunity to live in a more enlightened age,  and to study spectral sequences \ci{oxfordquotes}:

{\em Sir, [it] is like a dog's walking on his hinder legs.  It is not done well; but you are surprised to see it done at all.}

Spectral Sequences have an interesting history.  See for example \ci{dieudonne,aboutjeanleray}.
To conclude, here is a quote from  G. W. Whitehead (taken from \ci{mclearypage62}):

{\em "The machinery of spectral sequences, stemming from the algebraic work of Lyndon and Koszul, seemed complicated and obscure to many topologists.  Nevertheless it was successful..."}

\hspace{2.5in} {\bf Acknowledgments}
\vspace{.2in}

I thank J.C. Taylor for stimulating correspondence and conversations over many years.  Raymond Stora suggested that I learn spectral sequences, which was a good idea, although a challenging one, and it has certainly been an interesting journey.  Liam O'Raiffeartaigh expressed keen encouragement, as did Victor Elias, George Leibbrandt and Julius Wess, and I miss them.   I have benefited from  encouragement, healthy scepticism   and useful feedback, over the many years that this paper took to write,  from a large number of physicists, including Ben Allanach, Dick Arnowitt, Jonathan Bagger, Carlo Becchi,  Margaret Blair, Cliff Burgess, Friedemann Brandt,  Philip Candelas, James Dodd, Louise Dolan,  Mike Duff,  Keith Ellis,  Alon Faraggi,   Paul Frampton, Sebastian Frank,  M. K. Gaillard, Gian Giudice, Marc Grisaru,  Howard Haber, Marc Henneaux, Bob Holdom, Ron Horgan,  Andre Lucas, Tim Jones, Ramzi Khuri, Andre Lucas, Steve Martin, Ruben Minasian, Rob Myers,  Pran Nath, Hans Peter Nilles,  Hugh Osborne, Heath Pois, Joe Polchinski, Erich Poppitz,  Chris Pope, Pierre Ramond, Joachim Rahmfeld, Peter Scharbach, Ergin Sezgin, 
  Yuri Shifman, Richard Staltenburg,  Tom Steele, Kelly Stelle,, Mark Walton,  Steven Weinberg,  Peter West,  Ed Witten and Bruno Zumino and I thank them all.

\appendix

\section{Preliminary Matters}

\la{derivofe3chap}

\subsection{Counting Operators}

\la{countingopssubsection}

We define the Dimension so that the Action has dimension zero and the Lagrangian has dimension four: 
\be
N_{\rm Dim} =
  N_{\pa} -\fr{1}{2}
{N}_C 
-\fr{1}{2}
{ {N}}_{\ov C}
-  N_{\x}
\eb
+ \fr{3}{2}
N_{\y}
+ \fr{3}{2}
N_{\oy}
+ N_{A} + N_{\ov A} 
+ 2 N_{F} + 2 N_{\ov F} 
\eb
+ \fr{5}{2}
N_{Y}
+ \fr{5}{2}
N_{\ov Y}
+ 3
N_{\G}
+ 3 N_{\ov \G}
+  2 N_{\ov \Lambda}
+ 
2 N_{\Lambda}
+  N_{m}
\ee
The Zinn counting operator is
\be
N_{\rm Zinn} =
N_{Y}
+ 
N_{\ov Y}
+ 
N_{\G}
+  N_{\ov \G}
+   N_{\ov \Lambda}
+ 
 N_{\Lambda}
\ee

We define the ghost number so that  the action has ghost number zero.  This means that the Lagrangian has ghost number $-4$, because it gets integrated with the 4-form $\int d^4 x$ which has ghost number 4: 
\be
N_{\rm Ghost} =
- 4 + 4 N_{\rm dx} 
+
{N}_C 
+
{ {N}}_{\ov C}
+
  N_{\x}
-
 N_{\rm Zinn}
\la{ghostnumber}
\ee
It is also useful to define Form number to be 
\be
N_{\rm Form} =
N_{\rm Ghost} +4
\la{formnumber}
\ee
The Grading is defined by (\ref{NewGrading}), which we repeat here for convenience:
\be
{N}_{\rm  Grading} 
= 
{N}_C
+
{ {N}}_{\ov C}
+
2  N_{\x}
+
2  N_{\rm Fields}
+
2  N_{\rm Zinn}
+
4  N_{\rm m}
\ee

The grading satisfies the following:
\be
\lt [N_{\rm Grading} , \d_r \rt ] = r \d_r
\la{meaningofgrading}
\ee
\be
\lt [N_{\rm Grading} , d_r \rt ] = r d_r
\ee
Most relevant operators commute with the dimension:
\be
\lt [N_{\rm Dim} , \d \rt ] =0
\ee
\be
\lt [N_{\rm Dim} , \d_r \rt ] = 0
\ee
\be
\lt [N_{\rm Dim} , d_r \rt ] = 0
\ee
Many relevant operators have ghost charge one:
\be
\lt [N_{\rm Ghost} , \d \rt ] = \d
\ee
\be
\lt [N_{\rm Ghost} , \d_r \rt ] = \d_r
\ee
\be
\lt [N_{\rm Ghost} , d_r \rt ] = d_r
\ee

There is a mapping
\be
{\rm Map}: \;
E_{\infty} \Ra
{\cal H}
\ee
This mapping  conserves all the quantum numbers that are not violated by the grading, such as Baryon number, Lepton number, spin etc. So this mapping satisfies
\be
\lt [N_{\rm Ghost} , {\rm Map} \rt ] = 0
\ee
\be
\lt [N_{\rm Dim} , {\rm Map} \rt ] = 0
\ee
for example.

\subsection{Conversion to spinor indices}
It is very helpful to convert from Lorentz indices to Spinor Indices by the transformation:
\la{qfeqwfqeerghhrw}
\be
\pa_{\m_1}  \cdots \pa_{\m_n }A^i \equiv 
A^i_{\m_1 \cdots \m_n } \ra 
A^i_{\a_1 \dot \b_1 , \cdots \a_n  \dot \b_n} 
=
A^i_{\m_1 \cdots \m_n }   
\s^{\m_1}_{\a_1 \dot \b_1 }
\cdots 
\s^{\m_n}_{\a_n  \dot \b_n} 
\ee
and the symmetry
\be
A^i_{\m_1 \m_2 \cdots \m_n } =
A^i_{\m_2 \m_1 \cdots \m_n }  
\ee
becomes
\be
A^i_{\a_1 \dot \b_1 ,\a_2 \dot \b_2 , \cdots \a_n  \dot \b_n} 
= 
A^i_{\a_2 \dot \b_2 ,\a_1 \dot \b_1 , \cdots \a_n  \dot \b_n} 
\ee
Then for example the equation
\be
[g^{\m_1 \m_2} A^i_{\m_1 \m_2 \cdots \m_n} ]^{\dag}
E_1=0
\mbox{ for } n \geq 2
\ee
is equivalent to
\be
[ \e^{\a_1 \a_2} \e^{\dot \b_1 \dot \b_2} A^i_{\a_1 \dot \b_1 , \cdots \a_n  \dot \b_n}]^{\dag}  E_1=0
\mbox{ for } n \geq 2
\ee
but because of the symmetry above this   is equivalent to
\be
[\e^{\a_1 \a_2}   A^i_{\a_1 \dot \b_1 , \cdots \a_n  \dot \b_n}
]^{\dag}  E_1=0
\mbox{ for } n \geq 2
\la{concateqcc}
\ee
and it is also equivalent to 
\be
[ \e^{\dot \b_1 \dot \b_2} A^i_{\a_1 \dot \b_1 , \cdots \a_n  \dot \b_n} ]^{\dag}  E_1=0
\mbox{ for } n \geq 2
\la{concateq}
\ee
Define
\be
A^i_{(\a_1 \ldots \a_n), (\dot \b_1 \ldots \dot \b_n)} 
= \fr{1}{n!} \S_{ {\rm Permutations} (1 \cdots n) \ra (j_1 \cdots j_n)}   A^i_{\a_{j_1} \dot \b_1 , \a_{j_2} \dot \b_2 ,   \cdots \a_{j_n}  \dot \b_n} 
\ee

Then evidently we have
\be
 \e^{ \g_1  \g_2}  A^i_{(\g_1 \ldots \g_n), (\dot \d_1 \ldots \dot \d_n)} 
=0
\mbox{ for } n \geq 2
\ee
Denote the set of all such symmetrized variables like
$
A^i_{(\a_1 \ldots \a_n), (\dot \b_1 \ldots \dot \b_n)} $ by
$A_{\rm Sym}$.
\be
A_{\rm Sym}=
A^i 
,
A^i_{\a_1, \dot \b_1} 
\cdots
A^i_{(\a_1 \ldots \a_n), (\dot \b_1 \ldots \dot \b_n)} 
\cdots
\ee
Then the general solution of  equation  
(\ref{concateq}) or (\ref{concateqcc}) is 
\be
E_1 = E_1 \lt [  A_{\rm Sym} \rt ]
\ee

These symmetrized variables will be used frequently.

\section{The Operator $d_0$ and the  Space $E_1=\ker d_0 \cap \ker d_0^{\dag} \cap E_0$} 

\la{descripofe1subsection}
This first operator contains the  structure constants of supersymmetry and the kinetic terms for the fields.  
\be
d_{0} \equiv \d_{0} = \d_{{\rm Structure}}  +
\d_{0 \;{\rm Matter}} 
\ee
where
\be
\d_{{\rm Structure}}  =
 -   C_{\a} {\ov C}_{\dot \b}
\fr{\pa  }{\pa { \x}^{\a \dot    \b}} \equiv
 -   C_{\a} {\ov C}_{\dot \b}
  \x^{\a \dot    \b \dag} 
\la{structurediff}
\ee
 \be
\d_{0 \;{\rm Matter}} =
-\int d^4 x \;
\lt (
  \fr{1}{2} \pa_{ \a \dot \b  }       \pa^{ \a \dot \b  }        {\ov  A}_{i} 
 \fr{\d  }{\d \G_i } 
+
  \pa^{\a \dot \b  }   
{\ov \y}_{i   \dot \b}
\fr{\d }{\d Y_{i}^{ \a} } 
+ \ov F_i 
\fr{\d }{\d \Lambda_{i}} 
+*
 \rt )
\la{kineticstuff}
\ee
The various kinetic terms help to define the physically meaningful parts of the various fields. The above form  is for the superfield operator.  For the physical operator the last term $+ \ov F_i 
\fr{\d }{\d \Lambda_{i}} 
$ is not present.

\subsection{The form of  $\D_0$ and the form of $E_1$  }

\la{deltazerochap}

First we note that the operator
\be
\d= \int d^4 x \; \ov F_i 
\fr{\d }{\d \Lambda_{i}} 
\ee has trivial cohomology. The adjoint is
\be
\d^{\dag} =\int d^4 x \; \Lambda_i 
\fr{\d }{\d \ov F_{i}} 
\ee
and the Laplacian is
\be
\D= (\d+\d^{\dag})^2= \int d^4 x \; 
\lt \{
\ov F_i 
\fr{\d }{\d \ov F_i } +
 \Lambda_i 
\fr{\d }{\d \Lambda_i } 
\rt \}=  N_{\oF} + N_{\Lambda} 
\ee
The kernel of this Laplacian is independent of 
$\ov F_i $ and $\Lambda_i$, so the kernel is trivial and the cohomology is trivial.  The complex conjugates work the same way, of course. 

This operator $\int d^4 x \; \ov F_i 
\fr{\d }{\d \Lambda_{i}} 
$ only appears for the superfield approach, but one sees that the superfield approach  is quickly reduced to the  physical approach from the above elimination of the auxiliary fields and their Zinn sources from the space $E_1$.  

We need to do more work for the physical fields.

The Laplacian $ \D_{\rm Kinetic} $ has a simple structure when it is expressed in terms of the variables
\be
\pa_{\a_1 \dot \b_1} \cdots \pa_{\a_n \dot \b_n} A^i \equiv
A^i_{\a_1 \dot \b_1,  \cdots \a_n \dot \b_n}, \; {\rm etc.}
\ee
Then we have 
\be
\d_{0 } = 
 \d_{0\; {\rm Matter} } +  \d_{0\; {\rm Structure} }
\la{qwrweegetg}
\ee
where 
\be 
 \d_{0\; {\rm Matter} } =  \sum_{n=0}^{\infty} 
{\ov \y}^{\dot \b  }_{i  \a \dot \b   \a_1 \dot \b_1 \cdots  \a_n \dot \b_n   }   
{Y}^{    \dag}_{i  \a,  \a_1 \dot \b_1 \cdots  \a_n \dot \b_n   }   + {\rm c.c.}
\ee 
\be
+\sum_{n=2}^{\infty} 
 \ve^{ \a_1 \a_2}  \ve^{ \dot \b_1   \dot \b_2   } {\ov A}_{i \a_1 \dot \b_1   \a_2 \dot \b_2    \a_3 \dot \b_3 \cdots  \a_n \dot \b_n }
{\G}_{i  \a_3 \dot \b_3 \cdots  \a_n \dot \b_n }^{\dag}
+ {\rm c.c.}
\ee

and
\be
 \d_{{\rm Structure} }=  C_{\a}  {\ov C}_{\dot \b}  \x_{\a \dot \b}^{\dag}
\ee

\subsection{First form of $\d_0^{\dag}$ }

The adjoint of the above is
\be
\d_{0 }^{\dag} = 
 \d_{0\; {\rm Matter} }^{\dag}  +  \d_{{\rm Structure} }^{\dag}
\ee
where
\be 
 \d_{0\; {\rm Matter} }^{\dag} =  \sum_{n=0}^{\infty} 
{Y}^{  }_{i  \a,  \a_1 \dot \b_1 \cdots  \a_n \dot \b_n   } {\ov \y}^{\dot \b    \dag}_{i  \a \dot \b   \a_1 \dot \b_1 \cdots  \a_n \dot \b_n   }     + {\rm c.c.}
\ee 
\be
+\sum_{n=2}^{\infty} 
{\G}_{i  \a_3 \dot \b_3 \cdots  \a_n \dot \b_n }
 \ve_{ \a_1 \a_2}  \ve_{ \dot \b_1   \dot \b_2   }  {\ov A}^{\dag}_{i \a_1 \dot \b_1   \a_2 \dot \b_2    \a_3 \dot \b_3 \cdots  \a_n \dot \b_n } + {\rm c.c.}
\ee
and
\be
\d_{{\rm Structure} }^{\dag} =
  \x_{\a \dot \b}C_{\a}^{\dag}  {\ov C}_{\dot \b}^{\dag} 
\ee
Note that the operators $\d_{0\; {\rm Matter} } $   and $\d_{{\rm Structure} }$  anticommute with each other and with the adjoint of each other, because they are made from completely different fields:
Thus:
\be
\lt \{ \d_{0\; {\rm Matter} } ,\d_{{\rm Structure} } \rt \}
=
\lt \{ \d_{0\; {\rm Matter} } ,\d_{{\rm Structure} }^{\dag} \rt \}
=0
\la{qgrhwrhhrt1}
\ee
Define the Laplacians
\be
\D_{0\; {\rm Matter} } = 
\lt \{ \d_{0\; {\rm Matter} } ,\d_{0\; {\rm Matter} }^{\dag} 
\rt \}
\ee
\be
\D_{{\rm Structure} } = 
\lt \{ \d_{{\rm Structure} },\d_{{\rm Structure} }^{\dag} \rt \}
\ee
The anticommutation rules above 
in (\ref{qgrhwrhhrt1})
mean that  the Laplacians commute
\be
\lt [ \D_{0\; {\rm Matter} } ,\D_{{\rm Structure} } \rt ]
=0
\ee
This implies that the Laplacian of the operator 
(\ref{qwrweegetg}) 
is the direct sum of these two Laplacians:
\be
\D_0 = 
 \D_{0\; {\rm Matter} }     +  \D_{{\rm Structure} } 
\ee
This means that the space $E_1$ is the intersection of the kernels of these two operators:
\be
E_{1} = \ker 
\D_0 = 
\ker \D_{0\; {\rm Matter} }  \cap  
\ker \D_{{\rm Structure} } 
\ee

\subsection{The form of the Laplacian $ \D_{0\; {\rm Matter} } $ }

\la{ghwhhrthrthrthrt}

From
\be 
 \d_{0\; {\rm Matter} } =  \sum_{n=0}^{\infty} 
{\ov \y}^{\dot \b  }_{i  \a \dot \b   \a_1 \dot \b_1 \cdots  \a_n \dot \b_n   }   
{Y}^{    \dag}_{i  \a,  \a_1 \dot \b_1 \cdots  \a_n \dot \b_n   }   + {\rm c.c.}
\ee 
\be
+\sum_{n=2}^{\infty} 
 \ve^{ \a_1 \a_2}  \ve^{ \dot \b_1   \dot \b_2   } {\ov A}_{i \a_1 \dot \b_1   \a_2 \dot \b_2    \a_3 \dot \b_3 \cdots  \a_n \dot \b_n }
{\G}_{i  \a_3 \dot \b_3 \cdots  \a_n \dot \b_n }^{\dag}
+ {\rm c.c.}
\ee
and
\be 
 \d_{0\; {\rm Matter} }^{\dag} =  \sum_{n=0}^{\infty} 
{Y}^{  }_{i  \a,  \a_1 \dot \b_1 \cdots  \a_n \dot \b_n   } {\ov \y}^{\dot \b    \dag}_{i  \a \dot \b   \a_1 \dot \b_1 \cdots  \a_n \dot \b_n   }     + {\rm c.c.}
\ee 
Consider the following terms in 
\be 
\lt \{
 \d_{0\; {\rm Matter} } , \d_{0\; {\rm Matter} }^{\dag}\rt \}
\ee

namely
\be
\lt \{
{\ov \y}^{\dot \b  }_{i  \a \dot \b   \a_1 \dot \b_1 \cdots  \a_n \dot \b_n   }   
{Y}^{    \dag}_{i  \a,  \a_1 \dot \b_1 \cdots  \a_n \dot \b_n   }  ,
{Y}^{  }_{j  \g,  \g_1 \dot \d_1 \cdots  \g_n \dot \d_n   } {\ov \y}^{\dot \d    \dag}_{j  \g \dot \d   \g_1 \dot \d_1 \cdots  \g_n \dot \d_n   }  
\rt \}
\ee 
The ${\ov \y}{\ov \y}^{\dag}$ term is the sum of terms like:
\be
{\ov \y}^{\dot \b  }_{i  \a \dot \b   \a_1 \dot \b_1 \cdots  \a_n \dot \b_n   }  {\ov \y}^{\dot \d    \dag}_{j  \g \dot \d   \g_1 \dot \d_1 \cdots  \g_n \dot \d_n   }  
\lt (
 \d^{i}_{j} \d^{  \a}_{\g} \d^{ \a_1 \dot \b_1, \cdots  \a_n \dot \b_n   }_{ \g_1 \dot \d_1, \cdots  \g_n \dot \d_n   } 
\rt )
\ee 
\be
=
{\ov \y}^{\dot \b  }_{i  \a \dot \b   \a_1 \dot \b_1 \cdots  \a_n \dot \b_n   }  {\ov \y}^{\dot \d    \dag}_{i  \a \dot \d   \a_1 \dot \b_1 \cdots  \a_n \dot \b_n   }  
\ee 
This is a sum of terms each of which is positive definite, with the result shown for the kernel of the Laplacian shown in subsection \ref{gqthhtrjhrtwert}. 

The $YY^{\dag}$ term is the sum of terms like:
\be
{Y}^{  }_{j  \g,  \g_1 \dot \d_1 \cdots  \g_n \dot \d_n   } 
 {Y}^{    \dag}_{i  \a,  \a_1 \dot \b_1 \cdots  \a_n \dot \b_n   }
\d_{ \dot \d}^{ \dot \b}
\d_{i}^{j} \eb
\lt (
 \d_{  \a \dot \b}^{ \g \dot \d} \d_{ \a_1 \dot \b_1, \cdots  \a_n \dot \b_n   }
^{  \g_1 \dot \d_1 ,\cdots  \g_n \dot \d_n   } 
+n
 \d_{  \a_1 \dot \b_1}^{ \g \dot \d} \d_{ \a \dot \b,   \a_2 \dot \b_2 \cdots  \a_n \dot \b_n   }
^{  \g_1 \dot \d_1 ,\cdots  \g_n \dot \d_n   } 
\rt )
\eb= 
{Y}^{  }_{i  \g,  \g_1 \dot \d_1 \cdots  \g_n \dot \d_n   } 
 {Y}^{    \dag}_{i  \a,  \a_1 \dot \b_1 \cdots  \a_n \dot \b_n   }
  \eb
\lt (
2 \d_{  \a  }^{ \g  } \d_{ \a_1 \dot \b_1, \cdots  \a_n \dot \b_n   }
^{  \g_1 \dot \d_1 ,\cdots  \g_n \dot \d_n   } 
+n
 \d_{  \a_1  }^{ \g  } \d_{ \a \dot \b_1,   \a_2 \dot \b_2 \cdots  \a_n \dot \b_n   }
^{  \g_1 \dot \d_1 ,\cdots  \g_n \dot \d_n   } 
\rt )
\eb= 
2 {Y}^{  }_{i  \a,  \a_1 \dot \b_1 \cdots  \a_n \dot \b_n   } 
 {Y}^{    \dag}_{i  \a,  \a_1 \dot \b_1 \cdots  \a_n \dot \b_n   }
  \eb
+ n  {Y}^{  }_{i  \a_1,  \a \dot \b_1 \cdots  \a_n \dot \b_n   } 
 {Y}^{    \dag}_{i  \a,  \a_1 \dot \b_1 \cdots  \a_n \dot \b_n   }
\ee 
This is not so obviously a sum of terms each of which is positive definite.  However it can be rewritten so that we get  the result shown for the kernel of the Laplacian shown in subsection \ref{gqthhtrjhrtwert}. 

This is done by examining each term by itself.  For example  for $n=0$ we have:
\be 
2 {Y}^{  }_{i  \a    } 
 {Y}^{    \dag}_{i  \a   }
\ee 

For $n=1$ we need to do some careful symmetrization:
\be 
2 {Y}^{  }_{i  \a,  \a_1 \dot \b_1   } 
 {Y}^{    \dag}_{i  \a,  \a_1 \dot \b_1    }
  \eb
+  {Y}^{  }_{i  \a_1,  \a \dot \b_1    } 
 {Y}^{    \dag}_{i  \a,  \a_1 \dot \b_1    }
\eb=
\lt ( 2 {Y}^{  }_{i  \a,  \a_1 \dot \b_1   } 
+  {Y}^{  }_{i  \a_1,  \a \dot \b_1    } 
\rt )
 {Y}^{    \dag}_{i  \a,  \a_1 \dot \b_1    }
\eb=
   {Y}^{  }_{i  \a,  \a_1 \dot \b_1   }  {Y}^{    \dag}_{i  \a,  \a_1 \dot \b_1    }+
\lt (
  {Y}^{  }_{i  \a,  \a_1 \dot \b_1   } +  {Y}^{  }_{i  \a_1,  \a \dot \b_1    } 
\rt )
 {Y}^{    \dag}_{i  \a,  \a_1 \dot \b_1    }
\eb=
   {Y}^{  }_{i  \a,  \a_1 \dot \b_1   }  {Y}^{    \dag}_{i  \a,  \a_1 \dot \b_1    }
\eb
+
\fr{1}{2}
\lt (
  {Y}^{  }_{i  \a,  \a_1 \dot \b_1   } 
+  {Y}^{  }_{i  \a_1,  \a \dot \b_1    } 
\rt )
\lt (
 {Y}^{    \dag}_{i  \a,  \a_1 \dot \b_1    }
+ {Y}^{    \dag}_{i  \a_1,  \a \dot \b_1    }
\rt )
\ee
and now it can be seen that this is a sum of positive terms with the result in subsection \ref{gqthhtrjhrtwert}.
The rest of the demonstration is similar.

\subsection{Equations   from $ \D_{0\; {\rm Matter} }  E_1 = 0$ and the symmetrized fields 
${\cal F}_{\rm Sym}= {\cal F}   \cup{\ov {\cal F}}$
}
\la{gqthhtrjhrtwert}

The form of the Laplacian $ \D_{0\; {\rm Matter} } $ is discussed in Appendix \ref{ghwhhrthrthrthrt}, where we outline a method to  show that the resulting equations are of the form:

\be 
{Y}^{    \dag}_{i  \a, \a_1\dot \b_1, \cdots \a_n\dot \b_n }E_1 = 0
\mbox{ for } n \geq 0 \mbox{ (and c.c.)} 
\ee
\be
  {\G}_{i  \a_1\dot \b_1, \cdots \a_n\dot \b_n}^{\dag} E_1 =0
\mbox{ for } n \geq 0  \mbox{ (and c.c.)}
\ee

\be  
( \ve^{\dot \b\dot \b_1} {\ov \y}_{j , \dot \b, \a_1\dot \b_1, \cdots \a_n\dot \b_n  })^{\dag }
E_1 = 0
\mbox{ for } n \geq 0  \mbox{ (and c.c.)}
\ee
\be
(\ve^{\a_1  \a_2} \ve^{\dot \b_1\dot \b_2} {\ov A}_{j, \a_1\dot \b_1\cdots \a_n\dot \b_n})^{\dag}E_1 = 0
\mbox{ for } n \geq 0
 \mbox{ (and c.c.)}
\ee

These equations have the form discussed in subsection
 \ref{qfeqwfqeerghhrw} and consequently the solution is that one must define the symmetrized fields, and then $E_1$ is a function of those symmetrized fields. 

\be
A^i_{\a_1 \dot \b_1,  \cdots \a_n \dot \b_n} \ra 
A^i_{\a_1 \cdots \a_n, \dot \b_1 \cdots \dot \b_n} 
\equiv 
A^i_{(\a_1 \cdots \a_n), (\dot \b_1 \cdots \dot \b_n)} 
\la{physfield}
\ee
In other words, contractions such as 
\be
A^i_{a \dot \b , \g \dot \d} \e^{\dot \b \dot \d}
\ee
are not permitted in $E_1$. This makes the work after $E_1$ much easier. Similar considerations apply to the other fields. 
We shall denote the following collection of variables by the notation 
 ${\cal F} $ and we shall call them the chiral physical  fields:
\be
{\cal F}: \; A^i_{(\a_1 \ldots \a_n), (\dot \b_1 \ldots \dot \b_n)} 
\mbox{ for } n \geq 0
\eb
{\ov \y}_{i, (\dot \a \dot \a_1 \ldots \dot \a_n),(\b_1 \ldots \b_n)} 
\mbox{ for } n \geq 0
\la{chiralfields2}
\ee
We shall denote the following collection of variables by the notation 
 $ {\ov {\cal F}}$ and we shall call them the antichiral  physical  fields:

\be
{\ov {\cal F}}: \;
{\ov A}_{i, (  \dot \a_1 \ldots \dot \a_n),(\b_1 \ldots \b_n)} 
\mbox{ for } n \geq 0
\eb 
\y^i_{(\a \a_1 \ldots \a_n), (\dot \b_1 \ldots \dot \b_n)}
\mbox{ for } n \geq 0
\la{antichiralfields2}
\ee
 
The reason for these peculiar definitions will appear when we study the regular equations for the subspace $E_2$ below.

\la{physfields}
 The auxiliary fields and all the Zinns are eliminated by the Laplacian.  The spinor fields and the scalar fields survive into $E_1$, but only when totally symmetrized as explained above.

This holds for both the physical and the superfield approach, and it is a good start on understanding why these both have the same spectral sequence.

 We define the symmetric fields ${\cal F}_{\rm Sym}$ as the combination of the fields defined above:
\be
{\cal F}_{\rm Sym}= {\cal F}   \cup{\ov {\cal F}}
\ee

\subsection{Structure of $E_1$}

\la{formofe1sec}

The  cohomology of $\d_{ {\rm Structure}}$ was worked out in detail in \ci{dixmin}, and those results are used to write the following forms.  We can write the form of $E_1$ as follows:

\be
E_1 = 
  {\cal P} 
 + {\cal Q}  
   +  {\cal R}  
+
{\ov {\cal P}}  
+  
{\ov {\cal Q}}  
+  
{\ov {\cal R}}  
+  
{\cal S}_3  
\ee
where
\be
 {\cal P} = \sum_{n=0}^{\infty} 
{\cal P}_n
\ee
\be
 {\cal Q} = \sum_{n=0}^{\infty} 
{\cal Q}_{2+n}
\ee
\be
 {\cal R} = \sum_{n=0}^{\infty} 
{\cal R}_{4+n}
\ee
In this notation, the subscripts refer to the ghost charge of the object.  Thus for example ${\cal Q}_{2+n}$ has ghost charge ${2+n}$. 

${\cal R}_{4+n}$ etc. 
have the following more explicit forms in terms of the Ghosts:
 \be
\begin{tabular}{|c|c|}
\hline
\multicolumn{2}
{|c|}{Table \ref{termsinE1}: Terms in $E_1$ }
\\
\multicolumn{2}
{|c|}{  Subscripts as in
${\cal P}_n$ are Ghost Number}
\\
\hline
{\rm Symbol} 
& \begin{tabular}{c}
Form with Explicit Ghosts
\end{tabular}
\\
\hline
${\cal P}_n;n=0,1,\cdots $
& \begin{tabular}{c}
${\cal P}_{\a_1 \cdots \a_n} C^{\a_1}\cdots  C^{\a_n}$
\end{tabular}
\\
\hline
${\cal {\ov P}}_n;n=0,1,\cdots $
& \begin{tabular}{c}
${\cal {\ov P}}_{\dot \a_1 \cdots \dot \a_n} \oC^{\dot \a_1}\cdots  
\oC^{\dot \a_n}$
\end{tabular}
\\
\hline
${\cal Q}_{n+2};n=0,1,\cdots $
& \begin{tabular}{c}
${\cal Q}_{\dot \b \a_1 \cdots \a_n} ( \x \cdot C)^{\dot \b}  C^{\a_1}\cdots  C^{\a_n}$
\end{tabular}
\\
\hline
${\cal \oQ}_{n+2};n=0,1,\cdots $
& \begin{tabular}{c}
${\cal \oQ}_{\b \dot \a_1 \cdots \dot \a_n} ( \x \cdot \oC)^{  \b}  \oC^{\dot \a_1}\cdots  \oC^{\dot \a_n}$
\end{tabular}
\\
\hline
${\cal R}_{n+4};n=0,1,\cdots $
& \begin{tabular}{c}
$({  C} \x^2 {  C}){\cal R}_{ \a_1 \cdots \a_n}    C^{\a_1}\cdots  C^{\a_n}$
\end{tabular}
\\
\hline
${\cal \ov R}_{n+4};n=0,1,\cdots $
& \begin{tabular}{c}
$({\ov C} \x^2 {\ov C}) {\cal \ov R}_{ \dot \a_1 \cdots \dot \a_n}  \oC^{\dot \a_1}\cdots  \oC^{\dot \a_n}$
\end{tabular}
\\
\hline
${\cal S}_{3}$
& \begin{tabular}{c}
$( C \x \oC) {\cal S}_0 $
\end{tabular}
\\
\hline
\hline
\end{tabular}
\la{termsinE1}
\ee

Another, more explicit, way to write  the space $E_1$ is:
\[
E_1 = 
  {\cal P}_0\left [{\cal F}_{\rm Sym}\right ] 
+ ( C \x  {\ov C} )
{\cal S}_0\left [{\cal F}_{\rm Sym}\right ] 
\]
\[
+ {\cal P}
\left [{\cal F}_{\rm Sym},C\right ] 
+ (\x   C)^{\dot \b} 
{\cal Q}_{\dot \b}\left [{\cal F}_{\rm Sym},C\right ] 
  + ( C \x^2  C) {\cal R} \left [{\cal F}_{\rm Sym},C\right ] 
\]
\be
+
{\ov {\cal P}} 
\left [{\cal F}_{\rm Sym},\ov C\right ] 
+ (\x {\ov C})^{\b} 
{\ov {\cal Q}}_{  \b}  
\left [{\cal F}_{\rm Sym},\ov C\right ] 
+ ({\ov C} \x^2 {\ov C}) \;
{\ov {\cal R}} 
\left [{\cal F}_{\rm Sym},\ov C\right ] 
\la{simpexfore1}
\ee

From the above discussion it follows that the various terms are  the most general  arbitrary local functions of
 ${\cal F}_{\rm Sym}$ and $C$ or $\oC$ as shown.   

Note that the space  $E_1$ has the following properties:
\ben
\item
There are no Zinn sources or auxiliary fields   in $E_1$.
\item
All the dependence on the fields in $E_1$ is through the symmetric variables ${\cal F}_{\rm Sym}$ defined in subsection  \ref{physfields}. 
\item

The $\x$ dependence of  $E_1$ is  restricted  to zero, one or two $\x$. 
\item
The $C$ and $\oC$ and $\x$  dependence are tied together  in a detailed way by contracting indices, as in the example $ (C \x)^{\dot \b} \equiv C_{\a} \x^{\a \dot \b}$.
\een

\section{The Operator $d_1$ and the  Space $E_{2{\rm Normal}}=\ker d_1 \cap \ker d_1^{\dag} \cap E_1\cap{\rm Normal}$} 

\la{regchap}

\subsection{Regular and Irregular Equations from  $d_1$  in the space $E_1$} 

The operator $d_1$ arises from  the supersymmetry translations:
\be
\d_1= \int d^4 x\; \lt (
\y^{i}_{  \b} {C}^{  \b}  \fr{\d}{\d  A^i}
+
\pa_{ \a \dot \b }  A^{i} {\ov C}^{\dot \b}  
\fr{\d }{ \d \y_{\a}^i  } 
+ \cdots + *
\la{supertrans}
\rt \}
\ee

Since we know that $E_1$ is independent of the Zinn sources and the auxiliary, the $\P_1$ orthogonal projection operator eliminates the auxiliary and Zinn terms in (\ref{supertrans}), and so the $d_1$ operator  simplifies to
\be
d_1 =
\P_1 
\lt \{
C^{\a} \na_{\a} + {\ov C}^{\dot \b}  {\ov \na}_{\dot \b} 
\rt \}
\P_1
\ee
where
\be
\na_{\a}
\equiv 
 \int d^4 x \;
\lt \{
 {  \y}^{i}_{  \a} 
\fr{\d}{\d A^{i} } 
+
 \pa_{\a \dot  \b} \A_i
\fr{\d}{\d {\oy}_{i \dot \b} } 
\rt \}
\la{rfweerhhjyetjyt}
\ee
This is true for both the physical and the superfield approach. We can write the operator $\na_{\a}$ in terms of the symmetrized physical variables as follows:
\be
\na_{\a } =  \sum_{n = 0}^{\infty }  \y^i_{\a \a_1 \cdots \a_n , \dot \b_1 \cdots \dot \b_n } {  A}_{ \a_1 \cdots \a_n , \dot \b_1 \cdots \dot \b_n }^{i \dag} \eb
+\sum_{n = 0}^{\infty }  \ov A_{i \a \a_1 \cdots \a_n , \dot \b \dot \b_1 \cdots \dot \b_n } { \ov \y}_{i \a_1 \cdots \a_n , \dot \b \dot \b_1 \cdots \dot \b_n }^{\dag} 
\ee
where symmetrization of the dotted and undotted indices is understood.  Because of the detailed structure of the cohomology of $C \oC \x^{\dag}$, the application of $d_1$ to $E_1$ breaks down to a number of different problems.  
First we have a set of equations which we call regular.  These take over for large values of the ghost number:
 \be
\begin{tabular}{|c|c|}
\hline
\multicolumn{2}
{|c|}{Table \ref{eqsforE2}: Regular equations  for $E_2$ }
\\
\multicolumn{2}
{|c|}{  Subscripts as in
${\cal P}_{n+1}$ are Ghost Number}
\\
\hline
{\rm Equation} 
& \begin{tabular}{c}
Mapping to:
\end{tabular}
\\
\hline
$ \P_1 ( C \na) {\cal P}_n =0;n=0,1,\cdots$ 
& \begin{tabular}{c}
$\P_1 {\cal P}_{n+1}$
\end{tabular}
\\
\hline
$\P_1  ( C \na) {\cal Q}_{2+n} =0;n=0,1,\cdots$ 
& \begin{tabular}{c}
$\P_1 {\cal Q}_{3 +n}$
\end{tabular}
\\
\hline
$ \P_1 ( C \na) {\cal R}_{4+n} =0;n=0,1,\cdots$ 
& \begin{tabular}{c}
$\P_1 {\cal R}_{5 +n}$
\end{tabular}
\\
\hline
$ \P_1 ( C \na)^{\dag} {\cal P}_n =0;n=2,3,\cdots$ 
& \begin{tabular}{c}
$\P_1 {\cal P}_{n-1}$
\end{tabular}
\\
\hline
$ \P_1 ( C \na)^{\dag} {\cal Q}_{4+n} =0;n=0,1,2,\cdots$ 
& \begin{tabular}{c}
$\P_1 {\cal Q}_{3 +n}$
\end{tabular}
\\
\hline
$ \P_1 ( C \na)^{\dag} {\cal R}_{4+n} =0;n= 1,2\cdots$ 
& \begin{tabular}{c}
$\P_1 {\cal R}_{3 +n}$
\end{tabular}
\\
\hline
\hline
\end{tabular}
\la{eqsforE2}
\ee
These are simple because the structure of $E_1$ is simple for the relevant sectors.

Then we have a set of equations which we call irregular.  These occur for low values of the ghost number where the structure of $E_1$ has a lot of detail:
 \be
\begin{tabular}{|c|c|}
\hline
\multicolumn{2}
{|c|}{Table \ref{irregeqsforE2}: Irregular equations  for $E_2$ }
\\
\multicolumn{2}
{|c|}{  Subscripts as in
${\cal P}_{n+1}$ are Ghost Number}
\\
\hline
{\rm Equation} 
& \begin{tabular}{c}
Mapping to:
\end{tabular}
\\
\hline
$ \P_1 \lt \{
( \oC \ov \na) {\cal Q}_{2}+
( C \na) {\cal \ov Q}_2 \rt \}
 =0 $ 
& \begin{tabular}{c}
$\P_1 {\cal S}_{3}$
\end{tabular}
\\
\hline
$ \P_1 \lt \{( C \na)^{\dag} {\cal P}_1 + ( \oC \ov \na)^{\dag} 
{\cal \ov P}_{1}\rt \}
=0$ 
& \begin{tabular}{c}
$\P_1 {\cal P}_{0}$
\end{tabular}
\\
\hline
$ \P_1 \lt \{( C \na)^{\dag} {\cal Q}_{3} + ( \oC \ov \na)^{\dag} 
{\cal S}_{3}\rt \}=0 $ 
& \begin{tabular}{c}
$\P_1 {\cal Q}_{2}$
\end{tabular}
\\
\hline
$ \P_1 \lt \{( C \na)^{\dag} {\cal S}_{3} + ( \oC \ov \na)^{\dag} 
{\cal \ov Q}_{3}\rt \}
=0$ 
& \begin{tabular}{c}
$\P_1 {\cal \ov Q}_{2}$
\end{tabular}
\\
\hline
\hline
\end{tabular}
\la{irregeqsforE2}
\ee
Sometimes it is useful to redefine some of the above as follows:
\be
{\cal Q}_{2 +n}= (\x C)_{\dot \b}
Q_{n}^{\dot \b}\; n=0,1,2\cdots
\ee
\be
{\cal S}_{3}= (C \x \oC)
S_{0}
\ee
\be
{\cal R}_{4 +n}= (C \x^2 C)
R_{n}\; n=0,1,2\cdots
\ee
If we do this then we can usually remove the $\P_1$ above, so long as we remember to use the symmetrized variables.
 
We call a term $T$ regular if it has two equations that apply to it:
\be
( C \na)T=0
\ee
\be
( C \na)^{\dag} T=0
\ee
and irregular if either of these is missing or changed.  
So in summary,  the following are regular:
  \be
{\cal P}_n \;
 {\rm for} \;n \geq 2;\;
{  Q}_{n}^{\dot \b} \;
 {\rm for}\; n \geq 2;\;
{  R}_n \;
 {\rm for} \;n \geq 1
\ee
and the following are irregular:
  \be
{\cal P}_0 \;
{\cal P}_1 \;
{  Q}_{0}^{\dot \b} \;
{  Q}_{1}^{\dot \b} \;
{  S}_0 \;
{  R}_0 \;
\ee
The regular equations can be solved in detail.  However this does require  a detailed  analysis.  That analysis follows in the rest of this section, and the results are given below in subsection \ref{regpartofe2}, starting with equation 
(\ref{firstoftheseaboute2reg}). The irregular equations tend to be  more difficult, and they must be treated one by one. Some solutions for them are contained in Appendix
\ref{simplifiedequas}.  They are done later, because the solution uses results from the intervening work.

\subsection{The operator $
 \na_{\a}$ expressed in terms of the multispinor variables} 

\la{delone}

 \subsubsection{  Normalization of Physical Variables}
\la{appendixnormphysvar}

\la{normsubsection}

There is some freedom in choosing the coefficients $a_n$ and 
 $x_n$ in the following:
\be
\Big [ A^{q  \dag }_{ \a_1 \a_2  \cdots \a_{n} , \dot \b_1 \cdots \dot \b_n} ,
A^{p }_{ \g_1 \g_2  \cdots \g_{n} , \dot \d_1 \cdots \dot \d_n} 
\Big ]
=
a_n \d^{p}_{q} \d^{ \a_1 \a_2  \cdots \a_{n}  }_{ \g_1 \g_2  \cdots \g_{n}  }
\d^{  \dot \b_1 \cdots \dot \b_n}_{ \dot \d_1 \cdots \dot \d_n} 
\la{firstcomm}
\ee
and
\be
\Big [ \y^{q  \dag }_{\a \a_1 \a_2  \cdots \a_{n} , \dot \b_1 \cdots \dot \b_n} ,
\y^{p }_{\g \g_1 \g_2  \cdots \g_{n} , \dot \d_1 \cdots \dot \d_n} 
\Big ] = x_n \d^{p}_{q}
  \d^{\a \a_1 \a_2  \cdots \a_{n}  }_{\g \g_1 \g_2  \cdots \g_{n}  }
\d^{  \dot \b_1 \cdots \dot \b_n}_{ \dot \d_1 \cdots \dot \d_n} 
\la{secondcomm}
\ee
Here we define the delta tensors to be  totally symmetric 
\be \d^{  \a_1 \a_2  \cdots \a_{n}  }_{  \g_1 \g_2  \cdots \g_{n}  } =
\d^{(  \a_1 \a_2  \cdots \a_{n}  )}_{(  \g_1 \g_2  \cdots \g_{n} ) } 
\ee
with unit weight, which means that:
\be 
\d^{ \a_1 \a_2  \cdots \a_{n}  }_{  \g_1 \g_2  \cdots \g_{n}  } \d^{  \g_1 \g_2  \cdots \g_{n}  }_{  \d_1 \d_2  \cdots \d_{n}  } 
= \d^{  \a_1 \a_2  \cdots \a_{n}}_{  \d_1 \d_2  \cdots \d_{n}  }
\la{unitweight}
\ee
There are $n!$ different kinds of terms   $\d^{ \a_1}_{  \g_n     }d^{ \a_2}_{  \g_m     }d^{ \a_3}_{  \g_p     }$ to completely symmetrize each of these tensors, and  there is a factor of $\fr{1}{n!}$ in front of the sum. For example
\be \d^{ \a_1 \a_2    }_{  \g_1 \g_2    } 
=
\fr{1}{2!}\lt \{
\d^{ \a_1}_{  \g_1     }  \d^{ \a_2    }_{    \g_2    } 
+
\d^{ \a_1}_{  \g_2     }  \d^{ \a_2    }_{    \g_1    } 
\rt \}
\ee

The equation (\ref{unitweight}) 
means that this tensor is idempotent, and so it plays a role in the projection operator:
\be
\P_1 A^{p }_{ \g_1 \dot \d_1,  \g_2 \dot \d_2, \cdots , \g_{n}  \dot \d_n} 
=
A^{p }_{ \g_1 \g_2  \cdots \g_{n} , \dot \d_1 \cdots \dot \d_n} 
\eb
=\d^{ \a_1 \a_2  \cdots \a_{n}  }_{  \g_1 \g_2  \cdots \g_{n}  } \d^{  \dot \b_1 \dot \b_2  \cdots \dot \b_{n}  }_{  \dot \d_1 \dot \d_2  \cdots \dot \d_{n}  } 
A^{p}_{ \a_1 \dot \b_1 ,\a_2 \dot \b_2,  \cdots ,  \a_{n}  \dot \b_n}
\ee

The following is a good choice for the normalization in  
(\ref{firstcomm}) and 
(\ref{secondcomm}), 
because it simplifies the algebra immensely:
\be
a_n = n! n!
\ee
\be
x_n = n! ( n+1)! 
\ee
This means that
we are removing the $\fr{1}{n!}$  for the unit weight tensors.  For example this results in
\be
\Big [ A^{q  \dag }_{ \a_1 \a_2  , \dot \b_1  \dot \b_2} ,
A^{p }_{ \g_1 \g_2    , \dot \d_1  \dot \d_2} 
\Big ]
=
\eb
\lt (
\d^{ \a_1}_{  \g_1     }  \d^{ \a_2    }_{    \g_2    } 
+
\d^{ \a_1}_{  \g_2     }  \d^{ \a_2    }_{    \g_1    } 
\rt )
 \lt (
\d^{  \dot \b_1}_{ \dot \d_1} 
\d^{  \dot \b_2}_{ \dot \d_2} 
+
\d^{  \dot \b_1}_{ \dot \d_2} 
\d^{  \dot \b_2}_{ \dot \d_1} 
\rt )\ee
We shall see how this choice simplifies the algebra  below.  

\subsubsection{The operator $
 \na_{\a}$} 

We have defined the operator $\na_{\a}$  by:
\be 
\P_1 C^{\a} \na_{\a}  \P_1
\eb
= \P_1  
\int d^4 x \;
\lt \{
  {  C}^{  \a  } {  \y}^{i}_{  \a} 
\fr{\d}{\d A^{i} } 
+
  {  C}^{  \a  } \pa_{\a \dot \b}  {\ov A}_{i} 
\fr{\d}{\d {\ov  \y}_{i  \dot  \b} }  
\rt \}
\P_1
\la{funcderivform}
\ee
Then we have
\be
d_1 = C^{\a} \na_{\a}  
+
{\ov C}^{\dot \a} 
{\ov \na}_{\dot \a} 
\ee

Given our choice of normalization in subsection
 \ref{normsubsection}, we have the multi-index form: 
\[
  \na_{\a}  =   \sum_{n=0}^{\infty}   \lt \{
 \fr{1}{n! n!} {  \y}^{i}_{  \a \a_1 \cdots \a_n, \dot \b_1 \cdots \dot \b_n} 
{  A}^{i \dag}_{    \a_1 \cdots \a_n, \dot \b_1 \cdots \dot \b_n} 
\rt.
 \]
\be  +   \fr{1}{(n+1)! n!} 
{\ov  A}_{ i , \dot \b_1 \cdots \dot \b_{n+1}, \a  \a_1 \cdots \a_{n}} 
{\ov  \y}^{\dag}_{i , \dot \b_1 \cdots \dot \b_{n+1},     \a_1 \cdots \a_n} 
 \la{multiderivform}
 \ee
This choice of normalization, together with the normalization chosen in subsection (\ref{normsubsection}), 
 ensures that the operator 
(\ref{multiderivform}) has the same effect as the functional derivative form 
(\ref{funcderivform}).

The hermitian conjugate is:
\be
  \na_{\a}^{\dag}   =   \sum_{n=0}^{\infty}   \lt \{
 \fr{1}{n! n!} 
{  A}^{i}_{    \a_1 \cdots \a_n, \dot \b_1 \cdots \dot \b_n} {  \y}^{i \dag}_{  \a \a_1 \cdots \a_n, \dot \b_1 \cdots \dot \b_n} 
\rt.
  \ee
\be
  +   \fr{1}{(n+1)! n!} 
{\ov  \y}_{i , \dot \b_1 \cdots \dot \b_{n+1},     \a_1 \cdots \a_n} 
{\ov  A}^{\dag}_{ i , \dot \b_1 \cdots \dot \b_{n+1}, \a  \a_1 \cdots \a_{n}}   \ee
and the complex conjugate is
\be
  {\ov \na}_{\dot \a}  =   \sum_{n=0}^{\infty}   \lt \{
 \fr{1}{n! n!} {  \oy}^{}_{i  \dot \a \dot  \a_1 \cdots \dot\a_n,  \b_1 \cdots \b_n} 
{  \A}^{ \dag}_{i \dot   \a_1 \cdots \dot\a_n,  \b_1 \cdots  \b_n} 
\rt.
  \ee
\be
  +   \fr{1}{(n+1)! n!} 
{  A}^{i}_{   \b_1 \cdots   \b_{n+1}, \dot \a  \dot \a_1 \cdots \dot \a_{n}} 
{  \y}^{i \dag}_{   \b_1 \cdots   \b_{n+1},   \dot  \a_1 \cdots \dot \a_n} 
  \ee
with hermitian conjugate
\be
  {\ov \na}_{\dot \a}^{\dag}  =   \sum_{n=0}^{\infty}   \lt \{
 \fr{1}{n! n!}  
{  \A}^{ }_{i \dot   \a_1 \cdots \dot\a_n,  \b_1 \cdots  \b_n} {  \oy}^{\dag}_{i  \dot \a \dot  \a_1 \cdots \dot\a_n,  \b_1 \cdots \b_n}
\rt.
  \ee
\be
  +   \fr{1}{(n+1)! n!}  
{  \y}^{i }_{   \b_1 \cdots   \b_{n+1},   \dot  \a_1 \cdots \dot \a_n} {  A}^{i\dag}_{   \b_1 \cdots   \b_{n+1}, \dot \a  \dot \a_1 \cdots \dot \a_{n}}
  \ee

\subsubsection{Explicit First Terms}
\la{firsttermssubsubsection}
Sometimes insight is gained by looking at the  first few terms, which  are
\be
\na_{\a} = 
\fr{1}{0! 0!}\y^i_{\a } A^{i \dag}
+ 
\fr{1}{1! 1!}\y^i_{\a \b \dot \g} A^{i \dag}_{\b \dot \g}
\eb
+ 
\fr{1}{1! 0!}{\ov A}_{ i \dot \b \a  } {\ov \y}_{i \dot \b }^{\dag}    + 
\fr{1}{2! 1!} 
{\ov A}_{ i \dot \b_1 \dot \b_2 \a \a_1  } 
{\ov \y}_{i \dot \b_1 \dot \b_2   \a_1   }^{\dag} + 
\cdots 
\ee
\be
{\ov \na}_{\dot \a} = 
\fr{1}{0! 0!}{\ov \y}_{i \dot \a } {\ov A}_i^{\dag}
+ 
\fr{1}{1! 0!}{  A}^i_{ \b \dot \a  } { \y}_{\b }^{i \dag}   + \cdots 
\ee

For actual calculations it is less confusing if one  divides this up in a form like this:

\be
\na_{\a}  = \na_{ \a ( \y A^{\dag})} + \na_{ \a (\A \oy^{\dag})}\ee
where
\be
\na_{ \a ( \y A^{\dag})} =  \sum_{n=0}^{\infty} 
 \fr{1}{n! n!} \na_{ \a,n(\y A^{\dag})}
\ee
\be
\na_{ \a ( \A \oy^{\dag})} =  \sum_{n=0}^{\infty} 
 \fr{1}{(n+1)! n! } \na_{ \a,n(\A \oy^{\dag})}
\ee

and we define the terms like this:
\be
  \na_{ \a,n(\y A^{\dag})}=
 {  \y}^{i}_{  \a \a_1 \cdots \a_n, \dot \b_1 \cdots \dot \b_n} 
{  A}^{i \dag}_{    \a_1 \cdots \a_n, \dot \b_1 \cdots \dot \b_n} 
  \ee
\be
 \na_{ \a,n(\A \oy^{\dag})}=   
{\ov  A}_{ i , \dot \b_1 \cdots \dot \b_{n+1}, \a  \a_1 \cdots \a_{n}} 
{\ov  \y}^{\dag}_{i , \dot \b_1 \cdots \dot \b_{n+1},     \a_1 \cdots \a_n} 
  \ee
The hermitian conjugates are
\be
  \na_{ \b,n( A \y^{\dag})}^{\dag}=
{ A}^{i}_{    \g_1 \cdots \g_n, \dot \d_1 \cdots \dot \d_n}  {\y}^{i \dag}_{  \b \g_1 \cdots \g_n, \dot \d_1 \cdots \dot \d_n} 
  \ee
\be
 \na_{ \b,n( \oy \A^{\dag})}^{\dag}=   
{\ov  \y}_{i , \dot \d_1 \cdots \dot \d_{n+1},     \g_1 \cdots \g_n} 
 {\ov  A}^{\dag}_{ i , \dot \d_1 \cdots \dot \d_{n+1}, \b  \g_1 \cdots \g_{n}} \ee

\subsection{Algebra of $\na_{\a},{\ov \na}_{\dot \b}$ and $\pa_{\a \dot \b}$ }

Since the multi-index form mimics the behaviour of the functional derivative form, one anticipates that they will have the same algebra, and they do.   Using methods similar to the above,  we can derive the following relations:

\be
 \pa_{\g \dot \d} = \sum_{n=0}^{\infty} 
  \left ( \fr{1}{n! n!} \pa_{\g \dot \d,n,A}
+ \fr{1}{n! (n+1)!}\pa_{\g \dot \d,n,\y} \right )
+*\ee
where
we define:
\be
\pa_{\g \dot \d,n,A} = 
 A^{a }_{\g \g_1   \cdots \g_{n} , \dot \d \dot \d_1 \cdots \dot \d_n} 
 A^{a \dag}_{ \g_1   \cdots \g_{n} , \dot \d_1 \cdots \dot \d_n} 
\ee
\be
\pa_{\g \dot \d,n,\y} = 
\y^{a  }_{ \g \g_1  \cdots \g_{n+1} ,\dot \d \dot \d_1 \cdots \dot \d_n}
\y^{a \dag }_{  \g_1  \cdots \g_{n+1}  , \dot \d_1 \cdots \dot \d_n}
\ee

Then we find that:

\be
  \lt \{
\na_{\a} , {\ov \na}_{\dot \a} 
\rt \}
 = \pa_{\a \dot \a}
\ee
and
\be
  \lt [
\na_{\a} , \pa_{\b \dot \g} 
\rt ]
=  \lt [
\ona_{\dot \a} , \pa_{\b \dot \g} 
\rt ]
=0
\ee
Note that
\be
\na_{\a  }   \na_{\b}  
 = - \fr{1}{2} \e_{ \a  \b} \na^{\g} \na_{\g} 
 = - \fr{1}{2} \e_{ \a  \b} ( \na )^2 
 \ee
Then
\be
  \lt [
( \na )^2 
 ,  {\ov \na}_{\dot \b}
\rt ]
=
 \lt \{ \na^{\a} ,
\lt \{ \na_{\a} 
 ,  {\ov \na}_{\dot \b}
\rt \}
\rt \}
 = 
2 { \na}^{\a}
 \pa_{ \a \dot \b}   \ee
The complex conjugate is
\be
  \lt [
 { \na}_{\b}
,
(\ov \na )^2 
 \rt ]
 = 
- 2 {\ov \na}^{\dot \a}
 \pa_{ \b \dot \a}    = 
 2 {\ov \na}_{\dot \a}
 \pa_{ \b}^{\;\; \dot \a}   \ee
A more complicated relation is
\be
  \lt [
( \na )^2 
 ,  ({\ov \na})^2 \rt ]
=
  \na^{\a} 
\lt [ \na_{\a} 
 ,  ({\ov \na})^2 \rt ]
 + 
\lt [ \na^{\a} 
 ,  ({\ov \na})^2 \rt ] \na_{\a} 
 \ee
\be
=
  \na^{\a} 
\lt [ \na_{\a} 
 ,  ({\ov \na})^2 \rt ]
 - 
\lt [ \na_{\a} 
 ,  ({\ov \na})^2 \rt ] \na^{\a} 
=
 \lt [ \na^{\a} ,
\lt [ \na_{\a} 
 ,  ({\ov \na})^2 \rt ]
\rt ]
\ee
\be
=  \lt [ \na^{\a} ,
- 2 {\ov \na}^{\dot \a}
 \pa_{ \a  \dot \a} 
\rt ]
 = 
- 2  \pa_{ \a \dot \a} \pa^{ \a \dot \a}  + 4 \pa_{ \a \dot \a} 
{\ov \na}^{\dot \a}  
 \na^{ \a }  
\ee
\be
= 
2  \pa_{ \a \dot \a} \pa^{ \a \dot \a}  - 4 \pa_{ \a \dot \a} 
 \na^{ \a }
 {\ov \na}^{\dot \a}  \ee

\subsection{Physical Anticommutator $ 
  \lt \{
\na_{\a  } ,  \na_{\b}^{\dag}
\rt \}
= \d_{ \a}^{ \b} { N} - 
\s_{ \a}^{i \b} R^i 
$}

\la{multiderivchap}

The difference between the functional derivative form and the multi-index form is of course that the latter has an adjoint form. We will use this in the formula
\be
\D_1 = \lt ( d_1 + d_1^{\dag} \rt )^2
\ee
to find the space $E_2$. 
First we need to derive the following important relation:
\be
  \lt \{
\na_{\a  } ,  \na_{\b}^{\dag}
\rt \}
= \d_{ \a}^{ \b} { N} - 
\s_{ \a}^{i \b} R^i 
\la{thebasicidentity}
\ee
The operators on the right are defined in subsection 
\ref{somedefsofops}.

We can start this calculation as follows:
\be
  \lt \{
\na_{\a  } ,  \na_{\b}^{\dag}
\rt \}
=
\lt \{
 \na_{ \a ( \y A^{\dag})} 
,  \na_{ \b ( A \y^{\dag})}^{\dag}  
\rt \}
+
\lt \{
   \na_{ \a (\A \oy^{\dag})}
,    \na_{ \b (\oy \A ^{\dag})}^{\dag}
\rt \}
\la{separatedstuff}
\ee

The first term in (\ref{separatedstuff}) is:
\be
\lt \{
 \na_{ \a ( \y A^{\dag})} 
,  \na_{ \b ( A \y^{\dag})}^{\dag}  
\rt \}
\ee
\be
= 
\sum_{n=0}^{\infty} 
 \fr{1}{n! n!} \fr{1}{n! n!} 
\eb
\lt \{
 {  \y}^{i}_{  \a \a_1 \cdots \a_n, \dot \b_1 \cdots \dot \b_n} 
{  A}^{i \dag}_{    \a_1 \cdots \a_n, \dot \b_1 \cdots \dot \b_n} 
,
{ A}^{i}_{    \g_1 \cdots \g_n, \dot \d_1 \cdots \dot \d_n}  {\y}^{i \dag}_{  \b \g_1 \cdots \g_n, \dot \d_1 \cdots \dot \d_n} 
 \rt \}
 \ee
Now  we can use the following inside the expression above
\be
\lt \{
 {\y}^{i \dag}_{  \b \g_1 \cdots \g_n, \dot \d_1 \cdots \dot \d_n} 
, {  \y}^{i}_{  \a \a_1 \cdots \a_n, \dot \b_1 \cdots \dot \b_n} 
\rt \}
\eb
= n! n! \d_{\a}^{\b} \d^{  \g_1 \cdots \g_n}_{\a_1 \cdots \a_n}
\d^{ \dot \d_1 \cdots \dot \d_n}_{   \dot \b_1 \cdots \dot \b_n} 
+ n n! n!\d_{\a}^{\g_1} \d^{  \b \g_2\cdots \g_n}_{\a_1 \cdots \a_n}
\d^{ \dot \d_1 \cdots \dot \d_n}_{   \dot \b_1 \cdots \dot \b_n} 
\ee
to get
\be
= 
\sum_{n=0}^{\infty} 
 \fr{1}{n! n!} \fr{1}{n! n!} 
\eb
\lt \{
n! n! {  \y}^{i}_{  \a \a_1 \cdots \a_n, \dot \b_1 \cdots \dot \b_n} 
  {\y}^{i \dag}_{  \b \a_1 \cdots \a_n, \dot \b_1 \cdots 
\dot \b_n} 
\ebp
+
 \lt ( n! n! \d_{\a}^{\b} \d^{  \g_1 \cdots \g_n}_{\a_1 \cdots \a_n}
\d^{ \dot \d_1 \cdots \dot \d_n}_{   \dot \b_1 \cdots \dot \b_n} 
+ n n! n!\d_{\a}^{\g_1} \d^{  \b \g_2\cdots \g_n}_{\a_1 \cdots \a_n}
\d^{ \dot \d_1 \cdots \dot \d_n}_{   \dot \b_1 \cdots \dot \b_n} 
\rt )
\ebp
{ A}^{i}_{    \g_1 \cdots \g_n, \dot \d_1 \cdots \dot \d_n} 
{  A}^{i \dag}_{    \a_1 \cdots \a_n, \dot \b_1 \cdots \dot \b_n} 
 \rt \}
 \ee
and this is
\be
= 
\sum_{n=0}^{\infty} 
 \fr{1}{n! n!} \fr{1}{n! n!} 
\eb
\lt \{
n! n! {  \y}^{i}_{  \a \a_1 \cdots \a_n, \dot \b_1 \cdots \dot \b_n} 
  {\y}^{i \dag}_{  \b \a_1 \cdots \a_n, \dot \b_1 \cdots 
\dot \b_n} 
\ebp
+
 n! n! \d_{\a}^{\b}  
{ A}^{i}_{    \g_1 \cdots \g_n, \dot \d_1 \cdots \dot \d_n} 
{  A}^{i \dag}_{    \g_1 \cdots \g_n, \dot \d_1 \cdots \dot \d_n} 
\ebp
+ n n! n! 
{ A}^{i}_{    \a \g_2 \cdots \g_n, \dot \d_1 \cdots \dot \d_n} 
{  A}^{i \dag}_{  \b  \g_2 \cdots \g_n, \dot \d_1 \cdots 
\dot \d_n} 
 \rt \}
 \ee
or
\be
= 
\sum_{n=0}^{\infty} 
\eb
\lt \{
 \fr{1}{n! n!}  
 {  \y}^{i}_{  \a \a_1 \cdots \a_n, \dot \b_1 \cdots \dot \b_n} 
  {\y}^{i \dag}_{  \b \a_1 \cdots \a_n, \dot \b_1 \cdots 
\dot \b_n} 
\ebp
+
 \fr{1}{n! n!}  
 \d_{\a}^{\b}  
{ A}^{i}_{    \g_1 \cdots \g_n, \dot \d_1 \cdots \dot \d_n} 
{  A}^{i \dag}_{    \g_1 \cdots \g_n, \dot \d_1 \cdots \dot \d_n} 
\ebp
+  \fr{1}{(n-1)! n!}  
{ A}^{i}_{    \a \g_2 \cdots \g_n, \dot \d_1 \cdots \dot \d_n} 
{  A}^{i \dag}_{  \b  \g_2 \cdots \g_n, \dot \d_1 \cdots 
\dot \d_n} 
 \rt \}
 \ee

Performing the same exercise for the second term on the right side of (\ref{separatedstuff}) 
results in the following total expression:
\be
  \lt \{
\na_{\a  } ,  \na_{\b}^{\dag}
\rt \}
 = \d_{ \a}^{ \b}   N_{\cal F} + R_{ \a}^{ \b} 
\la{nearlythere}
\ee
where: 
\[
N_{  {\cal F}} =
\sum_{n=1}^{\infty} 
\left \{
\fr{1}{n! n!} {A}^i_{  \a_1 \a_2    \cdots \a_{n} , \dot \b_1 \cdots \dot \b_n} 
{A}^{i \dag}_{ \a_1 \a_2  \cdots \a_{n} , \dot \b_1 \cdots \dot \b_n}
\right.
\]
\be
\left.
+
\fr{1}{(n+1)! n!} {\ov \y}_{i ,  \dot \b_1 \cdots \dot \b_{n+1},  \a_1 \a_2    \cdots \a_{n} } 
{\ov \y}^{\dag}_{i ,  \dot \b_1 \cdots \dot \b_{n+1},  \a_1 \a_2    \cdots \a_{n} } 
\right.
\la{nfcounter}
\ee
\[
N_{ \ov {\cal F}} =
\sum_{n=1}^{\infty} 
\left \{
\fr{1}{n! n!} {\A}_{i , \a_1 \a_2    \cdots \a_{n} , \dot \b_1 \cdots \dot \b_n} 
{\A}^{ \dag}_{i ,\a_1 \a_2  \cdots \a_{n} , \dot \b_1 \cdots \dot \b_n}
\right.
\]
\be
\left.
+
\fr{1}{(n+1)! n!} {\y}^i_{  \dot \b_1 \cdots \dot \b_{n+1},  \a_1 \a_2    \cdots \a_{n} } 
{\y}^{i \dag}_{ \dot \b_1 \cdots \dot \b_{n+1},  \a_1 \a_2    \cdots \a_{n} } 
\right.
\la{novfcounter}
\ee
and
\[
R_{\a}^{\;\;\b} =
\sum_{n=1}^{\infty} 
\left \{
\fr{1}{(n-1)! n!} {A}^i_{  \a \a_2    \cdots \a_{n} , \dot \b_1 \cdots \dot \b_n} 
{A}^{i \dag}_{ \b \a_2  \cdots \a_{n} , \dot \b_1 \cdots \dot \b_n}
\right.
\]
\[\left.
+
\fr{1}{n! (n-1)!} {\ov A}_{i ,  \dot \b_1 \cdots \dot \b_{n},  
\a \a_2    \cdots \a_{n} } 
{\ov A}^{\dag}_{i ,  \dot \b_1 \cdots \dot \b_{n},  \b \a_2    \cdots \a_{n} } 
\right.
\]
\[
+
\left.
\fr{1}{n! n!} {\y}^i_{  \a \a_1    \cdots \a_{n} , \dot \b_1 \cdots \dot \b_n} 
{\y}^{i \dag}_{i \b \a_1  \cdots \a_{n} , \dot \b_1 \cdots \dot \b_n}
\right.
\]
\be
\left.
+
\fr{1}{(n+1)! (n-1)!} {\ov \y}_{i ,  \dot \b_1 \cdots \dot \b_{n+1},  
\a \a_2    \cdots \a_{n} } 
{\ov \y}^{\dag}_{i ,  \dot \b_1 \cdots \dot \b_{n+1},  \b \a_2    \cdots \a_{n} } 
\right.
\la{fwewerfgwgwte4}
\ee

\subsection{Definitions of $N,N_{\cal F},N'$ and $R^i$}
\la{somedefsofops}.

As noted and derived above,  we have the important relation
(\ref{nearlythere}):
\be
  \lt \{
\na_{\a  } ,  \na_{\b}^{\dag}
\rt \}
 = \d_{ \a}^{ \b}   N_{\cal F} + R_{ \a}^{ \b} 
\ee
The operator 
$ N_{\cal F} $ 
counts the number of  $A$ and $\oy$ fields, but $R_{ \a}^{ \b}$ is more obscure. To make this more comprehensible we  define the following operators:
 \be
N' 
=
\d^{\a}_{\b} R^{\;\;\;\b}_{\a} 
\la{fwewerfgwgwte5}
\ee
and
\be
R^i
= 
-\fr{1}{2}
\s_{\a}^{i \b} R_{\b}^{\;\; \a}   
\la{fwewerfgwgwte6} \ee
 The inverse is
 \be
R^{\;\;\;\b}_{\a} = \fr{1}{2}N' 
\d^{\a}_{\b} 
- 
 R^i
\s_{\a}^{i \b}  
 \ee
and so we have 
\[
  \lt \{
\na_{\a  } ,  \na_{\b}^{\dag}
\rt \}
 = \d_{ \a}^{ \b} \lt ( N_{\cal F} + \fr{1}{2} N' \rt ) - 
\s_{ \a}^{i \b} R^i 
\]
\be
= \d_{ \a}^{ \b} { N} - 
\s_{ \a}^{i \b} R^i 
\la{derivoffundrel}
\ee
Here we define the counting operator
\be
N= \lt ( N_{\cal F} + \fr{1}{2} N' \rt )
\ee

So now we have arrived at the expression (\ref{derivoffundrel}), which is the same as
(\ref{thebasicidentity}).
A look at equation
(\ref{fwewerfgwgwte4}) and the definitions 
(\ref{fwewerfgwgwte5}) and (\ref{fwewerfgwgwte6})  shows 
 that: 
\ben
\item
The operator $N'$ counts the number of undotted spinor indices in an expression made of the fields $A^{j }_{ \g_1 \g_2  \cdots \g_{n} , \dot \d_1 \cdots \dot \d_n} $, $\A^{ }_{j, \g_1 \g_2  \cdots \g_{n} , \dot \d_1 \cdots \dot \d_n} $, $\y^{j}_{\g \g_1 \g_2  \cdots \g_{n} , \dot \d_1 \cdots \dot \d_n} 
$ and $
\oy^{}_{j,\g \g_1 \g_2  \cdots \g_{n} , \dot \d_1 \cdots \dot \d_n} 
$, for all values of $n$.
\item
The operator $R^i
$ is the properly normalized rotation operator that rotates each undotted spinor index on the fields $A^{j }_{ \g_1 \g_2  \cdots \g_{n} , \dot \d_1 \cdots \dot \d_n} $, $\A^{ }_{j, \g_1 \g_2  \cdots \g_{n} , \dot \d_1 \cdots \dot \d_n} $, $\y^{j}_{\g \g_1 \g_2  \cdots \g_{n} , \dot \d_1 \cdots \dot \d_n} 
$ and $
\oy^{}_{j,\g \g_1 \g_2  \cdots \g_{n} , \dot \d_1 \cdots \dot \d_n} 
$, for all values of $n$, as a spinor under SU(2).
\een

\subsection{Angular Momentum and Counting Operators}

Now it is easy to derive the following relations:
\be
\lt [     R^i  ,
\y^{\b} 
\rt ]
 =   \fr{1}{2} \y^{\g}
 \s^{i \b }_{\g}  \ee
Also
\be
\lt [     R^i  ,
\na^{\b} 
\rt ]
 =   \fr{1}{2} \s^{i \b }_{\g}    \na^{\g}
\ee
Since by inspection we can see that:
\be
\na_{\a} \approx {\ov {\cal F}}_{\a} {\cal F}^{\dag} 
; \na_{\a}^{\dag}  \approx  {\cal F} {\ov {\cal F}}_{\a}^{\dag} 
\la{setofeasy}
\ee
\be
\ona_{\dot \a} \approx { {\cal F}}_{\dot \a} {\cal \oF}^{\dag} 
; \ona_{\dot \a}^{\dag} \approx {\cal \oF} { {\cal F}}_{\dot \a} ^{\dag} 
\la{compsetofeasy}
\ee
we can easily derive relations like
\be
\lt [    N_{\ov {\cal F}}  ,
\na^{\b} 
\rt ]
 =    \na^{\b}
\ee
and
\be
\lt [    N_{  \cal F}  ,
\na^{\b} 
\rt ]
 =   - \na^{\b}
\ee
The operator that counts undotted indices also has a simple commutator with ${\na}_{\a}$: 
\be
\lt [    N' ,
\na^{\b} 
\rt ]
 =    \na^{\b}
\ee
Putting these together yields
\be
\lt [    N  ,
\na^{\b} 
\rt ]
=\lt [    N_{\cal F} + \fr{1}{2} N'   ,
\na^{\b} 
\rt ]
 =   - \fr{1}{2} \na^{\b}
\ee

\subsection{The Expressions $Z_+$ and $Z_-$}
\la{someidetnites}
Next we note the important identity:
\be
\lt \{
\d_{\a}^{\b} \lt (  N + 1 \rt )  + \s_{\a}^{i \b} R^i
\rt \}
\lt \{
\d_{\b}^{\g} \lt (  N   \rt )  - \s_{\b}^{i \g} R^i
\rt \}
\ee
\be
=
\d_{\a}^{\g}  N  \lt (  N + 1 \rt )  - \s_{\a}^{i \g} R^i   - \s_{\a}^{i \b} \s_{\b}^{j \g } R^i   R^j
=
 \d_{\a}^{\g} Z_+
 \ee
and also the identity:
\be
\lt \{
\d_{\a}^{\b} \lt (  N - 1 \rt )  - \s_{\a}^{i \b} R^i
\rt \}
\lt \{
\d_{\b}^{\g} \lt (  N   \rt )  + \s_{\b}^{i \g} R^i
\rt \}
\ee
\be
=
\d_{\a}^{\g}  N  \lt (  N - 1 \rt )  - \s_{\a}^{i \g} R^i   - \s_{\a}^{i \b} \s_{\b}^{j \g } R^i   R^j
=
 \d_{\a}^{\g} Z_-
 \ee
where
\be
Z_+ = 
  N   \lt (  N + 1 \rt )  -  R^i
R^i 
\la{Zplus}
 \ee
\be
Z_- = 
  N   \lt (  N - 1 \rt )  -  R^i
R^i 
\la{Zminus}
 \ee

\subsection{Solution of the regular equations in $E_2$}
\la{regsubsection}

Now we will solve the regular equations for the regular part of $E_2$.

The equations
\be
C^{\a} \na_{\a} {\cal E} = 0
\ee
\be
C^{\a \dag} \na_{\a}^{\dag} {\cal E} = 0
\ee
are equivalent to the Laplacian form
\be
\D_{1,{\rm regular}} {\cal E} = 0
\ee
where the Laplacian is
\be
\D_{1,{\rm regular}}
=
\lt \{
C^{\a} \na_{\a} 
,
C^{\b \dag} \na_{\b}^{\dag} 
\rt \}
\ee
In order for this to be true, we have to be careful that the $\P_1$ that is implicit in the operator $\na_{\a}$ is taken care of properly, by expressing it in terms of the physical variables that are used to construct $E_1$.

Now we note that
\be
\D_{1,{\rm regular}}
=
C^{\a}C^{\b \dag}
\lt \{
 \na_{\a} 
,
 \na_{\b}^{\dag} 
\rt \}
+ \na_{\a}^{\dag} 
 \na_{\a} 
\ee
We  calculated   this anticommutator in Appendix 
\ref{multiderivchap}, and the result is
\be
  \lt \{
\na_{\a  } ,  \na_{\b}^{\dag}
\rt \}
= \d_{ \a}^{ \b} N- 
\s_{ \a}^{i \b} R^i 
\ee
where we use the shorthand
\be
 N
=
 N_{\cal F} + \fr{1}{2} N'
\ee
Then the Laplacian is
\be
 \D_{1, {\rm regular}} =    \na_{\a}^{\dag} \na_{\a}
+ 
N_c  \lt ( N_{\cal F} + \fr{1}{2} N' \rt ) - 
2 J^i R^i
\la{egerpfewfp}
\ee
where $J^i$ is the properly normalized rotation operator that rotates the undotted spinor index of the ghost $C_{\a}$ as a spinor under SU(2):
\be
J^i = 
\fr{1}{2} \s_{ \a}^{i \b} C_{\b}
 C_{\a}^{ \dag} 
=-
\fr{1}{2} \s_{ \a}^{i \b}  C^{\a} 
C^{\b  \dag} 
\ee

We claim that this is a sum of three positive semi-definite hermitian  operators, with the following consequences: 

\be
 \D_{1, {\rm regular}} {\cal E} = 0      
\Ra
\ee
\be  
 N_c  N_{\cal F} {\cal E} = 0
\la{qregehrthhrt1}
\ee
\be
\na_{\a}^{\dag} \na_{\a} {\cal E}= 0
\la{qregehrthhrt2}
\ee
\be
\lt ( \fr{1}{2} N_c  N'   - 
2 J^i R^i \rt ) {\cal E} = 0 
\la{qregehrthhrt3}
\ee
In order to demonstrate that this split makes sense, 
 we must show that
\be
\lt ( \fr{1}{2} N_c N' - 2 J^i R^i \rt )   
\ee
 is a positive operator. Here is the demonstration of this fact.
First note that we can write:
\be
\lt ( \fr{1}{2} N_c N' - 2 J^i R^i \rt )  =
 \fr{1}{2} N_c   N'  
- 
(J^i + R^i)
(J^i + R^i)
+J^i J^i +R^i R^i 
\ee
\be
=
Z[J+R]
-Z[J]
-Z[R]
\la{qregehrthhrt4}
\ee
where
\be
 Z \lt [ J+ R \rt]  = \lt [ \fr{( N' + N_c )}{2} \fr{( N' + N_c +2 )}{2}  -  (J+ R)^i(J+ R)^i   \rt ] 
\ee
\be
 Z \lt [ J \rt]  = \lt [ \fr{   N_c }{2} \fr{  N_c +2 }{2}  -  J^i J^i   \rt ] 
\ee
\be
 Z \lt [ R \rt]  = \lt [ \fr{   N' }{2} \fr{  N' +2 }{2}  -  R^i R^i   \rt ] 
\ee
Now we note the following:
\ben
\item 
Firstly, the following equation really implies that
the relevant undotted indices are all symmetrized:
\be
 Z \lt [ J \rt ]   {\cal E}  = 0
\ee
\be
\equiv \lt [ \fr{( N_C  )}{2} \fr{( N_C  +2 )}{2}  -  (J )^i(J )^i   \rt ]   {\cal E} = 0
\ee
This is automatically satisfied because the indices involved here are the indices on $C_{\a_1}\cdots C_{\a_n}$ and they are always symmetric.
\item 

Secondly, the following equation really implies that
the relevant undotted indices are all symmetrized:
\be
 Z \lt [ R \rt ]   {\cal E}  = 0
\ee
\be
\equiv \lt [ \fr{( N'  )}{2} \fr{( N'  +2 )}{2}  -  (R )^i(R )^i   \rt ]   {\cal E} = 0
\ee
This is not automatically satisfied because the indices involved here are  all the undotted indices on all the variables $A^{j }_{ \g_1 \g_2  \cdots \g_{n} , \dot \d_1 \cdots \dot \d_n} $, $\A^{ }_{j, \g_1 \g_2  \cdots \g_{n} , \dot \d_1 \cdots \dot \d_n} $, $\y^{j}_{\g \g_1 \g_2  \cdots \g_{n} , \dot \d_1 \cdots \dot \d_n} 
$ and $
\oy^{}_{j,\g \g_1 \g_2  \cdots \g_{n} , \dot \d_1 \cdots \dot \d_n} 
$,  for all values of $n$, and they can be contracted between different fields, so they are not always automatically symmetric.
\item 

Thirdly, the equation
\be
 Z \lt [ J+ R \rt]  {\cal E}  = \eb
\lt [ \fr{( N' + N_c )}{2} \fr{( N' + N_c +2 )}{2}  -  (J+ R)^i(J+ R)^i   \rt ] {\cal E}=0
\ee
says that all the undotted indices in any homogeneous part of ${\cal E}$, both those on $C_{\a}$ and those on all the variables $A,\y,\A,\oy$ with any number of indices, are symmetrized.
\een
Now we need to show that 
\be
Z[J+R]
- Z[J]
- Z[R]
\la{thisherething}
\ee
is a positive operator.  Suppose then that we have some expression ${\cal E} $ which is an eigenvector of the hermitian matrices above in (\ref{thisherething}), and that 
\be
 Z[J]=0
\ee
(since this is always true) and 
\be
 Z \lt [ R \rt ]   {\cal E}  = 
 \lt [ \fr{( N'  )}{2} \fr{( N'  +2 )}{2}  -   \fr{( r )}{2} \fr{(r  +2 )}{2}   \rt ]   {\cal E}  
\ee
and that
\be
 Z \lt [ J+ R \rt] {\cal E}  = \lt [ \fr{( N' + N_c )}{2} \fr{( N' + N_c +2 )}{2}  -  \fr{( l )}{2} \fr{(l +2 )}{2}   \rt ] {\cal E} 
\ee
From the theory of the addition of angular momentum in three dimensions we know that
\be
N'  \geq r \geq 0
\la{gergwiojwgejio}
\ee
and
\be
r + N_c \geq l \geq |r - N_c|
\la{wgiojgroe}
\ee
So we have
\be
Z[J+R]
- Z[J]
- Z[R]
\ee
\be
= \lt [ \fr{( N' + N_c )}{2} \fr{( N' + N_c +2 )}{2}  -  \fr{( l )}{2} \fr{(l +2 )}{2}   \rt ] {\cal E}
\eb
-
 \lt [ \fr{( N'  )}{2} \fr{( N'  +2 )}{2}  -   \fr{( r )}{2}
 \fr{(r  +2 )}{2}   \rt ]   {\cal E} 
\ee
\be
=   \fr{( N' + N_c )}{2} \fr{( N' + N_c +2 )}{2} {\cal E}
-
   \fr{( N'  )}{2} \fr{( N'  +2 )}{2} {\cal E}
\eb
 +   \fr{( r )}{2}
 \fr{(r  +2 )}{2}    {\cal E} 
 -  \fr{( l )}{2} \fr{(l +2 )}{2}    {\cal E}
\ee

The smallest value of this occurs when
$l$ is largest, so we see from (\ref{wgiojgroe}) that:
\be
{\rm Min\; Value\; of\;} \lt [ Z[J+R]
- Z[J]
- Z[R]\rt ] {\cal E}
\ee
\be
=   \fr{2 N'  }{2} \fr{N_C }{2} {\cal E}
+  
\fr{N_C  }{2} \fr{( N_c +2 )}{2} {\cal E}
\eb
 +   \fr{( r )}{2}
 \fr{(r  +2 )}{2}    {\cal E} 
 -  \fr{( r + N_c )}{2} \fr{(r + N_c +2 )}{2}    {\cal E}
\ee\be
=   \fr{2 N'  }{2} \fr{N_C }{2} {\cal E}
 -  \fr{2 r }{2} \fr{  N_c }{2}    {\cal E}
\ee
and it is clear from 
(\ref{gergwiojwgejio}) that this is greater
than zero, except in the case where $r = N'$, when it 
is zero.

So we have shown that this is a positive operator, except in the case where all the undotted indices, both on $C_{\a}$ and on all the variables $A^{j }_{ \g_1 \g_2  \cdots \g_{n} , \dot \d_1 \cdots \dot \d_n} $, $\A^{ }_{j, \g_1 \g_2  \cdots \g_{n} , \dot \d_1 \cdots \dot \d_n} $, $\y^{j}_{\g \g_1 \g_2  \cdots \g_{n} , \dot \d_1 \cdots \dot \d_n} 
$ and $
\oy^{}_{j,\g \g_1 \g_2  \cdots \g_{n} , \dot \d_1 \cdots \dot \d_n} 
$,  for all values of $n$, are totally symmetrized.  then the operator has its lowest possible value, which is zero.  So it is a positive semi-definite operator.

\subsection{Summary of the regular part of $E_2$ assuming that  $N_C\neq0$ }

\la{regpartofe2}

So here are the equations that govern the solutions for the regular part of $E_2$, assuming that  $N_C\neq0$, collected together and explained:

\ben
\item
Firstly, we have from (\ref{qregehrthhrt1}) that
\be
N_c N_{\cal F} {\cal E} = 0
\la{firstoftheseaboute2reg}
\ee
and assuming that  $N_C\neq0$, this implies that
\be
N_{\cal F} {\cal E} = 0
\ee
and that means that we have
\be
{\cal E} = 
{\cal E}[ {\ov {\cal F}},C]
\la{qghergergrgre}
\ee

\item

Secondly, we have from (\ref{qregehrthhrt2}) that
\be
\na_{\a}  {\cal E} = 0
\ee
but this adds nothing since any polynomial which satisfies 
equation (\ref{qghergergrgre}) automatically solves this equation too, because $\na$ has the form $\na \approx {\ov {\cal F}}{{\cal F}}^{\dag}$.
\item

Thirdly we have from   (\ref{qregehrthhrt3}) and
  (\ref{qregehrthhrt4}) and the discussion after that: 
\be
Z[J+R] {\cal E} = 0
\ee
\be
Z[J] {\cal E} = 0
\ee
\be
Z[R] {\cal E} = 0
\ee
and these equations mean that 
 all the undotted indices, both on $C_{\a}$ and on  the antichiral physical symmetrized variables  \be
{\ov {\cal F}} \equiv \A^{ }_{j, \g_1 \g_2  \cdots \g_{n} , \dot \d_1 \cdots \dot \d_n},\y^{j}_{\g \g_1 \g_2  \cdots \g_{n} , \dot \d_1 \cdots \dot \d_n}, 
\ee   for all values of $n$, are totally symmetrized. 
\een

As noted above, these solutions apply to the following regular parts, all of which have $N_C\neq0$:

 \be
{\cal P}_n \;
 {\rm for} \;n \geq 2;
{\cal Q}_n \;
 {\rm for}\; n \geq 4;
{\cal R}_n \;
 {\rm for} \;n \geq 5
\la{theregstuffatd2}
\ee

but not to the following irregular parts:
\be
{\cal P}_0, {\cal P}_1, 
{\cal Q}_2, {\cal Q}_3, 
{\cal S}_3, {\cal R}_4
\ee

It can be verified that these, combined with the results for $d_3$ below, and the results for the separated irregular parts of $R_0$ and $S_0$ in          subsection    \ref{sepforsqq} and
\ref{sepirregRsec}, yield the descriptions given in subsection 
\ref{breifsummsec}.

\section{The Operator $d_2$ and the  Space $E_{3{\rm Normal}}
=\ker d_2 \cap \ker d_2^{\dag} \cap E_2\cap {\rm Normal}$} 

\la{d2chap}

\subsection{Regular and Irregular Equations from  $d_2$  in the space $E_2$}

In general

\be
d_2 = \P_2 
\left \{
\d_2 - \d_1 \fr{\d_0^{\dag}}{\D_0} \d_1 
\right \}
\P_2
\la{d2stuff}
\ee
So it has the form:
\be
d_{ {\rm Trans}} = \P_2 \lt \{ \x \pa -
\lt (C \na + {\ov C} {\ov \na} \rt )
\fr{\x_{\a \dot \b}  C_{\a}^{\dag} {\ov C}_{\dot \b}^{\dag} }{\D_{0}} 
\lt (C \na + {\ov C} {\ov \na} \rt )
 \rt \} \P_2
\ee
which immediately reduces to
\be
d_{ {\rm Trans}} = \P_2 \lt \{ \x \pa -
\lt (C \na + {\ov C} {\ov \na} \rt )
\fr{ 1  }{\D_{0}} 
\ebp
\lt (\x_{\a \dot \b} {\ov C}_{\dot \b}^{\dag} \na^{\a} + \x_{\a \dot \b}{\ov \na}^{\dot \b} C_{\a}^{\dag}\rt )
 \rt \} \P_2
\ee
and its adjoint is
\be
d_2^{\dag} =\P_2 \lt \{  \x^{\dag} \pa^{\dag} - (C^{\dag} \na^{\dag} + \ov C^{\dag} \ov \na^{\dag})\fr{ C_{\a}  {\ov C}_{\dot \b} 
\x_{\a \dot \b}^{\dag} }{\D_{0}} 
(C^{\dag}\na^{\dag} + \ov C^{\dag}  \ov \na^{\dag} ) \rt \} 
\P_2
\ee

Again it is easy to see that there is no difference between the physical and the superfield approach at this stage. As promised, it is evident that an understanding  of the  $\na_{\a}$ operators is essential to understand $d_2$, because $\na_{\a}$ occurs in it as well as in $d_1$.
Again there is a useful distinction to be made between the  regular equations and those which are irregular. 
\be
\begin{tabular}{|c|c|}
\hline
\multicolumn{2}
{|c|}{Table \ref{regeqsforE3}: Regular equations  for $E_3$ }
\\
\multicolumn{2}
{|c|}{  Subscripts as in ${\cal P}_{n}$  signify Form Number n}
\\
\hline
{\rm Equation} 
& \begin{tabular}{c}
Mapping to:
\end{tabular}
\\
\hline
$  
\P_2 ({\x \pa})^{\dag}
  {\cal { R}}_{n+4}; n=0,1,2 \cdots     =0$ 
& \begin{tabular}{c}
$\P_2{\cal { Q}}_{n+3}$
\end{tabular}
\\
\hline
$  
\P_2({\x \pa})^{\dag}
  {\cal { Q}}_{n+3}; n=0,1,2 \cdots     =0$ 
& \begin{tabular}{c}
$\P_2{\cal { P}}_{n+2}$
\end{tabular}
\\
\hline
$  
\P_2({\x \pa}) 
  {\cal { P}}_{n+2}; n=0,1,2 \cdots     =0$ 
& \begin{tabular}{c}
$\P_2{\cal { Q}}_{n+2}$
\end{tabular}
\\
\hline
$  
\P_2({\x \pa}) 
  {\cal { Q}}_{n+4}; n=0,1,2 \cdots     =0$ 
& \begin{tabular}{c}
$\P_2{\cal { R}}_{n+5}$
\end{tabular}
\\
\hline
\hline
\end{tabular}
\la{regeqsforE3}
\ee

and 
 \be
\begin{tabular}{|c|c|}
\hline
\multicolumn{2}
{|c|}{Table \ref{irregeqsforE3}: Irregular equations  for $E_3$ }
\\
\multicolumn{2}
{|c|}{   Subscripts as in ${\cal P}_{n}$  signify Form Number n}
\\
\hline
{\rm Equation} 
& \begin{tabular}{c}
Mapping to:
\end{tabular}
\\
\hline
$\P_2\lt \{ (\x \pa) {\cal P}_1+ (C\cdot \x \cdot {\ov C}^{\dag} ) \na^2    {\cal {\ov P}}_1 \rt \}
=0$ 
& \begin{tabular}{c}
$\P_2{\cal Q}_{2}$
\end{tabular}
\\
\hline
$\P_2\lt \{  (\x \pa) {\cal {\ov P}}_1+ (C^{\dag}\cdot \x \cdot {\ov C} ) 
{\ov \na}^2    {\cal { P}}_1 \rt \}
=0$ 
& \begin{tabular}{c}
$\P_2{\cal {\ov Q}}_{2}$
\end{tabular}
\\
\hline
\hline
$\P_2\lt \{  (\x \pa)^{\dag} {\cal Q}_2 + (C\cdot \x^{\dag}  \cdot {\ov C}^{\dag} ) 
({\ov \na}^{\dag})^2    {\cal {\ov Q}}_2 \rt \}
=0$ 
& \begin{tabular}{c}
$\P_2 {\cal P}_{1}$
\end{tabular}
\\
\hline
$\P_2\lt \{  (\x \pa)^{\dag} {\cal {\ov Q}}_2+ (C^{\dag}\cdot \x^{\dag}  \cdot {\ov C} ) 
({\na}^{\dag})^2    {\cal { Q}}_2 \rt \}
=0$ 
& \begin{tabular}{c}
$\P_2 {\cal {\ov P}}_{1}$
\end{tabular}
\\
\hline
\hline
$ \P_2\lt \{ (\x \pa) {\cal Q}_3+ (C\cdot \x \cdot {\ov C}^{\dag} ) 
{\na}^2    {\cal S}_3 \rt \}
=0$ 
& \begin{tabular}{c}
$\P_2 {\cal R}_{4}$
\end{tabular}
\\
\hline
$ \P_2\lt \{ (\x \pa) {\cal {\ov Q}}_3+ (C^{\dag} \cdot \x \cdot {\ov C}) 
{\ov \na}^2    {\cal S}_3 \rt \}
=0$ 
& \begin{tabular}{c}
$\P_2  {\cal {\ov R}}_{4}$
\end{tabular}
\\
\hline
\hline
$  
\P_2\lt \{ ({\ov \na}^{\dag})^2 (C\cdot \x^{\dag}  \cdot {\ov C}^{\dag} )  {\cal {\ov R}}_4 +
({ \na}^{\dag})^2    (C^{\dag}\cdot \x^{\dag}  \cdot {\ov C} )    {\cal R}_4  \rt \}  =0$ 
& \begin{tabular}{c}
$\P_2 {\cal S}_{3}$
\end{tabular}
\\
\hline
\hline
\end{tabular}
\la{irregeqsforE3}
\ee

We call a term T regular for this $d_2$ operator if it was regular for the $d_1$ operator, and if it also satisfies equations of the form
\be
\P_2 \x^{\a \dot \b} \pa_{\a \dot \b} T \equiv 
\P_2  (\x \pa) T=0
\ee
or the form
\be
\P_2 (\x \pa)^{\dag} T=0
\ee
or both of these. 
According to this definition, the following forms are  regular for this $d_2$ operator:

 \be
{\cal P}_n \;
 {\rm for} \;n \geq 2;
{\cal Q}_n \;
 {\rm for}\; n \geq 4;
{\cal R}_n \;
 {\rm for} \;n \geq 5
\la{theregstuffatd2agaibn}
\ee
All others are irregular.  These are\footnote{The terms ${\cal P}_0$ are gone because we can show they are zero in $E_2$.  This is done in subsection \ref{posubsection}}: 
\be
 {\cal P}_1,  
{\cal Q}_2, {\cal Q}_3, 
{\cal S}_3, {\cal R}_4
\ee

This spectral sequence ends with $d_2$, as we show in Appendix \ref{proofofcollpseappendix}.  Consequently $E_3= E_{\infty}$ for this case of the free massless theory.

Again, it is possible to solve the regular equations  in detail.  Again this  requires  a detailed  analysis.  That analysis follows in the rest of this section, and the results are given below in subsection \ref{regpartofe3}, starting with equation 
(\ref{firstoftheseaboute3reg}).

As noted above, the irregular equations   must be treated one by one. Some solutions for them are contained in Appendix
\ref{simplifiedequas}.

\subsection{The Simple Commutator $\lt [  \na_{\a}  , \pa_{\b \dot \b}^{\dag} 
\rt ]
= - \d_{\a}^{\b}  {\ov \na}_{\dot \b}^{\dag}
 $} 
Using the above we have:
\be
\lt [  \na_{\a}  , \pa_{\b \dot \b}^{\dag} 
\rt ]
\ee
\be
=
\lt [  \na_{\a}  , \lt \{ \na_{\b}^{\dag} , \ona_{\dot \b}^{\dag} 
\rt \}
\rt ]
\ee
\be
=
\lt [  \na_{\a}  ,   \na_{\b}^{\dag}  \ona_{\dot \b}^{\dag} 
\rt ]
+
\lt [  \na_{\a}  ,    \ona_{\dot \b}^{\dag} \na_{\b}^{\dag} 
\rt ]
\ee
\[
=
 \lt \{ \na_{\a}  , \na_{\b}^{\dag}   
\rt \} \ona_{\dot \b}^{\dag}
-
\na_{\b}^{\dag}  \lt \{ \na_{\a}  ,   \ona_{\dot \b}^{\dag} 
\rt \}
\]
\be
+
 \lt \{ \na_{\a}  , \ona_{\dot \b}^{\dag} 
\rt \} 
\na_{\b}^{\dag}   
-
\ona_{\dot \b}^{\dag} \lt \{ \na_{\a}  ,   \na_{\b}^{\dag} 
\rt \}
\la{uptoastage}
\ee
Now it is easy to see from the expressions (\ref{setofeasy}) and 
(\ref{compsetofeasy}) 
that
\be
  \lt \{ \na_{\a}  ,   \ona_{\dot \b}^{\dag} 
\rt \}=0
\ee
So (\ref{uptoastage})
reduces to 
\be
\lt [
 \lt \{ \na_{\a}  , \na_{\b}^{\dag}   
\rt \}, \ona_{\dot \b}^{\dag}
\rt ]
=  \lt [
\d_{\a}^{\b} N  - \s_{\a}^{i \b} R^i
, {\ov \na}_{\dot \b}^{\dag}
\rt ]
 \ee
\be
=  \d_{\a}^{\b} \lt [
 N_{\cal F}   
, {\ov \na}_{\dot \b}^{\dag}
\rt ]
= - \d_{\a}^{\b}  {\ov \na}_{\dot \b}^{\dag}
 \ee
because ${\ov \na}_{\dot \b}^{\dag}$ does not change the number of undotted indices, and it has no free undotted indices.
Thus we have derived the simple and important identity: 
\be
\lt [  \na_{\a}  , \pa_{\b \dot \b}^{\dag} 
\rt ]
= - \d_{\a}^{\b}  {\ov \na}_{\dot \b}^{\dag}
\ee

\subsection{Evaluation and discussion of $\left [ 
 {\pa}_{a \dot \b}^{\dag} , {\pa}_{\g \dot \d}  \rt ]$ }

\la{appendixpapadag}

This important commutator can be simply derived using the above information:
\be
  \lt [ \pa_{\a \dot \b}^{\dag} ,\pa_{\g \dot \d} \rt ]
 = 
  \lt [ \lt \{ \na_{\a}^{\dag}, {\ov \na}_{ \dot \b}^{\dag} 
\rt \}
,\pa_{\g \dot \d} \rt ]
\ee
\be
  = 
 \na_{\a}^{\dag}  \lt [  {\ov \na}_{ \dot \b}^{\dag} 
,\pa_{\g \dot \d} \rt ]
+  {\ov \na}_{ \dot \b}^{\dag} 
 \lt [  \na_{\a}^{\dag} ,\pa_{\g \dot \d} \rt ]
+   \lt [ \na_{\a}^{\dag} ,\pa_{\g \dot \d} \rt ]
{\ov \na}_{ \dot \b}^{\dag} 
+   \lt [ {\ov \na}_{ \dot \b}^{\dag}  ,\pa_{\g \dot \d} \rt ]
\na_{\a}^{\dag}
\ee
\be
  = 
 \na_{\a}^{\dag}  \lt [  \d^{ \dot \b}_{ \dot \d} \na_{\g } \rt ]
+  {\ov \na}_{ \dot \b}^{\dag} 
 \lt [  \d^{\a}_{\g} {\ov \na}_{\dot \d} \rt ]
+   \lt [ \d^{\a}_{\g} {\ov \na}_{\dot \d}\rt ]
{\ov \na}_{ \dot \b}^{\dag} 
+   \lt [ \d^{ \dot \b}_{ \dot \d} \na_{\g } \rt ]
\na_{\a}^{\dag}
\ee
\be
  = 
 \d^{ \dot \b}_{ \dot \d} \lt \{ \na_{\a}^{\dag}  ,  \na_{\g } \rt \}
+   \d^{\a}_{\g}  \lt \{ {\ov \na}_{ \dot \b}^{\dag} ,  {\ov \na}_{\dot \d} \rt \}
\ee
\be
  = 
 \d^{ \dot \b}_{ \dot \d} \lt \{ \d_{ \g}^{ \a} N - 
\s_{ \g}^{i \a} R^i 
 \rt \}
+   \d_{ \g}^{ \a}  \lt \{ \d^{\dot \b}_{\dot \d}  {\ov N} - 
{\ov \s}_{\dot \d}^{i \dot \b} {\ov R}^i 
 \rt \}
\ee
This can be written in the form:
\be
\left [    \pa_{\g \dot \d}^{\dag}, \pa_{\a \dot \b}  \right ] = 
\d^{ \g}_{\a} \d^{ \dot \d}_{\dot \b}   {\cal D} 
- \d^{ \dot \d}_{\dot \b} \s^{i \b}_{\a} R^i
- \s^{i \dot \d}_{\dot \b} \d^{\b}_{\a} {\ov R}^i
\ee
where the dimension operator (for the fields) is
\be {\cal D}_{\rm Phys}  = N_{\cal F} + N_{\ov {\cal F}} 
+ \fr{1}{2} N'
+ \fr{1}{2} {\dot N}'
\ee
So if we contract the indices, we get 
\be
\left [    \pa_{\a \dot \b}^{\dag}, \pa_{\a \dot \b}  \right ] = 
4   {\cal D}_{\rm Phys} 
\ee
Note also the useful relations:
\be
\left [  R^i , \pa_{\a \dot \b} \right ] = - \fr{1}{2} 
\s^{i \b}_{\a} \pa_{\a \dot \b} 
\ee
\be
\left [ {\cal D}_{\rm Phys}, \pa_{\a \dot \b} \right ] =  \pa_{\a \dot \b} 
\ee

\subsection{Solution of the regular part of $E_3$ }

\la{rege3app}

The proof here is analogous to that in  subsection 
(\ref{posubsection}) below.  If we can show that
\be
\pa_{\a \dot \b} T=0
\ee
then it follows that
\be
T=0.
\ee
On the other hand 
\be
\pa_{\a \dot \b}^{\dag} T=0
\la{ddagger}
\ee
has plenty of non-trivial solutions.

Here is how we do this.  To be specific, let us start with the example:
\be
d_{ {\rm Trans}}{\cal P}_2 = \P_2 \lt \{ \x \pa -
\lt (C \na + {\ov C} {\ov \na} \rt )
\fr{\x_{\a \dot \b}  C_{\a}^{\dag} {\ov C}_{\dot \b}^{\dag} }{\D_{0}} 
\lt (C \na + {\ov C} {\ov \na} \rt )  
 \rt \} \P_2{\cal P}^{\a \b}_2 C_{\a} C_{\b} 
\ee
which immediately reduces to
\be
d_{ {\rm Trans}} = \P_2 \lt \{ \x \pa -
\lt (C \na + {\ov C} {\ov \na} \rt )
\fr{ 1  }{\D_{0}} 
\ebp
\lt (\x_{\a \dot \b} {\ov C}_{\dot \b}^{\dag} \na^{\a} + \x_{\a \dot \b}{\ov \na}^{\dot \b} C_{\a}^{\dag}\rt )
 \rt \} \P_2 {\cal P}^{\g \d}_2 C_{\g} C_{\d}  
\eb
=
( \x C)^{\dot \b} \pa_{\a \dot \b} 
{\cal P}^{\g \a}_2 C_{\g}
\ee
where we have used the information that
\be
{\cal P}^{\g \a}_2={\cal P}^{\g \a}_2({\cal \oF})
\ee
At first glance one might think that there was a contribution here from 
\be
d_{ {\rm Trans}} = \P_2 \lt \{   {\ov C} {\ov \na}  
\x_{\a \dot \b}{\ov \na}^{\dot \b} C_{\a}^{\dag} 
 \rt \} \P_2 {\cal P}^{\g \d}_2 C_{\g} C_{\d}  
\eb
=
( \x \oC)^{\b}  \P_2 \ov \na^2 
   {\cal P}^{\g \a}_2 C_{\g}
\eb
=
(C \x \oC)    \P_2 \ov \na^2 
   {\cal P}^{\g \a}_2  \ve_{\a \g}
\equiv 0
\ee
but that is identically zero as shown.

So the big operator above is really equivalent to simply

\be
d_{ {\rm Trans}}{\cal P}_2 = \P_2   \x \pa  {\cal P}^{\a \b}_2 C_{\a} C_{\b} =0
\ee
and this amounts to 
\be
\P_2  \pa_{\a \dot \b}   {\cal P}^{\a \g}_2  =0
\ee
The indices make this look more complicated than it is.  
Judicious use of the form of $E_2$ makes the argument simple here. Let us consider an example to show how this works:

Suppose we start with say
\be
p_2
= a_i 
C_{(\a} C_{\b}\y^i_{\g)}  \in {\cal P}_2  \cap E_2
\ee
This is explicitly in the form required for it to be in ${\cal P}_2  \cap E_2$.  

Then 
\be
\P_2  (\x \pa) p_2
=  \P_2 a_i  \x^{\d \dot \z} 
C_{(\a} C_{\b} \y^i_{\g \d  \dot \z } 
=  a_i   
( \x C)^{\dot \z} C_{(\a} \y^i_{\b \g ) \dot \z } 
\in {\cal Q}_3  \cap E_2
\ee
where we have performed the projection for this to be in $E_2$ again. 

Now it is clear that for this to be zero, we require that
\be
a_i=0
\ee
  Adding more fields changes nothing essential. The requirement that $d_2$ yields zero kills all possible expressions that belong to ${\cal P}_2$ here. 
This is easily generalized to the other stages   ${\cal P}_n$ for $n \geq 2$.

Analysis of the equations like (\ref{ddagger}) 
 leaves plenty of room for non-zero objects in the cohomology space however.  For example if one starts with a set of fields with no derivatives, it is automatically zero from the start.

Here is another example that is a little different from the above.  We could start with
\be
q_4  
= a_i 
(\x C)_{\dot \z} C_{(\a} C_{\b} \y^i_{\g)}  \in E_2 \cap Q_4
\ee
This is explicitly in the form required for it to be in $E_2$.  
Then 
\be
 d_2 q_4  
= b_i 
\P_2 \x \pa (\x C)_{\dot \z} C_{(\a} C_{\b} \y^i_{\g)}   
\eb
= b_i 
(C \x^2 C)     C_{(\a} \y^i_{\g \b) \dot \z}  
\in E_2 \cap R_5  
\ee
and for this to be zero we require that $b_i=0$.
This is easily generalized to all possible  stages   ${\cal Q}_n$ for $n \geq 4$ with any number of fields.

So we see that for the regular equations, starting with 
(\ref{theregstuffatd2agaibn}),
 only the following survive to live in $E_3$:
 \be
{\cal R}_n \;
 {\rm for} \;n \geq 5
\ee
and they are subject to
\be
\P_2 \pa_{\a \dot \b}^{\dag} {\cal R}_n =0
\;
{\rm for} \;n \geq 5
\ee
as well as the equations that qualify them to 
be in $E_2$, which can be found in subsection
 \ref{regpartofe2}. 

\subsection{The Space ${\cal R}_n \cap E_3  \;
{\rm for} \;n \geq 5 $ for the free massless theory}
So, in summary, we have the following results for the regular part of the space $ E_3 $  for the free massless theory:

\la{regpartofe3}

We have shown how to prove that
\be
{\cal P}_{n} \cap E_3 = 0\;
{\rm for} \;n \geq 2
\la{firstoftheseaboute3reg}
\ee
and
\be
{\cal Q}_{n} \cap E_3 = 0\;
{\rm for} \;n \geq 4
\ee
We define  
\be
{\cal R}_{n+4} = (C \x^2 C) R_{n}\;
{\rm for} \;n \geq 0
\ee
where
\be
N_C R_{n} = n R_{n}
\ee
Then, for $n\geq1$ we have shown that:
\be
N_{\cal F} \lt \{ R_{n}\cap E_3 \rt \} = 0
\ee
and we also have the results
\be
\lt \{ R_{n}\cap E_3 \rt \} = 
\lt \{ R_{n}\cap E_3 \rt \} [ {\ov {\cal F}},C]
\ee
\be
\P_2 \pa_{\a \dot \b}^{\dag} \lt \{ R_{n}\cap E_3 \rt \} =0
\la{ddageqforr}
\ee
\be
Z[J+R] \lt \{ R_{n}\cap E_3 \rt \} = 0
\ee
\be
Z[J] \lt \{ R_{n}\cap E_3 \rt \} = 0
\ee
\be
Z[R] \lt \{ R_{n}\cap E_3 \rt \} = 0
\ee
The   equation (\ref{ddageqforr}) 
has lots of solutions.  It means that $\lt \{ R_{n}\cap E_3 \rt \}$ is not a total derivative. The last three equations mean that 
 all the undotted indices, both on $C_{\a}$ and on  the antichiral physical symmetrized variables  \be
{\ov {\cal F}} \equiv \A^{ }_{j, \g_1 \g_2  \cdots \g_{n} , \dot \d_1 \cdots \dot \d_n},\y^{j}_{\g \g_1 \g_2  \cdots \g_{n} , \dot \d_1 \cdots \dot \d_n}, 
\ee   for all values of $n$, are totally symmetrized. 

So at this point we have done very little with the irregular part of $E_3$, which consists of:
\be
E_{3 {\rm Irregular}}= 
{\cal P}_1
\oplus
{\cal \oP}_1
\oplus
{\cal Q}_2
\oplus
{\cal \oQ}_2
\oplus
{\cal Q}_3
\oplus
{\cal \oQ}_3
\oplus
(C \x \oC) S_0
\oplus
\oplus
(C \x^2 C)R_0
\oplus
(\oC \x^2 \oC)  \oR_0
\la{irregbitsate3}
\ee
but we have completely solved the problem for the 
regular part of $E_3$, which consists of:
\be
E_{3 {\rm Regular}}= 
\sum_{n=1}^{\infty} 
\lt \{
(C \x^2 C) {R}_n
\oplus
(\oC \x^2 \oC) {\oR}_n
\rt \}
\la{regbitsate3}
\ee
subject to the above conditions.  We have not written
$ \lt \{ R_{n}\cap E_3 \rt \}$ above in equation (\ref{regbitsate3}) because it makes the formulae cluttered, but it should be understood here that we are talking about the part of $   R_{n} $ that survives to $E_3$. 
Recall that $R_n$ in the above has n ghosts $C$ in it. 
We are going to proceed to analyze the effect of interactions on this Regular part (\ref{regbitsate3}) first in Appendix \ref{specsumchapinter}.

\section{Solutions to the Simple Separated Equations arising from the Irregular Equations, and the Normal Sector $E_{3\; {\rm Normal}}$}

\la{simplifiedequas}

\subsection{Overview of the irregular equations }

From the previous two sections, we know that the remaining unsolved equations for $E_2$ and $E_3$ involve the following pieces of $E_1$:

\be
E_{3 {\rm Irregular}}= 
\lt \{ (C \x^2 C)R_0
\oplus
(\oC \x^2 \oC)  \oR_0
\rt \}_{\rm Class \;1}
\oplus
\lt \{
(C \x \oC) S_0
\oplus
{\cal Q}_3
\oplus
{\cal \oQ}_3
\rt \}_{\rm Class \;2}
\eb
\oplus
\lt \{
{\cal Q}_2
\oplus
{\cal \oQ}_2
\rt \}_{\rm Class \;3}
\oplus
\lt \{
{\cal P}_1
\oplus
{\cal \oP}_1
\rt \}_{\rm Class \;4}
\oplus
\lt \{
{\cal P}_0
\rt \}_{\rm Class \;5}
\la{irregbitsate32}
\ee

This section starts the treatment of these irregular equations that arise for the free massless chiral SUSY theory in 3+1 dimensions.  
In this section we write down these irregular equations in full form. 

The fifth class of equations is easily solved and shown to be empty.

 Then as a start, we separate all the irregular equations for the other four classes and consider the solutions for these separated irregular equations. These problems are easily solved using our results for the regular equations.

In this paper we shall not attempt to describe the solutions for  the more difficult problems that arise for the unseparated irregular equations.

We shall now discuss the relevant equations, starting with the easiest one, which is ${\cal P}_0
$:

\subsection{Fifth Class of irregular equations : ${\cal P}_0$}

\la{posubsection}

In this subsection we solve the equations for ${\cal P}_0$ and show that: 
\be{\cal P}_0=0\ee
 These equations are: 
 \be
\begin{tabular}{|c|c|}
\hline
\multicolumn{2}
{|c|}{Table \ref{eqsforP0}: All equations  for ${\cal P}_0$ }
\\
\hline
\hline
{\rm equations  for $E_2$} 
& \begin{tabular}{c}
Mapping to:
\end{tabular}
\\
\hline
$ \P_1 ( C \na) {\cal P}_0 =0 $ 
& \begin{tabular}{c}
$\P_1 {\cal P}_{1}$
\end{tabular}
\\
\hline
$\P_1 ( \oC \ov \na) {\cal P}_0 =0 $ 
& \begin{tabular}{c}
$\P_1 {\cal {\ov P}}_{1}$
\end{tabular}
\\
\hline
\hline
\end{tabular}
\la{eqsforP0}
\ee

The $\P_1$ does nothing here, and so we can immediately deduce that: 
\be
\lt \{ \na_{\a} ,
\ov \na_{\dot \a} \rt \}
 {\cal P}_0  
=0
\ee
This implies that
\be
\pa_{\a \dot \a} 
 {\cal P}_0[{\ov {\cal F}},{ {\cal F}} ]
=0
\ee
which means that 
\be
\left \{ 
\left [ 
 {\pa}_{a \dot \b}^{\dag} , {\pa}_{\a \dot \b}  \rt ]
 +  {\pa}_{\a \dot \b} {\pa}_{\a \dot \b}^{\dag} \rt \}
{\cal P}_0   =0
\la{papadag1n}
\ee
This can be written in the form:
\be
\left \{ 
4\lt (  N_{\cal F} + N_{ \ov {\cal F} } + \fr{1}{2} N'  
+ \fr{1}{2} \dot N'  \rt )
 +  {\pa}_{\a \dot \b} {\pa}_{\a \dot \b}^{\dag}  
  \rt \}
{\cal P}_0   =0
\la{papadag2n}
\ee
See Appendix \ref{appendixpapadag} for a derivation of  equation (\ref{papadag2n}) from equation (\ref{papadag1n}).
Since these are positive operators,  equation (\ref{papadag2n}) separates into the following equations
\be
   {\pa}_{\a \dot \b} {\pa}_{\a \dot \b}^{\dag}  
  {\cal P}_0   =0
\la{papadag3n}
\ee
\be 
 N_{\cal F}    {\cal P}_0   =0
\la{papadag4n}
\ee
\be 
 N_{\ov {\cal F}}    {\cal P}_0   =0
\la{papadag5n}
\ee
\be 
 N'    {\cal P}_0   =0
\la{papadag6n}
\ee
\be 
 {\dot N}'    {\cal P}_0   =0
\la{papadag7n}
\ee
 equations  (\ref{papadag4n}) and (\ref{papadag5n}) imply that ${\cal P}_0 $ is independent of the fields. Then equations (\ref{papadag3n}), (\ref{papadag6n}) and (\ref{papadag7n}) are automatically satisfied. 
So 
\be
 {\cal P}_0[{\ov {\cal F}},{ {\cal F}} ]
= {\rm  constant \;independent\; of } \; {\ov {\cal F}} 
\; {\rm and} \; { {\cal F}} 
\ee
and we can set it to zero for our purposes here.
\be
{\cal P}_0[{\ov {\cal F}},{ {\cal F}} ]=0
\ee

Now we shall jump to the First class:

\subsection{First Class of irregular equations : ${\cal R}_4$ and ${\cal {\ov R}}_4$}

The expressions ${\cal R}_4$ and ${\cal \oR}_4$ yield zero ghost charge expressions when they are matched to the corresponding integrated terms in the cohomology  space ${\cal H}$.  So we define zero ghost charge expressions $R_0$ as follows:
\be
(C \x^2 C) R_0 = {\cal R}_4  
\ee
\be
  (\oC \x^2 \oC)\oR_0 
=   {\cal \oR}_4  
\ee

The equations that govern this sector are as follows:
 \be
\begin{tabular}{|c|c|}
\hline
\multicolumn{2}
{|c|}{Table \ref{eqsforR4n}: All equations  for 
${\cal R}_4$ and ${\cal {\ov R}}_4$ for the free massless case}
\\
\hline
\hline
{\rm equations  for $E_2$} 
& \begin{tabular}{c}
Mapping to:
\end{tabular}
\\
\hline
$  \P_1  ( C \na) {\cal R}_4 =0 $ 
& \begin{tabular}{c}
$ \P_1{\cal R}_{5}$
\end{tabular}
\\
\hline
$ \P_1 ( \oC \ov \na) {\cal {\ov R}}_4 =0 $ 
& \begin{tabular}{c}
$\P_1 {\cal {\ov R}}_{5}$
\end{tabular}
\\
\hline
\hline
{\rm equations  for $E_3$} 
& \begin{tabular}{c}
Mapping to:
\end{tabular}
\\
\hline
$  
 \P_2   ({\x \pa})^{\dag}
  {\cal { R}}_{4}      =0$ 
& \begin{tabular}{c}
$ \P_2 {\cal { Q}}_{3}$
\end{tabular}
\\
\hline
$  
 \P_2({\x \pa})^{\dag}
  {\cal { \ov R}}_{4}      =0$ 
& \begin{tabular}{c}
$ \P_2 {\cal { \ov Q}}_{3}$
\end{tabular}
\\
\hline
$  
 \P_2 \lt \{(C \cdot \x^{\dag} \cdot {\oC}^{\dag})({\ov \na}^{\dag})^2 {\cal {\ov R}}_4 
+ (\oC \cdot \x^{\dag} \cdot {C}^{\dag}) ({ \na}^{\dag})^2      {\cal R}_4  \rt \}  =0$ 
& \begin{tabular}{c}
$ \P_2 {\cal S}_{3}$
\end{tabular}
\\
\hline
\hline
\end{tabular}
\la{eqsforR4n}
\ee

\subsubsection{The separated Solutions for   ${\cal R}_4= (C \x^2 C) {\cal R}_0$} 
\la{sepforR0}

It is generally a good idea to try to solve the `separated' equations first, because it is easier.  So we do that here.  The separated equations are as above except that we `separate' the   equation
\be
 \P_2 \lt \{(C \cdot \x^{\dag} \cdot {\oC}^{\dag})({\ov \na}^{\dag})^2 {\cal {\ov R}}_4 
+ (\oC \cdot \x^{\dag} \cdot {C}^{\dag}) ({ \na}^{\dag})^2      {\cal R}_4  \rt \}  =0
\la{tworequat}
\ee
 into two equations, as follows:
\be
  (\na^{\dag})^2 R_0=0
\la{nadagsqR}
\ee
\be
 (\ona^{\dag})^2 \oR_0=0
\la{onadagsqR}
\ee
Note that we have also removed the projection operators $\P_2$ from these.

Clearly all solutions of (\ref{nadagsqR}) and
(\ref{onadagsqR}) are solutions of (\ref{tworequat}). However there could be (and there are)  solutions of (\ref{tworequat}) that are not solutions of either (\ref{nadagsqR}) or
(\ref{onadagsqR}).

We start work on the separated equations by noting that 
\be
\na_{\a} R_0=0
\la{naR}
\ee
and we can derive from 
(\ref{naR})
 and (\ref{nadagsqR}) that:
\be
\na^{\a} \lt [\na_{\a}, (\na^{\dag})^2  \rt ]
 R_0=0
\Ra 
Z_+ R_0=0\la{zpluseqforr}
\ee
where $Z_+$ is defined by (\ref{Zplus}).  To get this we need to be careful of the $\P_2$ and use the formulae in subsubsection \ref{someidetnites}.

Similarly the complex conjugate also follows: 
\be
{\ov Z}_+ \oR_0=0
\ee
It is simple to show that 
(\ref{zpluseqforr}) implies that
\be
N_{\cal F} R_0=0
\la{nofhere}
\ee
and that
\be
Z_{R} R_0=0
\la{symundot}
\ee
The equation (\ref{nofhere}) implies that 
\be
  R_0= R_0\lt [ {\cal {\ov F}}\rt ] 
\la{onlyfbar}
\ee
and the second equation (\ref{symundot}) 
implies that all the undotted indices in $R_0$ are symmetrized.  So this implies that this part is of the form claimed in  (\ref{canonicalform3}).  The remaining equation is 
\be
  \P_2 {\pa}_{\a \dot \b}^{\dag}  
  {R}_0   =0
\la{padageq}
\ee
Equation (\ref{padageq})  affects only terms with derivatives, and the purpose of (\ref{padageq}), put simply, is to ensure that the no term in $ R_0$ is a total derivative. There are plenty of such solutions.

This describes the separated irregular part of the cohomology in this sector, and justifies the description  in subsection \ref{breifsummsec}.

\la{sepirregRsec}

\subsection{Second Class of irregular equations : ${\cal Q}_3 \oplus {\cal {\ov Q}}_3 \oplus{\cal S}_3$ }

The following are the general equations
 for these three sectors for the free massless case:

 \be
\begin{tabular}{|c|c|}
\hline
\multicolumn{2}
{|c|}{Table \ref{eqsforQ3n}: All equations  for 
${\cal Q}_3$, ${\cal {\ov Q}}_3$ and ${\cal S}_3$ for the free massless case}
\\
\hline
\hline
{\rm equations  for $E_2$} 
& \begin{tabular}{c}
Mapping to:
\end{tabular}
\\
\hline
$ \P_1( C \na) {\cal Q}_3 =0 $ 
& \begin{tabular}{c}
$\P_1 {\cal Q}_{4}$
\end{tabular}
\\
\hline
$ \P_1
( \oC \ov \na) {\cal {\ov Q}}_3 =0 $ 
& \begin{tabular}{c}
$\P_1 {\cal {\ov Q}}_{4}$
\end{tabular}
\\
\hline
$ \P_1 \lt \{( C \na)^{\dag} {\cal Q}_{3} + ( \oC \ov \na)^{\dag} {\cal S}_{3} \rt \}=0$ 
& \begin{tabular}{c}
$\P_1 {\cal Q}_{2}$
\end{tabular}
\\
\hline
$  \P_1 \lt \{( C \na)^{\dag} {\cal S}_{3} + ( \oC \ov \na)^{\dag} 
{\cal \ov Q}_{3}\rt \}
=0$ 
& \begin{tabular}{c}
$\P_1 {\cal \ov Q}_{2}$
\end{tabular}
\\
\hline
\hline
{\rm equations  for $E_3$} 
& \begin{tabular}{c}
Mapping to:
\end{tabular}
\\
\hline
$ \P_2 \lt \{
(\x \pa) {\cal Q}_3+ (C\cdot \x \cdot {\ov C}^{\dag} ) 
{\na}^2    {\cal S}_3 \rt \}
=0$ 
& \begin{tabular}{c}
$ \P_2 {\cal R}_{4}$
\end{tabular}
\\
\hline
$ \P_2 \lt \{ (\x \pa) {\cal {\ov Q}}_3+ (C^{\dag} \cdot \x \cdot {\ov C}) 
{\ov \na}^2    {\cal S}_3 \rt \}
=0$ 
& \begin{tabular}{c}
$ \P_2 {\cal {\ov R}}_{4}$
\end{tabular}
\\
\hline
$  \P_2 \lt \{(\x \pa)^{\dag} {\cal {Q}}_3 \rt \}
=0$ 
& \begin{tabular}{c}
$ \P_2 {\cal {P}}_{2}$
\end{tabular}
\\
\hline
$ \P_2 \lt \{ (\x \pa)^{\dag} {\cal {\ov Q}}_3\rt \}
=0$ 
& \begin{tabular}{c}
$ \P_2 {\cal {\ov P}}_{2}$
\end{tabular}
\\
\hline
\hline
\end{tabular}
\la{eqsforQ3n}
\ee

\subsubsection{Separated equations  for ${\cal S}_3 = (C \x \oC) S_0$}

\la{sepforsqq}

First we note that
a  naive derivation of these equations would include the following terms in the final two rows of table (\ref{eqsforQ3n}).
\be
\P_2  (C \cdot \x^{\dag} \cdot {\ov C}^{\dag}) 
({\ov \na}^{\dag})^2    {\cal S}_3
\ee
and
\be
\P_2  (\oC \cdot \x^{\dag} \cdot {C}^{\dag}) 
({ \na}^{\dag})^2    {\cal S}_3
\ee
but these terms are identically zero because of the identity
\be
(C \cdot \x^{\dag} \cdot {\ov C}^{\dag}) ( C \x \oC)= C^{\a} C_{\a} =0
\ee
and its complex conjugate.

To start with we will look for  `separated solutions' for these equations.  For the ${\cal Q}_3$ this means that ${\cal Q}_3=0$,  using arguments identical to those for ${\cal Q}_n$ in the analysis of the regular equations.
We can write
\be
{\cal S}_3 = ( C \x \oC) S_0
\ee
For the $S_0$ we then get the following `simple and separated equations':

\be
{\ov \na}_{\dot \b}^{\dag} S_0=
{ \na}_{  \a}^{\dag} S_0
\eb
=
\P_2\pa_{\a \dot \b}^{\dag} S_0 
=
\na^2 S_0 
=
{\ov \na}^2 S_0 
=0\ee

Using a  similar  derivation  to the one in  subsection 
\ref{sepforR0}, these equations imply that:
\be
Z_- S_0 =0
\ee
where 
\be
Z_- = 
  N   \lt (  N - 1 \rt )  -  R^i
R^i  \ee
and
\be
N= \lt ( N_{\cal F} + \fr{1}{2} N' \rt )
\ee
where $Z_-$ is defined by (\ref{Zminus}),
so that:
\be
Z_- = 
 \lt ( N_{\cal F} + \fr{1}{2} N' \rt )   \lt (  N_{\cal F} + \fr{1}{2} N'   - 1 \rt )  -  R^i
R^i  \ee
But we also have the complex conjugate equations which imply that:
\be
\oZ_- S_0 =0
\ee
where 
\be
\oZ_- = 
  \oN   \lt (  \oN - 1 \rt )  -  \oR^i
\oR^i  \ee
and
\be
\oN= \lt ( N_{\cal {\ov F}} + \fr{1}{2} \oN' \rt )
\ee
so that
\be
\oZ_- = 
 \lt ( N_{\cal {\ov F}}  + \fr{1}{2} \oN' \rt )   \lt (  N_{\cal {\ov F}}  + \fr{1}{2} \oN'   - 1 \rt )  -  \oR^i
\oR^i  \ee

We cannot have $N_{\cal {F}} \geq 2$ or $N_{\cal {\ov F}} \geq 2$
because for these we would get an impossible equation from one of the above.  So in order to have anything at all in the space we must have both  $N_{\cal {F}} =0,1$ and  $N_{\cal {\ov F}} = 0, 1$. If $N_{\cal {F}} =N_{\cal {\ov F}} =0$ there is nothing there at all.

So there are just three cases  of any interest, and they are all restricted by the above, plus the equation 
\be
\P_2 \pa_{\a \dot \b}^{\dag} S_0 
=0
\la{notaderiv}
\ee 
\ben
\item

{\bf Case  where $N_{\cal {\ov F}}=0$, $N_{\cal {F}} = 1$}
The only possibility is:
\be
(C\cdot  \x \cdot \oC) A \in {\cal S}_3  \subset E_{\infty}  
\la{simplestthingins3}
\ee
\item
 {\bf Case  where $N_{\cal {\ov F}}=0$, $N_{\cal {F}} = 1$}
The only possibility is:

 \be
(C \cdot \x \cdot \oC)  \A   \in {\cal S}_3  \subset E_{\infty}
\ee

\item
 {\bf Case  where $N_{\cal {\ov F}}=1$, $N_{\cal {F}} = 1$}

Here there are several possibilities.  First note that

If  $N_{\cal {F}} = 1$. 
\be
Z_- \Ra  
 \lt ( 1 + \fr{1}{2} N' \rt )   \lt (   \fr{1}{2} N'    \rt )  -  R^i
R^i  \equiv Z_R \ee

If  $N_{\cal {\ov F}} = 1$. 
\be
Z_- = 
 \lt (   \fr{1}{2} \oN' \rt )   \lt (   \fr{1}{2} \oN'   + 1 \rt )  -  \oR^i 
\oR^i = Z_{\ov R} \ee
So we see that for this case the solution must have all its undotted indices symmetrized and all of its dotted indices symmetrized also. Incorporating (\ref{notaderiv}) we get the following:

Simplest Example:
\be
S_{3,1}
=
(C \cdot \x \cdot \oC)  A \A \in {\cal S}_3  \la{morecompthingins3}
 \ee
Second Example:
 \be
S_{3,\a \dot \b}
=
(C \x \oC) \lt (
A^j  \A_{j\g \dot \d}-   A^j_{\g \dot \d} \A_j
-
\y^j_{\g}
\oy_{j \dot \d}
\rt ) 
\ee
We shall not look here for examples with more derivatives, but presumably they do exist.
\een

When the theory becomes interacting or massive, these simple Solutions get mapped into ${\cal R}_4$ by $d_4$, resulting in new constraints for them and for ${\cal R}_4$.  The same applies to $d_6$ and $d_8$.

\subsection{Third and Fourth  Classes of irregular equations : ${\cal P}_1 \oplus {\cal {\ov P}}_1 $ and  ${\cal Q}_2 \oplus {\cal {\ov Q}}_2 $}

\la{p1andq2subsection}

The last two classes of irregular equations are those which govern sectors of the theory with   ghost number 
${\cal G} = -3$ and ${\cal G} = -2$.  So the simplest possible examples of these would be something like
\be
{\cal \oP}_1 \in E_{\infty} \ra \int d^4 s \; 
\lt \{ Y Y \oY 
\rt \} \in {\cal H}
\ee
or
\be
{\cal \oQ}_2 \in E_{\infty} \ra \int d^4 s \; 
\lt \{ Y Y 
\rt \} \in {\cal H}
\ee
Nothing this simple seems to be present however, as one can verify by trying some examples. 

Here are the relevant full equations:
 \be
\begin{tabular}{|c|c|}
\hline
\multicolumn{2}
{|c|}{Table \ref{eqsforP1other}: All equations  for ${\cal P}_1$ and ${\cal {\ov P}}_1$}
\\
\hline
\hline
{\rm equations  for $E_2$} 
& \begin{tabular}{c}
Mapping to:
\end{tabular}
\\
\hline
$ \P_1 ( C \na) {\cal P}_1 =0 $ 
& \begin{tabular}{c}
$\P_1 {\cal P}_{2}$
\end{tabular}
\\
\hline
$ \P_1 ( \oC \ov \na) {\cal {\ov P}}_1 =0 $ 
& \begin{tabular}{c}
$\P_1 {\cal {\ov P}}_{2}$
\end{tabular}
\\
\hline
$ \P_1 \lt \{( C \na)^{\dag} {\cal P}_1 + ( \oC \ov \na)^{\dag} 
{\cal {\ov P}}_{1}\rt \} =0$ 
& \begin{tabular}{c}
$\P_1 {\cal P}_{0}$
\end{tabular}
\\
\hline
\hline
{\rm equations  for $E_3$} 
& \begin{tabular}{c}
Mapping to:
\end{tabular}
\\
\hline
$ \P_2 \lt \{(\x \pa) {\cal P}_1+ (C\cdot \x \cdot {\ov C}^{\dag} ) \na^2    {\cal {\ov P}}_1 \rt \}
=0$ 
& \begin{tabular}{c}
$\P_2 {\cal Q}_{2}$
\end{tabular}
\\
\hline
$ \P_2 \lt \{(\x \pa) {\cal {\ov P}}_1+ (C^{\dag}\cdot \x \cdot {\ov C} ) 
{\ov \na}^2    {\cal { P}}_1 \rt \}=0$ 
& \begin{tabular}{c}
$\P_2  {\cal {\ov Q}}_{2}$
\end{tabular}
\\
\hline
\hline
\end{tabular}
\la{eqsforP1other}
\ee

 \be
\begin{tabular}{|c|c|}
\hline
\multicolumn{2}
{|c|}{Table \ref{eqsforQ2next}: All equations  for ${\cal Q}_2$  and ${\cal {\ov Q}}_2$}
\\
\hline
\hline
{\rm equations  for $E_2$} 
& \begin{tabular}{c}
Mapping to:
\end{tabular}
\\
\hline
$ \P_1 ( C \na) {\cal Q}_2 =0 $ 
& \begin{tabular}{c}
$\P_1 {\cal Q}_{3}$
\end{tabular}
\\
\hline
$ \P_1( \oC \ov \na) {\cal {\ov Q}}_2 =0 $ 
& \begin{tabular}{c}
$\P_1 {\cal {\ov Q}}_{3}$
\end{tabular}
\\
\hline
$ \P_1\lt \{
( \oC \ov \na) {\cal Q}_{2}+
( C \na) {\cal \ov Q}_2 \rt \}=0 $ 
& \begin{tabular}{c}
$\P_1 {\cal S}_{3}$
\end{tabular}
\\
\hline
\hline
{\rm equations  for $E_3$} 
& \begin{tabular}{c}
Mapping to:
\end{tabular}
\\
\hline
$ \P_2\lt \{(\x \pa)^{\dag} {\cal Q}_2 + (C\cdot \x^{\dag}  \cdot {\ov C}^{\dag} ) 
({\ov \na}^{\dag})^2    {\cal {\ov Q}}_2 \rt \}
=0$ 
& \begin{tabular}{c}
$\P_2 {\cal P}_{1}$
\end{tabular}
\\
\hline
$ \P_2\lt \{(\x \pa)^{\dag} {\cal {\ov Q}}_2+ (C^{\dag}\cdot \x^{\dag}  \cdot {\ov C} ) 
({\na}^{\dag})^2    {\cal { Q}}_2 \rt \}
=0$ 
& \begin{tabular}{c}
$\P_2 {\cal {\ov P}}_{1}$
\end{tabular}
\\
\hline
\hline
\end{tabular}
\la{eqsforQ2next}
\ee

The separated equations here for ${\cal P}_1$ and ${\cal Q}_2$
are  similar to the regular equations, with some extra equations added.  So it seems likely that there are no solutions for the separated equations. 
However this is not yet proved in general.  It is easy to establish at low dimensions for the operators, because if there any solutions for these sectors, they must have ghost charge ${\cal G}=-3$ and ghost charge ${\cal G}=-2$ respectively, which implies that they must also have  a fairly high dimension.  It has not yet been  determined whether such examples exist or not.

\section{Summary of the Cohomology of the Free Massless Chiral SUSY Theory}

\la{freemasslesscohom}

\subsection{The Normal Part $E_{3\; {\rm Normal}}$  of the space $E_3$}

In this section we summarize the form of the Normal part $E_{3 \;{\rm Normal}}$  of the space $E_3$.  This is the final step for this part for the free massless theory.  To completely describe the BRS cohomology of the free massless theory, we need only to expand this to include the exceptional part of the space $E_3$.

However that is not solved in this paper for two reasons:
\ben
\item
It  is a long process that is not yet completed.
\item
Much of it has no relevance to what we need for present purposes.
\item
There is already much to think about just dealing with the Normal solutions.
\een

Once we have described this space $E_{3 \;{\rm Normal}}$ 
we will go on to consider what happens to it when there are interactions in the theory.

\la{specsumchapinter}

\subsection{The Space $E_{3 \;{\rm Normal}}=
E_{3 \;{\rm Simple}}\oplus
E_{3\; {\rm Regular}}$}

\la{e3stuffassem}

Here we assemble the information that we have derived in the foregoing.  The Regular solutions plus the separated irregular solutions up to the level of $E_3$ are summarized in the following form:

\be
E_{3 \;{\rm Normal}
} = 
  ( C \x  {\ov C} )
S_{0\;{\rm Simple}}
+
(C \x^2 C) {R}_{0\; {\rm Simple}}
+
(\oC \x^2 \oC) {\oR}_{0\; {\rm Simple}}
\eb
\oplus
\sum_{n=1}^{\infty} 
\lt \{
(C \x^2 C) {R}_{n\;{\rm Regular}}
+
(\oC \x^2 \oC) {\oR}_{n\;{\rm Regular}}
\rt \}
\la{normalfor3}
\ee
where the regular   solutions   $E_{3 \;{\rm Regular}}  $  have the   form 
\be
{R}_{p\;{\rm Regular}}
=
{ R}_{(\a_1 \cdots \a_{n+p})}
=
 T_{[i_1\cdots i_n]}^{(j_1\cdots j_m)}\A_{j_1}\cdots \A_{j_m}
\y^{i_1}_{(\a_1} \cdots 
\y^{i_n}_{\a_n}  C_{\a_{n+1}}\cdots C_{\a_{n+p})} 
\la{canonicalform3}
\ee
The complex conjugate of (\ref{canonicalform3}) is:
\be
 {\oR}_{p\;{\rm Regular}} =  \oR_{(\dot \a_1 \cdots \dot \a_{n+p})} =   \oT^{[i_1\cdots i_n]}_{(j_1\cdots j_m)} A^{j_1}\cdots A^{j_m}
\oy_{i_1(\dot \a_1} \cdots 
\oy_{i_n \dot \a_n}  \oC_{\dot \a_{n+1}}\cdots \oC_{\dot \a_{n+p})} 
\la{canonicalformcc}
\ee
The simple solutions   ${R}_{0\;{\rm Simple}} $  have the   form 
\be
{R}_{0\;{\rm Simple}}
=
{ R}_{(\a_1 \cdots \a_{n})}
=
 T_{[i_1\cdots i_n]}^{(j_1\cdots j_m)}\A_{j_1}\cdots \A_{j_m}
\y^{i_1}_{(\a_1} \cdots 
\y^{i_n}_{\a_n)}  
\la{canonicalform3nsimp}
\ee
The complex conjugate of (\ref{canonicalform3}) is:
\be
 {\oR}_{0\;{\rm Simple}} =  \oR_{(\dot \a_1 \cdots \dot \a_{n})} =   \oT^{[i_1\cdots i_n]}_{(j_1\cdots j_m)} A^{j_1}\cdots A^{j_m}
\oy_{i_1(\dot \a_1} \cdots 
\oy_{i_n \dot \a_n)}  
\la{canonicalformccsimp}
\ee

and
\be
{\cal S}_{3 \;{\rm Simple}}= 
(C \x \oC) 
S_{0 \;{\rm Simple}}
\eb= 
(C \x \oC) 
\lt \{
f_i A^i
+
\of^i \A_i 
+
f^i_j A^j \A_i 
+ f^i_{2,j}\lt (
A^i  \A_{j\g \dot \d}-   A^i_{\g \dot \d} \A_j
-
\y^i_{\g}
\oy_{j \dot \d}
\rt )+ \cdots
\rt \}
\la{defiverer}
\ee

We have simplified in the foregoing, because we have not included terms with derivatives, except in the last term 
(\ref{defiverer}).  These terms can be added easily be going through the equations.

In the main body of the paper, we discuss how these objects in $E_3$ give rise to objects in the cohomology space ${\cal H}$.  We shall not repeat that here, except to say that the objects 
(\ref{canonicalform3}) reappear as the solutions 
(\ref{canonicalformsupefield}) in ${\cal H}$.

\section{Summary of the      Spaces   $E_{r \;{\rm Normal}}, r=3, 4,5$   and the Differentials $d_r, r=3,4$  and  for the Interacting Massless Chiral SUSY Theory}

\la{interactingspectralse}

\subsection{The Operator $d_3$ and the  Space $E_4=\ker d_3 \cap \ker d_3^{\dag} \cap E_3$} 

\la{defofd3}

The spectral sequence ends at $E_3$ for the free massless theory and we have discussed how the cohomology space for that theory arises out of $E_3=E_{\infty}$ for that case.  However, for  the  interacting theory, it is not true that $E_3=E_{\infty}$.   Here we will assume that $g_{ijk} \neq 0$ but that $m = v^i =0$ in equation (\ref{thebigoperator}).  

In this section we shall examine what happens for the space
\be E_{3\; {\rm Normal}}=
E_{3 \;{\rm Regular}}
\oplus
E_{3\; {\rm Simple \; Irregular}}
\ee
described in the preceding section.

An examination of the possible $d_3$ operators leads to the conclusion that for the physical approach we get:
\be
d_3 = \P_3 
\d_3 \P_3
\ee
where
\be
\d_3= \int d^4 x\; \lt (
g_{ijk} A^j A^k 
{\ov C}_{\dot \b } \fr{\d }{\d 
\oy_{i \dot \b} } 
+  C_{\a}   
\ov g^{ijk} \A_j \A_k 
\fr{\d }{\d \y_{\a}^i } 
 \rt )  
\la{defofd3eq}
\ee
However the operator $\d_3$ in 
(\ref{defofd3eq}) does not exist for the superfield approach.
But we arrive at the same $d_3$ anyway using   a different route:
\be
d_3 = \P_3 
\d_2 \fr{\d_0^{\dag}}{\D_0} \d_1 \P_3
\approx
\P_3   \lt (  \int
g_{ijk} A^j A^k 
 \fr{\d }{\d  \Lambda_i}
\rt ) \fr{\lt (\int    \Lambda_j \fr{\d }{\d \ov F_j}
\rt )}{\D_0}
\lt (\int  
{\ov C}_{\dot \b } \ov F_k   \fr{\d }{\d 
\oy_{k \dot \b} } 
\rt )
\la{funnyroute}
\ee

This differential has the effect:
\be
{\cal R}_{n} \stackrel{d_3}{\lra} {\cal R}_{n+1}; n= 4, 5, \cdots
\ee
and
\be
{\cal \oR}_{n} \stackrel{d_3}{\lra} {\cal \oR}_{n+1}; n= 4, 5, \cdots
\ee
So we can treat one of these, and then the other follows.

\subsubsection{The Simplest Example for $d_3$:  $(C \x^2 C) f_i  \y^i_{\a}   $}

\la{simplestford3}

The simplest case is:
\be
(C \x^2 C) f_i \y^i_{\a}  \in E_{3} 
 \stackrel{d_3}{\lra} 
(C \x^2 C)  C_{\a}   
f_i  \ov g^{ijk} \A_j \A_k \in E_{3},  
\la{firstexamp}
  \ee
So this means that 
\be
(C \x^2 C) f_i \y^i_{\a}  \in E_{4} 
\;{\iff} 
f_i  \ov g^{ijk} =0
\la{eqdepdnonX}\ee
This means that, for each i, the  equation (\ref{eqdepdnonX}) is true if and only if the superpotential is independent of the field $\A_i$.  This is equivalent to:
\be
f_i \A_i^{ \dag} \lt \{
\og^{pqr} \A_p  \A_q  \A_r   \rt \}=   
3 f_p  
\og^{pqr}    \A_q  \A_r   =  0 
\ee  
The complex conjugate is also true of course:
\be
\of^i A^{i \dag} \lt \{
g_{pqr} A^p  A^q  A^r \rt \}=   
3 \of^p    g_{pqr}   A^q  A^r = 0
\ee  

The adjoint equation to the above (\ref{firstexamp}) 
is
\be
(C \x^2 C)  C_{\a}   
T^{ij}   \A_i \A_j \in E_{3}
 \stackrel{d_3^{\dag}}{\lra} 
(C \x^2 C) T^{ij} g_{ijk}\y^k_{\a}  \in E_{3} 
    \ee
This means that
\be
(C \x^2 C)  C_{\a}   
T^{ij}   \A_i \A_j \in E_{4}
\;{\rm \iff} \; T^{ij} g_{ijk}=0
\ee
A more convenient way to evaluate this requirement in practice for a given model is to write it in the form:
\be
(C \x^2 C)  C_{\a}   
T^{ij}   \A_i \A_j \in E_{4}
\;{\iff} \; 
T^{ij}  A^{i  \dag} A^{j  \dag}   \lt \{
g_{pqr} A^p A^q A^r  \rt \}=   
6T^{pq} 
g_{pqr} A^r 
=0
\la{convenientform}
\ee  
This implies the complex conjugate equation:
\be
(\oC \x^2 \oC)  \oC_{\dot \a}   
\oT_{ij}  A^i A^j \in E_{4}
\;{\iff} \; 
\oT_{ij}  \A_i^{ \dag} \A_j^{ \dag} \lt \{
\og^{pqr} \A_p  \A_q  \A_r   \rt \}=   
6 \oT_{ij}
\og^{ijr}      \A_r   
=0
\la{convenientformcc}
\ee  
The meaning of this becomes clearer in a given model. For example in the CSSM one gets a space of solutions here quite easily using this form 
(\ref{convenientform})
of the requirement.

\subsubsection{A Lie Algebra Invariance of the Superpotential:  The Next to Simplest Example for $d_3$: $(C \x^2 C) f_i^l \y^i_{\a} \A_l $}

\la{importexamp}

The next example is:
\be
(C \x^2 C) f_i^l \y^i_{\a} \A_l  \in E_3 \stackrel{d_3}{\lra} 
(C \x^2 C)  C_{\a}   
f_i^l  \ov g^{ijk} \A_j \A_k  \A_l \in E_3
\la{d3tothreea}
 \ee
This means that
\be
(C \x^2 C) f_i^l \y^i_{\a} \A_l   \in E_4 \;{\iff} \;
f_i^l  \ov g^{ijk} \A_j \A_k  \A_l =0
\ee
Again, in practice for a given model,  this requirement can be conveniently written in the form:\be
{\cal L}_{f} \lt ( \og^{ijk} \A_i \A_j \A_k  \rt )=0
\la{eqdepdnonL}
\ee
where we define
\be
{\cal L}_{f}= f_i^l  \A_l \A_{i}^{ \dag}
\la{depdnonL}
\ee
This equation (\ref{eqdepdnonL}) is true if the superpotential is invariant under the action of the Lie algebra generator specified by  (\ref{depdnonL}).  

So we see that a solution of the equation 
\be
(C \x^2 C) f_i^l \y^i_{\a} \A_l   \stackrel{d_3}{\lra} 
0 \ee
is generated by each Lie invariance of the superpotential. In other words there is an object of the form $(C \x^2 C) f_i^l \y^i_{\a} \A_l $ which survives to $E_4$ for each such Lie algebra generator.  In fact we will see that it survives to $E_{\infty}$ in general for the  massless interacting case. Similar remarks apply to the more complicated examples where there is more than one $\y$ or more than 
one $\A$, or when there are derivatives involved.

The adjoint of (\ref{d3tothreea}) 
 also creates a constraint:
\be
(C \x^2 C)  C_{\a}   
T^{ijk}  \A_i \A_j \A_k  \in E_3 
 \stackrel{d_3^{\dag}}{\lra} 
(C \x^2 C) T^{ijk} g_{sjk} \y^s_{\a} \A_i
  \in E_3  
\la{thebasicconADJ}
\ee
So in this case, we get 
\be
(C \x^2 C)  C_{\a}   
T^{ijk}  \A_i \A_j \A_k  \in E_4
\;\iff\;
  T^{ijk} g_{sjk} =0
\ee
Again, in practice for a given model,  this requirement can be conveniently written in the form:\be
T^{pqr} \A_p  A^{q \dag}A^{r \dag} \lt ( g_{ijk}A^i A^j A^k  \rt )=0
\la{eqdepdnonLNEW}
\ee
For a given tensor $g_{ijk}$, which specifies a given model, this generates a space of solutions $T^{pqr}$.  The complex conjugate follows as usual.

\subsection{The Operator $d_4$ and the  Space $E_5=\ker d_4 \cap \ker d_4^{\dag} \cap E_4$}

\la{e5subsection}

The next differential has the form

\be
d_4 = \P_4 
\d_3 \fr{\d_0^{\dag}}{\D_0} \d_1  \P_4+ *
\ee
It arises from the combination where we use $\d_3$ from 
(\ref{defofd3eq}), $\d_0$ from 
(\ref{structurediff}) and 
$\d_1$ from the first term in (\ref{supertrans}).
Schematically this is:
\be d_4 = \P_4 
\lt ( \int d^4 x_1 C_{\a} {\ov g}^{ijk} {\ov A}_j {\ov A}_k \fr{\d }{\d \y_{\a}^j}\rt )
\lt (  \x {\ov C}^{\dag} {C}^{\dag} \rt ) 
\int d^4 x_2 \lt ( C \cd \y^j \fr{\d }{\d \ov A^j}\rt )
\P_4 + *
\la{seconddefofd3detail}
\ee
We can usually ignore the complicating factor 
$\fr{1}{\D_0}$ in these calculations, because it only adds an irrelevant factor. Now $d_4$ has the explicit form:
\be d_4 = \P_4 ( C \x {\ov C}^{\dag} ) 
\lt \{
 \int d^4 x \;  {\ov g}^{ijk} {\ov A}_j {\ov A}_k
 \fr{\d}{\d A^{i }} \rt \}
\P_4 + *
\la{seconddefofd3}
\ee

To get this operator in the superfield approach we follow the reasoning in (\ref{funnyroute}), as follows:
\be
d_4 = \P_4 
 \d_2 \fr{\d_0^{\dag}}{\D_0} \d_1 \fr{\d_0^{\dag}}{\D_0} \d_1  \P_4+ *
\la{funnyrouteagain}
\ee
and more specifically this is:
\be
d_4 = 
\P_4   \lt (  \int
 g_{ijk} A^j A^k 
 \fr{\d }{\d  \Lambda_i}
\rt )  \lt (\int    \Lambda_j \fr{\d }{\d \ov F_j}
\rt ) 
\eb
\lt (\int  
{\ov C}_{\dot \b } \ov F_k   \fr{\d }{\d 
\oy_{k \dot \b} } 
\rt )
\lt (  \x {\ov C}^{\dag} {C}^{\dag} \rt ) 
 \lt ( \int \; \oC \cd \oy_j \fr{\d }{\d \ov A_j}\rt )
\P_4 
\eb
= \P_4 (\x \oC C^{\dag}) A A
 \A^{\dag}\P_4
\la{funnyroute4}
\ee

This acts to take
\be
{\cal S}_{3}  \stackrel{d_4}{\lra} {\cal R}_{4} 
\ee
and it eliminates various possible objects in both spaces. 

\subsubsection{The Simplest Example for $d_4$:  $(C \x  \oC) f_i  A^i    $}

\la{simplestford4}

Again here let us look at the low dimensional examples.  
\be 
( C \x {\ov C} )f^i \A_i  
\stackrel{d_4}{\lra} 
( \oC \x^2 \oC ) f^i g_{ijk} A^j A^k
\ee

\be 
(  C \x {\ov C} )
T_{jk} \og^{ijk} \A_i  
\stackrel{d_4^{\dag}}{\longleftarrow} 
( \oC \x^2 \oC ) T_{jk} A^j A^k
\ee
So we see that
\be
( C \x {\ov C} )f^i \A_i \in E_5\;
{\rm \iff}\;  f^i g_{ijk} =0\ee
 and
\be
( \oC \x^2 \oC ) T_{jk} A^j A^k
\in E_4 \;
{\rm \iff}\;  T_{jk} \og^{ijk}  =0
\la{thetgequationagain}
\ee

The above is very similar to the analysis in subsection
\ref{simplestford3}.

\subsubsection{The Next to Simplest Example $(C \x  \oC) f_i^j A^i \A_j $for $d_4$: A Lie Algebra Invariance of the Superpotential}

Again here let us look at the low dimensional examples.  
\be 
( C \x {\ov C} ) f_i^j A^i \A_j
\stackrel{d_4}{\lra} 
( \oC \x^2 \oC ) f_i^s A^i   g_{sjk} A^j A^k
\ee
and
\be 
( C \x {\ov C} )
 \og^{sjk} T_{tjk} A^t \A_s  
\stackrel{d_4^{\dag}}{\longleftarrow} 
( \oC \x^2 \oC ) T_{ijk} A^i A^j A^k
\ee
So we see that
\be
( C \x {\ov C} )  f_i^j A^i \A_j\in E_5 
{\rm \iff}\;   f_i^s A^i   g_{sjk} A^j A^k  =0\ee
 and
\be
( \oC \x^2 \oC ) T_{ijk} A^i A^j A^k
\in E_5 \iff \og^{sjk} T_{tjk} A^t \A_s  =0
\ee

The above is very similar to the analysis in subsection
\ref{simplestford3}.

\section{Summary of the   Differentials $d_r, r=5,6,7,8$  and  Spaces   $E_{r}, r=6,7,8,9$     for the Massive Interacting  Chiral  SUSY Theory}

\la{chapfore9}

\subsection{The Operator $d_5$ and the  Space $E_6=\ker d_5 \cap \ker d_5^{\dag} \cap E_5$} 

For the massless interacting theory the spectral sequence ends at $E_5$. 
When the mass is non-zero, we must augment the space with a dependence on the parameter $m$ and then we find that  there is another differential:

\be
d_5 = \P_5 
\d_5 \P_5
\la{defofd5}
\ee
where
\be
\d_5= 2 \int d^4 x\; \lt (
 g_{ijk}m v^j A^k 
{\ov C}_{\dot \b } \fr{\d }{\d 
\oy_{i \dot \b} } 
+  C_{\a}   
\ov g^{ijk}  m \ov v_j \A_k 
\fr{\d }{\d \y_{\a}^i } 
 \rt )  
\la{funnyroute5}\ee
To get this operator in the superfield approach we follow the reasoning in (\ref{funnyroute}) again, as follows:
\be
d_5 = \P_5 
 \d_4 \fr{\d_0^{\dag}}{\D_0} \d_1 \P_5+ *
\la{funnyrouteagainfor5}
\ee
and more specifically this is:
\be
d_5 = 
\P_5   \lt (  \int
 g_{ijk} m v^j A^k 
 \fr{\d }{\d  \Lambda_i}
\rt )  \lt (\int    \Lambda_j \fr{\d }{\d \ov F_j}
\rt ) 
\lt (\int  
{\ov C}_{\dot \b } \ov F_k   \fr{\d }{\d 
\oy_{k \dot \b} } 
\rt ) \P_5
\la{funnyroutefor5}
\ee

This acts to take
\be
{\cal R}_{n} \stackrel{d_5}{\lra} {\cal R}_{n+1}; n= 4, 5, \cdots
\ee
and
\be
{\cal \oR}_{n} \stackrel{d_5}{\lra} {\cal \oR}_{n+1}; n= 4, 5, \cdots
\ee

So here we get

\be
(C \x^2 C) f_i^l \y^i_{\a} \A_l   \stackrel{d_5}{\lra} 
\P_5 (C \x^2 C)  C_{\a}   
f_i^l  \ov g^{ijk} m \ov v_j \A_k  \A_l 
\la{dfeqyatuibn}
\ee

Now we also know from multiplying the argument in subsubsection 
\ref{simplestford4} by m, that  
\be
(C \x^2 C)  C_{\a}m T^{jk}\A_j  \A_k  \in E_5 \;{\rm \iff}\;
m T^{jk} g_{ijk}=0
\la{E5SPACEfull}
\ee
So the effect of the $\P_5$ in (\ref{dfeqyatuibn})   is to ensure that
the equation (\ref{dfeqyatuibn}) is projected onto the subspace defined by
\be
T^{jk} g_{ijk}=0
\la{E5SPACE}
\ee
In practice, in a given explicit model like the CSSM, it is easy to implement this.  One simply writes down the general form of the space defined by 
(\ref{E5SPACEfull}), and then looks at the part of the equation 
(\ref{dfeqyatuibn}) which survives in this space.
This is much clearer with an example and a non-symmetric notation.

\subsection{The Operator $d_6$ and the  Space $E_7=\ker d_6 \cap \ker d_6^{\dag} \cap E_6$}

Similarly we have:
\be
d_6= \P_6 ( C \x {\ov C}^{\dag} ) 
\lt \{
 \int d^4 x \;   {\ov g}^{ijk} m {\ov v}_j {\ov A}_k
 \fr{\d}{\d A^{i }} 
\rt \}\P_6 + *
\la{defofd6}
\ee
This arises in the superfield approach by again using the reasoning in (\ref{funnyrouteagain}).

This acts to map the `separated solutions' of the following spaces into each other:
\be
{\cal S}_{3}    \stackrel{d_6}{\lra} {\cal R}_{4}  
\ee
Again, these become more meaningful in a specific model like the CSSM.  The analysis here is similar to those given above.\

\subsection{The Operator $d_7$ and the  Space $E_8=\ker d_7 \cap \ker d_7^{\dag} \cap E_7$} 

\la{dsevensubsection}

At the next level it is rather tricky to find a further differential.  However, after a search, we find that there is a non-zero differential of the following form (for the physical formulation):
\be
d_7 = 
\P_7 \d_5 \fr{d_3^{\dag}}{\D_3} \d_5\P_7
\approx
\P_7 m^2  \og g \og \A C_{\a} \A^{\dag} \y_{\a}^{\dag}
\P_7
+ *
\la{whatadiff}\ee
where
\be
\d_5 \approx m \og^{ijk} \ovv_j \A_k C_{\a} \y^{i \dag}_{\a}
\ee
\be
\d_3 \approx  \og^{ijk} \A_j \A_k C_{\a} \y^{i \dag}_{\a}\ee
One sees here the advantage of starting this problem with the physical formulation, because this operator in the superfield formulation is quite a complicated thing, of a form similar to (\ref{funnyroute}) and (\ref{funnyroute5}):
\be
d_7 = \P_7 
\lt ( \d_4 \fr{d_2^{\dag}}{\D_2} \d_4 \fr{\d_0^{\dag}}{\D_0} \d_1
\rt ) \P_7 +*
\la{dsevenforsups3}
\eb
\approx
\P_7\lt ( g m A \Lambda^{\dag} \rt )
\lt ( \og \Lambda A^{\dag} A^{\dag}  \rt )
\lt ( g m A \Lambda^{\dag}  \rt )
\lt ( \Lambda {\ov F}^{\dag}  \rt )
\lt (   \ov F \oC \oy^{\dag}  \rt )
\P_7+*\eb
\approx
\P_7  \; g \og g \; m^2 
\; A \oC_{\dot\a} 
A^{\dag} 
\oy_{\dot\a}^{ \dag}  
\P_7+*\ee
If one started with the superfield formulation it might be harder to get the insight necessary to find the above form (\ref{dsevenforsups3}).

This differential yields the important mapping
\be
(C \x^2 C) \y_{\a} \A \stackrel{d_7}{\lra} 
(C \x^2 C)m^2  \A C_{\a}
\ee

More generally it acts to take
\be
{\cal R}_{n} \cap E_7 \stackrel{d_7}{\lra} {\cal R}_{n+1}  \cap E_7 \; n= 4,5\cdots 
\ee
and it eliminates various  objects from these spaces. 
Its effect is best understood in a detailed model like the CSSM, and we discuss it in subsection 
\ref{d7ssmsubsection}.

\subsection{The Operator $d_8$ and the  Space $E_9
=\ker d_8 \cap \ker d_8^{\dag} \cap E_8$} 

\la{deightsubsection}

We find that there is a non-zero differential of the form:
\be
d_8 = 
\P_8 \d_5 \fr{d_3^{\dag}}{\D_3} \d_5
\fr{\d_0^{\dag}}{\D_0} \d_1 
\P_8
\la{whatadiff8}
\ee
This comes from
\be
\d_5 \approx m \og^{ijk} \ovv_j \A_k C_{\a} \y^{i \dag}_{\a}
\ee
\be
\d_3 \approx  \og^{ijk} \A_j \A_k C_{\a} \y^{i \dag}_{\a}\ee
\be
\d_0^{\dag}
\approx (\x C^{\dag} \oC^{\dag}) 
\eb
\d_1 
\approx  (\oC \oy \A^{\dag}) + (C \y A^{\dag})
\ee-
The superspace version arises easily once has found  $d_7$:
\be
d_8 = \P_8 
\lt ( \d_4 \fr{\d_2^{\dag}}{\D_2} \d_4 \fr{\d_0^{\dag}}{\D_0} \d_1\fr{\d_0^{\dag}}{\D_0} \d_1
\rt ) \P_8 +*
\la{dsevenforsups4}
\eb
\approx
\P_8\lt ( g m A \Lambda^{\dag} \rt )
\lt ( \og \Lambda A^{\dag} A^{\dag}  \rt )
\lt ( g m A \Lambda^{\dag}  \rt )
\lt ( \Lambda {\ov F}^{\dag}  \rt )
\lt (   \ov F \oC \oy^{\dag}  \rt )
(\x C^{\dag} \oC^{\dag}) 
 (\oC \oy \A^{\dag})
\P_8+*
\ee
Either way one gets something of the form\footnote{As usual we are showing only the first terms here, out of an infinite series with more derivatives.}
\be
d_8 \approx
\P_8  \lt ( m^2   g \og g  ( \oC \x C^{\dag} )A  \A^{\dag} A^{\dag}+ m^2  \og g \og ( C \x {\ov C}^{\dag} )\A  \A^{\dag} A^{\dag}\rt ) 
\P_8
+ *
\la{whatadiff28}\ee

This differential has the general effect
\be
(C \x \oC) A \A \stackrel{d_7}{\lra} 
\lt \{
(C \x^2 C)m^2  \A
+
(\oC \x^2 \oC)m^2  
A  
\rt \}
\ee

This acts to take
\be
{\cal S}_{3} \cap E_8 \stackrel{d_8}{\lra} {\cal R}_{4}  \cap E_8 
\ee
and it eliminates various  objects from ${\cal S}_{3} \cap E_8 $ and ${\cal R}_{4} \cap E_8 $. 
Again, this is best understood in a detailed model like the CSSM.

\subsubsection{The differentials $d_7$ and $d_8$ are not   antiderivations}

\la{notaderivation}

For all $d_r$ up to $r=6$, as regards the fields\footnote{There are some double destruction operators if one includes the ghosts. The symbol $[X]=1$ if $X$ is a fermion, and  
$[X]=0$ if $X$ is a boson.}, the differential  $d_r$ 
satisfies the `antiderivation' identity:
\be
d_r (X Y) = 
(d_r X) (Y) +  
(-1)^{[X]} ( X) (d_r Y) 
\ee
but this does not hold for the differential $d_7$, because it involves two field destruction operators $\A^{\dag} \y_{\a}^{\dag}$ 
in  (\ref{whatadiff}). The differential $d_8$ is similar.  This kind of feature is common for $d_r^{\dag}$, but   $d_r, r\leq 6$ are all derivations as regards the fields, because they have only one field destruction operator.    This means that one needs to be careful when applying  $d_7$ or $d_8$ to objects with more than two fields in them.  Objects with more than two fields  are common in (\ref{canonicalform3}) of course.  We shall not attempt to discuss  this further here.

\subsection{Unseparated Irregular Equations, and the End of the Spectral Sequence}

In the above, we have   discussed the regular and separated irregular sectors.  The unseparated irregular sectors are not solved here.
In many cases, particularly for the regular and separated irrregular sectors, it is possible to show that the spectral sequence collapses here at $E_9 = E_{\infty}$ for specific dimensions and index structures. 
This is discussed in  Appendix \ref{collapseappendix}.
  Noting, however, that 
the operator $d_7$ is not a derivation, and that it arises in a rather obscure way, makes it a somewhat daunting task to ensure that one has all the differentials for the general case.  For the time being, it seems sufficiently challenging to proceed to try to understand  the significance of the  cohomology that has already been found.  Furthermore, much of the cohomology relates to the gauge theory in various ways that require a paper on that subject.

\section{Collapse of the Spectral Sequences}

\la{collapseappendix}

\subsection{Collapse of the Spectral Sequence at $E_3$ for the  free massless stage }
\la{proofofcollpseappendix}

In this subsection we shall prove, for the  free massless stage , that 
\be
d_r =0 \;{\rm for} \; r \geq 3
\ee
so that
\be
E_3 = E_{\infty}\; { \rm for\; the \; free\;massless \;stage }
\ee

{\bf Proof}:
We have: 
\be
N_{\rm Zinn} E_r=0\;{\rm for} \; r \geq 1
\ee
Hence we have
\be
N_{\rm Ghost} \equiv
{N}_C 
+
{ {N}}_{\ov C}
+
  N_{\x}
\;{\rm for} \; r \geq 1
\ee
So for the free massless case, for $r \geq 1$, we have
\be
{N}_{\rm  Grading} 
= 
N_{\rm Ghost} 
+  N_{\x}
\la{someGrading}
\ee
Now since 
\be
\lt [N_{\rm Ghost} , \d_r \rt ] = \d_r
\ee
and
\be
\lt [N_{\rm Grading} , \d_r \rt ] = r \d_r
\la{meaningofgrading2}
\ee
and
\be
\lt [N_{\rm Grading} , d_r \rt ] = r d_r
\ee

It follows that
\be
\lt [N_{\x} , d_r \rt ] = 
(r-1) d_r \;{\rm for} \; r \geq 1
\ee

So for r=4 we need
\be
\lt [N_{\x} , d_r \rt ] = 
3 d_r  
\ee
and this is obviously impossible because there are no terms with $N_{\x} = 3$, given the result for the form of $E_2$, so 
\be
d_r =0 \;{\rm for} \; r \geq 4
\ee
Hence
\be
E_4 = E_{\infty}
\ee
for this free case.

So for r=3 we need to prove that
\be
\lt [N_{\x} , d_3 \rt ] = 
2 d_3   
\ee
does not happen for the free massless theory.  For this we need to look more carefully, since there are terms with $N_{\x} = 2$ in $E_2$.

The differential $ d_3$ can only link pieces of $E_3$ that differ by two factors of $\x$ and ghost charge one.  We have proved that
\[
E_3 = {\cal P}_1 \oplus{\cal \oP}_1 \oplus{\cal Q}_2 \oplus{\cal \oQ}_2
 \]
\[
\oplus {\cal Q}_3 \oplus{\cal \oQ}_3 \oplus{\cal S}_3
\]
\[
 \oplus{\cal R}_4 \oplus{\cal \oR}_4  
\]
\be
\oplus \sum_{n=5}^{\infty}
\lt \{
{\cal R}_n \oplus{\cal \oR}_n  
\rt \}
\la{notcomplete}
\ee

Even though we do not know everything about the terms 
in (\ref{notcomplete}), we can easily verify by inspection that there are no terms that differ by ghost charge one and two factors of $\x$ here.  
The only terms\footnote{In subsection \ref{p1andq2subsection} we noted that we have not succeeded in eliminating the possibility that there are terms $ {\cal P}_1 \oplus{\cal \oP}_1 $  (and $ {\cal Q}_2 \oplus{\cal \oQ}_2 $)  in $E_{\infty}$ for the massless free theory.} with $N_{\x}=0$ are $ {\cal P}_1 \oplus{\cal \oP}_1 $, and these have $N_{\rm Ghost}= -3$, whereas the only terms with $N_{\x}=2$ are $ {\cal R}_n \oplus{\cal \oR}_n $, and these have $N_{\rm Ghost}\geq 0$. 
So the spectral sequence collapses here and we have, for the free massless case:
\be
E_3 = E_{\infty}
\ee

We will not attempt to prove collapse for the interacting or massive cases, because we do not even know the form of the solutions for the unseparated irregular equations. My conjecture is that the interacting case collapses at $E_5$ and that the massive case collapses at $E_9$, in accord with Table  
\ref{summaryofdiffs}.

\subsection{Envoi}

   Although progress has been made, there is still work to be done, and interesting results to be derived, and possible errors to discover.  The irregular unseparated sectors   need to be finished.

Also it should be noted that some of the interesting phenomena do not require a complicated analysis to show that the spectral sequence collapses.  For example, we found in subsection \ref{d7ssmsubsection}
 that the simplest dotspinor Quarks and Leptons get  removed from $E_8$ by $d_7$.  So there cannot be higher $d_r$ for this sector, because there is nothing left for it to work on.  The interesting phenomenon for this sector  is not what is left in the cohomology space--it is the removal of these operators from the cohomology space when the internal symmetry breaks.  That means  that when these operators are coupled to external sources, the symmetry breaking changes the SUSY realization by mixing the effective dotspinors with elementary Quarks and Leptons.

\tableofcontents
{\tiny \articlenumber}
\end{document}